\documentclass[11pt,a4paper]{article}

\usepackage{amsthm,amsmath,multirow,graphicx,subfigure,booktabs,url,mathtools,wrapfig,mathrsfs,bm,relsize,caption,xcolor, enumitem, fullpage,algorithm}

\RequirePackage[authoryear]{natbib}
\RequirePackage[psamsfonts]{amssymb}
\usepackage[colorlinks=true,
linkcolor=blue,
urlcolor=blue,
citecolor=blue]{hyperref}

\usepackage[capitalize]{cleveref}

\usepackage{preamble}
\usepackage{algorithm,algpseudocode}

\numberwithin{equation}{section}
\newtheorem{conj}{Conjecture}
\newtheorem{thm}[conj]{Theorem}
\newtheorem{cor}[conj]{Corollary}
\newtheorem{prop}[conj]{Proposition}
\newtheorem{lemma}[conj]{Lemma}
\newtheorem{ass}{Assumption}

\newtheorem{definition}{Definition}

\crefname{ass}{Assumption}{Assumptions}
\crefname{thm}{Theorem}{Theorems}

\newtheorem{innercustomgeneric}{\customgenericname}
\providecommand{\customgenericname}{}
\newcommand{\newcustomtheorem}[2]{%
  \newenvironment{#1}[1] % #1 = env name, #2 = display name
  {%
    \renewcommand\customgenericname{#2}%
    \renewcommand\theinnercustomgeneric{##1}%
    \innercustomgeneric
  }%
  {\endinnercustomgeneric}
}

\newcustomtheorem{customass}{Assumption}  
\theoremstyle{definition}
\newtheorem{remark}{Remark}

\def\marten{\color{red}}

\def\EE{\mathbb{E}}
\newcommand*{\dd}{\, {\rm d}}
\def\diag{{\rm diag}}

\def\Dt{\Delta}

\def\S{\mathcal S}

\def\marten{\color{red}}

\def\tr{\rm tr}
\def\e{{\mathbf e}}

\def\PP{\mathbb{P}}
\def\RR{\mathbb{R}}
\def\EE{\mathbb{E}}
\def\tr{{\rm tr}}
\def\op{{\rm op}}
\def\wh{\widehat}
\def\eps{\varepsilon}
\def\N{\mathcal N}
\def\E{\mathcal E}
\def\F{\mathcal F}
\def\cX{\mathcal X}
\def\T{\top}
\def\cB{\mathcal B}

\def\bbO{\mathbb O}
\def\Var{{\rm Var}}
\def\i{\infty} 
\def\bX{\mathbf X}
\def\Sp{\mathbb S}
\def\KL{{\rm KL}} 
\def\Dtmin{\Dt_{\min}}
\def\Dtmax{\Dt_{\max}}

\crefname{ass}{Assumption}{Assumptions}
\Crefname{ass}{Assumption}{Assumptions}
 
\newcommand{\norm}[1]{\left\lVert#1\right\rVert}
\newcommand{\abs}[1]{\left\lvert#1\right\rvert}
 
\newcommand{\EEpi}{\mathbb E_\pi}

\title{Near-Purity Screening and Dominant-Atom Estimation in Softmax Mixtures and Their Ensembles}

\usepackage{authblk}

\author[1,2]{Xin Bing}
\author[3]{Florentina Bunea}
\author[3]{Marten Wegkamp}
\author[2]{Weicheng Wu}

\affil[1]{Department of Statistics, University of Illinois Urbana-Champaign}
\affil[2]{Department of Statistical Sciences, University of Toronto}
\affil[3]{Department of Statistics and Data Science, Cornell University}
\begin{document}
	\maketitle

\begin{abstract}
We study near-purity screening and dominant-atom estimation
in softmax mixtures, whose components share a common set of
category features. A mixture is called $\varepsilon$-pure if
one of its mixing weights is at least $1-\varepsilon$.
We introduce the Trace of Latent Covariance Criterion
($\mathrm{TLC}^{2}$), a direct estimator of the trace of
the covariance matrix of the latent mixing measure,
constructed without estimating its atoms or weights, or
knowing the number of components. Unlike existing
moment-based approaches, our estimator is constructed
for the softmax family and does not require knowledge
or estimation of the feature distribution.

We establish finite-sample guarantees for $\mathrm{TLC}^{2}$,
showing that it achieves a parametric convergence rate up
to an additive approximation error of order $p^{-1/2}$,
without requiring minimum separation between mixture atoms.
We also study estimation of the dominant atom in near-pure
mixtures by fitting a single softmax component. We prove
that the resulting maximum likelihood estimator has an
error bound consisting of the usual parametric term and a
bias term governed by the weighted distances between the
dominant and remaining atoms, revealing adaptation to the
geometry of the mixture.

Finally, we extend near-purity screening and dominant-atom
estimation to ensembles of softmax mixtures with shared
atoms and mixture-specific weights. We apply $\mathrm{TLC}^{2}$
separately to each ensemble member to screen for near-purity
without knowing the number of components. Under additional
assumptions, clustering the resulting dominant-atom
estimators yields finite-sample guarantees for estimating
the shared atoms. An application to the Million Song Dataset illustrates how the estimated softmax directions reveal interpretable structure in the data.

\end{abstract}
\newpage 

 \tableofcontents

\section{Introduction}
%\section*{Softmax mixtures and near-purity screening}

A \textbf{softmax mixture} is a discrete mixture of softmax-parametrized probability vectors. 
%It provides a flexible model for categorical probabilities when categories have associated feature vectors whose interactions with latent sources determine their probabilities.
Such models arise naturally when observed categories have known attributes and their probabilities reflect the influence of several latent sources with different preferences or effects. Examples include discrete choice models, where categories represent alternatives described by their attributes and latent sources capture heterogeneous preferences \citep{mcfadden2000mixed}, and biological applications, where categories represent genes or genomic features and latent sources reflect distinct biological processes or populations \citep{GonzalezBlas2019}. Related softmax structures also arise in multi-head attention mechanisms underlying large language models: when token representations and other model parameters are held fixed, each attention head induces a softmax distribution over tokens, and convex combinations of these distributions yield softmax mixtures \citep{vaswani2017attention}. In these settings, the softmax parametrization captures interactions between latent sources and category features, while the mixing weights describe their relative contributions.

Formally, let $\pi\in\Delta^{p-1} := \{v\in [0,1]^p: \sum_{i=1}^p v_i = 1\}$. For each category $j\in[p]:=\{1,2,\ldots,p\}$, let $x_j\in\RR^L$ denote its feature vector. We  let $Y$ be a discrete random vector supported on $\mathcal X=\{x_1,\ldots,x_p\}$, with $\PP(Y=x_j)=\pi(x_j)$ for $j\in [p]$. Let $\theta_1,\ldots,\theta_K\in\RR^L$ be distinct latent parameter vectors, referred to as \emph{mixture atoms}, and let $\alpha=(\alpha_1,\ldots,\alpha_K)^\T\in\Delta^{K-1}$ be the mixing weights.

The softmax mixture model specifies the mixture components
\begin{align}\label{soft}
A_{\theta_k}(x_j):=\frac{\exp(x_j^\T\theta_k)}{\sum_{\ell=1}^p\exp(x_\ell^\T\theta_k)},\qquad k\in[K],
\end{align}
and the mixture distribution
\begin{align}\label{mod-def}
\PP(Y=x_j)= \pi(x_j) =\sum_{k=1}^K\alpha_k A_{\theta_k}(x_j),\qquad j\in[p].
\end{align}
The support points in $\cX$ are a realization of feature vectors $X_1,\ldots,X_p$ drawn i.i.d. from an unknown distribution $\mu$  on $\RR^L$ with density $f_\mu$  with respect to the Lebesgue measure. An equivalent model formulation is 
\[
\pi(x)=\int A_\theta(x)\,dG(\theta),\qquad x\in\mathcal X,
\]
where
\begin{equation}\label{G}
G:=\sum_{k=1}^K\alpha_k\delta_{\theta_k}
\end{equation}
is the mixing measure. We assume that a finite $K$ exists, but its value need not be known.

In many applications, it is useful to determine whether the observed mixture distribution is driven primarily by a single latent source, even when several sources may contribute to it. Such distributions admit a simpler, more interpretable representation, in which one softmax component accounts for most of the probability mass. Identifying these distributions can reveal dominant preferences, mechanisms, or sources of variation, and can facilitate subsequent estimation and interpretation of the corresponding latent parameters.
Our focus is on screening softmax mixtures for \textbf{near-purity}, formally defined as follows. Let $e_1,\ldots,e_K$ denote the canonical basis vectors of $\RR^K$.
\begin{definition}\label{hm}
For $0\leq\varepsilon<1/2$, the mixing measure $G$ is called \textbf{$\varepsilon$-pure} if
\[
\min_{k\in[K]}\|\alpha-e_k\|_1\leq 2\varepsilon.
\]
The corresponding distribution $\pi$ is called an \textbf{$\varepsilon$-pure mixture}. When $\varepsilon=0$, the mixing measure and mixture are called \textbf{exactly pure}.
\end{definition}
\noindent Writing $\alpha_{(1)}=\max_{k\in[K]}\alpha_k$, we have
\[
\min_{k\in[K]}\|\alpha-e_k\|_1=2(1-\alpha_{(1)}).
\]
Thus, a mixture is $\varepsilon$-pure if and only if $\alpha_{(1)}\geq1-\varepsilon$. We refer to the corresponding component $A_{\theta_{(1)}}$ as the \emph{dominant component} and the atom $\theta_{(1)}$ as the \emph{dominant atom}. An exactly pure mixture reduces to a single softmax component, while increasing $\varepsilon$ allows greater departures from purity.

We develop a new procedure for $\varepsilon$-purity screening based on observations $Y_1,\ldots,Y_N$ drawn i.i.d. from $\pi$. Our procedure requires neither estimation of the mixing measure $G$ nor knowledge of $K$.  We also study dominant-atom estimation in near-pure mixtures, and further study  near-purity screening and estimation in  ensembles of softmax mixtures. Section \ref{our} summarizes our contributions and the organization of the paper.

\subsection{Main contributions and related work}\label{our}
\paragraph{Screening for $\varepsilon$-purity.}
Section \ref{sec_sharp} introduces the Trace of Latent Covariance Criterion ($\mathrm{TLC}^{2}$), a criterion for near-purity screening based on direct estimation of the trace of the covariance matrix of the latent mixing measure $G$.

The case $\varepsilon=0$ corresponds to the exactly pure mixture. For univariate finite mixtures, the variance of the mixing distribution has long been used to characterize purity \citep{Usp37,Lin89}. Motivated by this observation, we consider its multivariate counterpart, the trace of the covariance matrix of $G$. Specifically, let
\[
\bar\theta:=\sum_{k=1}^K\alpha_k\,\theta_k,\qquad
\Sigma_G:=\sum_{k=1}^K\alpha_k\,\theta_k\theta_k^\T-\bar\theta\bar\theta^\T
\]
denote the mean and covariance matrix of $G$, henceforth called latent mean and latent covariance. We consider the latent functional
\begin{align}\label{def_S}
S(G)&:=\tr(\Sigma_G) =\sum_{k=1}^K\alpha_k\|\theta_k\|_2^2-\|\bar\theta\|_2^2. 
\end{align}
Proposition \ref{prop_score} of  Section \ref{sec_sharp} shows  that $S(G)$ serves as a suitable criterion for screening $\varepsilon$-pure softmax mixtures.

Before explaining how we make use of this criterion for near-purity screening,  we briefly mention, for completeness,  the related literature on testing homogeneity in finite mixtures,  which in our terminology is the problem of testing whether a mixture is exactly pure.  Methods include likelihood-ratio tests \citep{DacunhaCastelleGassiat1999,ChenChenKalbfleisch2001}, 
as well as score-based and likelihood-ratio approaches \citep{GuKoenkerVolgushev2018}.
Existing methods do not directly apply to the softmax mixture model considered here, at least not without substantial extensions. More importantly, homogeneity testing and $\varepsilon$-purity screening are distinct inferential problems. The former seeks to distinguish a single-component distribution from a nonhomogeneous mixture, whereas the latter seeks to identify mixtures whose mixing weights are sufficiently concentrated on one component. In particular, failure to reject homogeneity does not establish near-purity, and rejection of homogeneity does not preclude it either. 

We now return to $\mathrm{TLC}^{2}$, which is  the estimator $\widehat S$ of $S(G)$ used for near-purity screening. Our first main contribution is the construction of this estimator directly from the data, without estimating the atoms or weights of the mixing distribution $G$, or knowing $K$. Although inspired by classical moment matching, our approach differs fundamentally from it. We explain this distinction below, as it is central to the challenge of screening for near-purity in softmax mixtures. To appreciate the difficulty of this problem, consider the basic problem of latent mean estimation in a general finite mixture
\[
f(y)=\int f(y;\theta)\,dG(\theta).
\]
For Gaussian location mixtures with $\theta_k$ being the mean vector of the $k$-th component, the latent mean $\bar\theta=\sum_k\alpha_k\theta_k$ coincides with the observable mean $\EE_f(Y)$ and can therefore be estimated directly by the sample mean. More generally, direct moment-based estimation of $\bar\theta$ would require a function $g$ satisfying
\[
\EE_\theta[g(Y)]=\theta
\]
for every component parameter $\theta$, where $\EE_\theta$ denotes expectation under a mixture component $f(\cdot;\theta)$. Consequently, $\EE_f[g(Y)]=\sum_{k=1}^K\alpha_k\theta_k=\bar\theta$. Such identities, based on explicit expressions of $g$, are available in certain classical exponential families, including families with quadratic variance  functions; see \cite{Lin89} and \cite{morris1982natural}. They do not, however, extend directly to the softmax mixture model considered here. Indeed, \cite{BBNW} shows that, for the softmax mixture $\pi$ in (\ref{mod-def}), no function $g$ can satisfy $\EE_\pi[g(Y)]=\bar\theta$ for every configuration of support points $\mathcal X$. Nevertheless, the authors establish the existence of a function $h$ such that
\[
\EE_\pi[h(Y)]\to\bar\theta,\qquad p\to\infty,
\]
almost surely with respect to the feature distribution $\mu$. This construction of $h$ depends explicitly on the unknown density $f_\mu$, and therefore does not directly yield a data-driven estimator. The main contribution of Section \ref{sec_score_est} is to resolve this issue. 

The key intuition is developed in Section \ref{main-ideas}. To summarize it here, suppose temporarily that the feature density $f_\mu$ is known. Let $h_\mu:\RR^L\to\RR^L$ and $H_\mu:\RR^L\to\RR^{L\times L}$ be the functions defined in Theorem \ref{Riesz}, both depending on $f_\mu$.  
As shown in Section \ref{main-ideas},
\begin{equation}\label{app_intro}
 \sum_{j=1}^p \pi_j\, \tr(H_{\mu}(x_j)) - \biggl\|\sum_{j=1}^p \pi_j\, h_\mu(x_j)\biggr\|_2^2 ~ \to~  S(G),
\qquad p\to\infty,
\end{equation}
$\mu$-almost surely. This suggests estimating $S(G)$ by replacing $\pi$ with the vector of empirical category frequencies. Since the support points are observed, the resulting estimator is fully specified when $h_\mu$ and $H_\mu$ are known. Importantly, this construction estimates $S(G)$ directly, without estimating $G$ or specifying $K$. The convergence above describes the {\em approximation} of the target $S(G)$ as $p\to\infty$; sampling error is addressed separately.

Building on these ideas,  we develop an approach that requires neither knowledge nor estimation of $f_\mu$. In Section \ref{sec_score_est}, we show how to approximate the action of $h_\mu$ and $H_{\mu}$, at each support point in $\cX$, on the mixture components, leading to our proposed estimator
$\widehat S (G)$, defined in (\ref{finalS-est}). %(\ref{beta-hat}), and (\ref{gamma-hat}).

This direct approach contrasts with the plug-in strategy of first estimating $G$ by $\wh G$ and then using the plug-in estimator $S(\widehat G)$. Without a lower bound on the separation between mixture atoms, the minimax rate for estimating the mixing measure in Wasserstein distance can be slower than the parametric $N^{-1/2}$ rate, even when the number of atoms is bounded. See \cite{H-K18} for univariate mixtures, \cite{WuYan20} for univariate Gaussian location mixtures, and \cite{doss2023optimal} for multivariate Gaussian location mixtures. See also \cite{nguyen2026geometry} for a recent characterization of the dependence of estimation rates on component separation. Under additional separation conditions, EM-based procedures can attain parametric rates for estimating mixture parameters when $K$ is known; see \cite{BKL25} for Gaussian mixtures and \cite{BBNW} for softmax mixtures.

In contrast, direct estimation of $S(G)$ does not require estimating $G$, nor its individual atoms. As a scalar functional of their collective dispersion, $S(G)$ may be estimable without the separation conditions or knowledge of $K$ required for full parameter  recovery. Theorem \ref{thm_rate_score} confirms this intuition: under regularity assumptions on $\mu$, and  assuming that the atoms are distinct, but without imposing any minimum atom separation, our estimator achieves a parametric convergence rate, up to an additive term of order $p^{-1/2}$, arising from the approximation step in \eqref{app_intro}.
A modification of the general procedure of Section \ref{sec_score_est} can lead to improvements in the order of magnitude of the constants involved in this approximation error, and becomes  particularly relevant when $L$ is large, but the mixture atoms only span a sub-space in $\RR^L$ of much smaller dimension.   This modification is presented in Section \ref{sec_prac_SS}, and is based on the construction of an estimator  of the subspace generated  by the mixture atoms, a result that may be of independent interest. 

Finally, Section \ref{sec_method_screen} describes screening for $\varepsilon$-purity by thresholding $\mathrm{TLC}^{2}$, denoted by $\wh S$. The order of magnitude of the thresholding level follows from  the rate analysis in Theorem \ref{thm_rate_score}, while fully data-driven thresholding levels can be found by  a simple bootstrap procedure stated in \cref{sec_method_screen}.

%\emph{[To add: State the TLC$^2$ screening threshold and its theoretical guarantee. Maybe the subspace ? Uniform rates ? Say we solve linear programs ? ]}

\paragraph{Dominant atom estimation.}

Section \ref{sec_rate_atom} studies estimation of the dominant atom $\theta^*:= \theta_{k^*}$ in an $\varepsilon$-pure softmax mixture. Writing $\varepsilon=1-\alpha_{(1)}$, we have
\[
\pi=(1-\varepsilon)A_{\theta^*}+\varepsilon\, Q_\varepsilon,
\qquad
Q_\varepsilon:=\sum_{k\neq k^*}\frac{\alpha_k}{\varepsilon}A_{\theta_k},
\]
with the exactly pure case ($\eps =0$) understood separately. Rather than fitting the full mixture, we fit a single softmax component by cross-entropy minimization, equivalently maximum likelihood, and analyze the resulting estimator $\widehat\theta$ of $\theta^*$.

Our second contribution is a finite-sample analysis of this estimator. Informally, Theorem \ref{thm:softmax} establishes that, with probability at least $1-\delta$,
\[
\max\left\{
\|A_{\widehat\theta}-A_{\theta^*}\|_1,\,
\|\widehat\theta-\theta^*\|_2
\right\}
\lesssim
\sqrt{\frac{L\log(1/\delta)}{N}}
+
\sum_{k\neq k^*}\alpha_k\|\theta_k-\theta^*\|_2,
\]
where
$
\sum_{k\neq k^*}\alpha_k\|\theta_k-\theta^*\|_2\leq 2\varepsilon.
$
The implicit constants may depend on quantities related with the Fisher information matrix, as specified in Theorem \ref{thm:softmax}.
The bound consists of the usual parametric estimation term and a bias term induced by the remaining mixture components. The latter depends on the weighted distances between the dominant and non-dominant atoms, revealing that the MLE  {\em adapts} to the geometry of the mixture. In particular, when the atoms are sufficiently close, the bias is of parametric order, even when the impurity level $\eps$ is not negligible, subject to the conditions of Theorem \ref{thm:softmax}.

Dominant-atom estimation is related to parametric estimation under Huber's contamination model \citep{huber1964robust}, with the dominant component playing the role of the target distribution and the remaining mixture components acting as contamination. In our setting, however, the contamination is structured, as it consists of other softmax components. This structure allows the usual single-component MLE to achieve the stated guarantees without modifying the cross-entropy loss or estimating the remaining mixture components.

In the limiting case where all atoms coalesce, the mixture reduces to a single softmax component. The classical parametric rate, previously established asymptotically by \cite{Agresti}, is recovered in finite samples as a special case of our general Theorem \ref{thm_softmax_general} in Appendix \ref{sec_contami}, and its consequence, Theorem \ref{thm:softmax}. The precise results are given in Section \ref{sec_rate_atom}.

\paragraph{Extension to softmax mixture ensembles}

Section \ref{sec_ensemble} extends our results to ensembles of softmax mixtures that share a common set of atoms but have mixture-specific weights, when data consist in a collection of samples from these mixtures.   The ensemble procedure has two steps.  The  first  step applies $\mathrm{TLC}^{2}$ to each ensemble mixture to determine  those that are near-pure; we use E-$\mathrm{TLC}^2$ to refer to the ensemble screening step.   The second step  estimates  dominant atoms of the detected near-pure mixtures. 
%by fitting a single softmax component to each selected sample.
%It can be followed by a dominant atom estimation step, as explained below. 
%then estimates their dominant atoms by fitting a single softmax component to each selected sample.

The screening step requires neither knowledge of $K$ nor additional structural assumptions on the ensemble. Moreover, the quantities used to construct $\mathrm{TLC}^{2}$ depend only on the common support points and need to be computed once for the entire ensemble. Thus, applying $\mathrm{TLC}^{2}$ to multiple mixtures incurs only a modest additional computational cost.

The $\varepsilon$-purity detection step can be followed by a dominant atom estimation step, performed as in Section \ref{sec_rate_atom}, if near-purity is detected. Since  the ensemble can contain several mixtures concentrated on the same atom,  grouping of dominant-atom estimators requires now specifying a value of $K$. Under Assumption \ref{ass_quasi}, we establish finite-sample guarantees for estimating the shared atoms by clustering the dominant-atom estimators obtained from the selected mixtures. These guarantees build on our single-mixture estimation results. The precise procedure and its theoretical guarantees are presented in Section \ref{sec_ensemble}.

\paragraph{Simulation studies and a real-data analysis.}
Simulation studies in \cref{sec_numeric} examine estimation error of TLC$^2$, dominant-atom estimation, and near-pure mixture screening in single and ensemble softmax mixtures, together with the finite-sample performance of the bootstrap confidence intervals. Additional experiments in \cref{app_sim} illustrate the benefits of subspace assistance and compare E-$\mathrm{TLC}^2$ with existing screening methods. 

In \cref{sec_real_data}, we apply the procedure to the Million Song Dataset and its associated user listening records, identifying near-pure users and three interpretable listening profiles: indie/folk, electronic/dance, and pop/R\&B/rap. Representative songs and distinct audio-feature patterns support these interpretations.\\

\noindent {\em Notation.} For a vector $v$, $\|v\|_q$ denotes its $\ell_q$-norm.
For matrices, $\|\cdot\|_\infty$, $\|\cdot\|_{\infty,2}$, and
$\|\cdot\|_{\op}$ denote the maximum absolute entry, the maximum
rowwise $\ell_2$-norm, and the operator norm, respectively.
For a symmetric $d\times d$ matrix $M$, we order its eigenvalues as
$\lambda_1(M)\ge\cdots\ge\lambda_d(M)$ and write  $M\succ0$  when $M$ is  positive definite.
We write $a\vee b=\max\{a,b\}$ and $a\wedge b=\min\{a,b\}$.
For nonnegative quantities, $a\lesssim b$ means $a\le Cb$ for a
positive constant whose dependence is specified in context, and
$a\asymp b$ means both $a\lesssim b$ and $b\lesssim a$.
Generic constants $c,C>0$ may vary from line to line.

\section{Screening for near-purity in softmax mixtures}\label{sec_sharp}

This section introduces  a procedure for detecting whether a $K$-softmax mixture \eqref{mod-def} is $\varepsilon$-pure. We begin by showing, in \cref{prop_score}, that the $\varepsilon$-purity of a finite measure $G$ in \eqref{G} can be characterized by the functional $S(G)$ in \eqref{def_S}.  This motivates performing $\varepsilon$-purity detection  by comparing an estimator $\wh S $ of $S(G)$ with an appropriate threshold. 

The key challenge in applying this natural idea is to estimate $S(G)$ from observations of the softmax mixture. In \cref{main-ideas}, we review related results on moment-matching-based estimation and highlight the challenges specific to softmax mixtures. In \cref{sec_score_est}, we show how these challenges can be overcome, and introduce our estimator, $\mathrm{TLC}^{2}$.

Let $G$ be a  mixing measure supported on $\theta_1,\ldots,\theta_K$ as in \eqref{G}. It is clear that $S(G) = 0$ if $K=1$. We focus on $K \ge 2$ throughout the rest of the paper. Define its smallest and largest squared atom separations as
\begin{equation}\label{def_Delta} 
     \Delta_{\min} :=
    \min_{k\ne \ell}\|\theta_k-\theta_\ell\|_2^2,
    \qquad
    \Delta_{\max} :=
     \max_{k\ne \ell}\|\theta_k-\theta_\ell\|_2^2.
\end{equation}
The following proposition shows that $S(G)$ in \eqref{def_S} can be used as a criterion for $\eps$-purity of $G$. Its proof is stated in  \cref{Propscore}. Recall \cref{hm}.
\begin{prop}\label{prop_score} 
 %, with $\Delta_{\min}$ given in (\ref{def_Delta}), and let  $\Dt_{\max}  = \max_{k\ne \ell}   \| \theta_k-\theta_\ell\|^2.$ 
    The following results hold for any $0\le \varepsilon<1/2$. 
   \begin{enumerate}[itemsep=0mm]
     \item If $S(G) \leq  \varepsilon \Delta_{\min}/2$, then $G$ is $\eps$-pure. %\min_{k\in[K]} \|\alpha-e_k\|_1 \le 2\varepsilon$ .
    \item  If $G$ is $\eps$-pure, then $S(G) \leq \varepsilon \Delta_{\max}$. % for any $0\le \varepsilon\le 1$.
      \item  Consequently, if $\Dt_{\min}>0$, then $ S(G)=0$  if and only if $G$ is exactly pure. % $\min_{k\in[K]} \|\alpha-e_k\|_1 = 0$.  
     \end{enumerate}   
\end{prop}    
The first part of the proposition suggests using $S(G)$ as a diagnostic for near-purity, while the first two jointly show that, for approximately equally separated mixtures ($\Delta_{\min}\approx\Delta_{\max}$), a small value of $S(G)$ is essentially equivalent to near-purity.

The final statement provides a necessary and sufficient condition for exact purity, particularly transparent when $L=1$. Indeed, for a random vector $Z\sim G$, we have $S(G)=\sum_{\ell=1}^L\Var(Z_\ell)$. When $L=1$, this reduces to $S(G)=\Var(Z)=0$ if and only if $Z$ is supported on a single atom in ${\theta_1,\ldots,\theta_K}$, with mixture weight one. The extension to multi-dimensions and to $\varepsilon$-purity for $\varepsilon>0$ follows from two simple but powerful lemmas (\cref{lem_iff,lem_norms}), stated and proved in \cref{Propscore}.

%As mentioned in the Introduction, this criterion is new.  It  can be viewed as a generalization  to high dimensions of the ideas first introduced by Uspensky (1937), and revisited  in  Lindsay 1989, {\color{red} Add to references}   who advocated the use of determinants of  latent moment matrices  for determining the number of components in a mixture. 

Proposition \ref{prop_score}  depends only on the mixing measure $G$,  and not on the choice of the  parametric family (here, the softmax model $A_\theta$) in the mixing distribution $\pi$ in \eqref{mod-def}. The estimation of $S(G)$ in the softmax mixture setting is the subject of the following sections.

%and  \begin{equation}\label{def_Delta_max} 
%		 \Delta_{\max} := %\Delta_{\max}(\theta_1,\ldots,\theta_K) := 
   %      \max_{k\ne \ell}\|\theta_k-\theta_\ell\|_2^2.
  % \end{equation} 
%\begin{ass}\label{Delta-poz}
%    The minimal squared Euclidean distance between the mixture atoms is strictly positive, 
%\begin{equation}\label{def_Delta} 
%		 \Delta_{\min} :=
      %   \min_{k\ne \ell}\|\theta_k-\theta_\ell\|_2^2 > 0.
  % \end{equation}
 %  \end{ass}
 % This is a very general assumption, stating only  that the mixture has   $K$ distinct atoms, for a value $K$ that does not have to be known.   We do not a-priori assume any other  lower bound on $\Delta_{\min}$.  We also define the maximal squared separation 
 % \begin{equation}\label{def_Delta_max} 
%		 \Delta_{\max} := %\Delta_{\max}(\theta_1,\ldots,\theta_K) := 
  %       \max_{k\ne \ell}\|\theta_k-\theta_\ell\|_2^2.
  % \end{equation} 

\subsection{Preliminaries}\label{main-ideas}
 In this subsection we explain the idea behind the estimation of $S(G)$. 
 Recall that the support points of the softmax mixture $\pi$ given by (\ref{mod-def}) are a  sample from a measure $\mu$ on $\RR^L$.  Theorem \ref{Riesz}, established in \citet[Proposition 1]{BBNW}, and reproduced below for clarity of exposition, provides  a key step in the construction of an estimator of  $S(G)$,   when $\mu$ is {\it known}. Moreover, Theorem \ref{Riesz} motivates  the Trace of Latent Covariance Criterion ($\mathrm{TLC}^{2}$)  given in Section \ref{sec_score_est}, when $\mu$ is {\it not known} and belongs to certain probability classes.

 We need the following modeling assumptions.  Denote  the unit ball in $\RR^L$ by 
\[
    \cB:= B_L(0,1) :=\left\{ u\in\mathbb{R}^L:\|u\|_2\le1 \right\}.
\]

\begin{ass}\label{ass:theta} 
The softmax mixture atoms $\theta_k$, $k\in[K]$, are distinct and belong to $\cB$.
\end{ass}
\noindent The choice of unit radius is made for clarity of exposition.
 The normalization of the atoms is standard in softmax-parametrized models  and prevents pathological spiking behavior  \citep{arora2016latent}.
 Note that  Assumption \ref{ass:theta} implies  $\Delta_{\min} > 0$.

The next assumption, used in \citet[Proposition 1]{BBNW},  imposes some regularity conditions on the density of $\mu$ as well as its partial derivatives. 
\begin{ass}\label{ass:mu}
The probability measure $\mu$ has a strictly positive density $f_\mu\in C^2(\RR^L)$ with finite moment generating function everywhere. Its density and all partial derivatives through order two decay super-exponentially at infinity.
%, that is,  for every \(a>0\), there is \(C_a<\infty\) such that
%\begin{equation}\label{eq:density-decay}
%\sup_{x\in\RR^L}e^{a\norm{x}_2}
%\bigl(f_\mu(x)+\norm{\nabla f_\mu(x)}_2+
%\norm{\nabla^2 f_\mu(x)}_F\bigr)\le C_a.
%\end{equation}
\end{ass}

\begin{thm}[Existence]\label{Riesz}
%Riesz representation
Under \cref{ass:theta,ass:mu}, there exist measurable mappings
\(h_\mu:\RR^L\to\RR^L\) and \(H_\mu:\RR^L\to\RR^{L\times L}\)  such that, for all \(\theta\in\cB\),
\begin{equation}\label{key-for-est-1}
\frac{\EE[H_\mu(X)\exp(X^\T\theta)]}{\EE[\exp(X^\T\theta)]}
=\theta\theta^\T,
\qquad
\frac{\EE[h_\mu(X)\exp(X^\T\theta)]}{\EE[\exp(X^\T\theta)]}
=\theta.
\end{equation}
The representers are given  by
\begin{equation}\label{eq:explicit-riesz}
h_\mu(x)=-\frac{\nabla f_\mu(x)}{f_\mu(x)},
\qquad
H_{\mu}(x)=\frac{\nabla^2 f_\mu(x)}{f_\mu(x)}.
\end{equation}
\end{thm}
% \begin{proof}
%     See \citet[Proposition 1]{BBNW} for the proof. 
% \end{proof} 
Theorem \ref{Riesz} is a constructive existence result, providing explicit functions of the form (\ref{eq:explicit-riesz}) that satisfy (\ref{key-for-est-1}). In \cite{BBNW}, this result was a key ingredient in a method-of-moments procedure for estimating the softmax mixture atoms. Its implications extend beyond atom estimation: in particular, it motivates the estimation of the latent functionals defining $S(G)$ in (\ref{def_S}), as we explain below.

For future reference, we write 
\begin{equation}\label{one-point}
\begin{aligned}
    \beta_{\mu} &~ :=~  (\tr(H_{\mu}(x_1)), \ldots, \tr(H_{\mu}(x_p)))^\T\in \RR^p,\\
\Gamma_{\mu} &~ :=~  (h_{\mu}(x_1), \ldots, h_{\mu}(x_p))^\T \in \RR^{p\times L}.
\end{aligned}
\end{equation}
For any $\theta\in \cB$, recall $A_{\theta} :=  (A_{\theta}(x_1), \ldots, A_{\theta}(x_p))^\T\in \Delta^{p-1}$ with $A_{\theta}(x_j)$ given in (\ref{soft}). By  (\ref{key-for-est-1}), the strong law of large numbers  ensures that,  $\mu$-almost surely as $p\to \i$,
 \begin{equation}\label{key-for-our-est}
 \beta_\mu^\T A_\theta =  \sum_{j=1}^p    \tr\left(H_{\mu}(x_j) \right)A_{\theta}(x_j)  
   ~ \rightarrow ~  
   \| \theta \|_2^2  ,  \quad  \quad  %\sum_{j=1}^p
  \Gamma_\mu^\T A_\theta = \sum_{j=1}^p   h_{\mu}(x_j) A_{\theta}(x_j)  ~ \rightarrow ~   \theta.
\end{equation} 
Therefore, by recalling $\pi$ from \eqref{mod-def}, using \eqref{key-for-our-est} for $\theta = \theta_k$ gives that
\begin{equation}
\begin{aligned}
    \beta_\mu^\T \pi  -  \|\Gamma_\mu^\T \pi\|_2^2 &~ =~  \sum_{k=1}^K  \alpha_k \beta_\mu^\T A_{\theta_k} - \Bigl\|\sum_{k = 1}^{K} \alpha_k   \Gamma_\mu^\T A_{\theta_k} \Bigr\|_2^2 ~ := ~ S_1(\beta_{\mu}) - S_2(\Gamma_{\mu}) \\
    &~ \rightarrow ~  \sum_{k = 1}^{K}\alpha_k\|\theta_k\|_2^2 - \| \bar{\theta} \|^2_2 ~ \overset{\eqref{def_S}}{=} ~  S(G),
\end{aligned}\label{firstS-est}
\end{equation} 
$\mu$-almost surely  as $p\to\infty$. 
For samples $Y_1,\ldots,Y_N$ i.i.d. from $\pi$, let $\wh \pi = (\wh \pi_1,\ldots,\wh \pi_p)^\T$ with 
\begin{equation}\label{pi-hat}
    \wh \pi_j = \frac{1}{N}\sum_{i = 1}^{N} 1(Y_i = x_j)
\end{equation}
denote the empirical frequencies of each $x_j$.  \cref{firstS-est} justifies estimating $S(G)$ by the following estimator, when $\mu$ is {\it known}, and hence when $h_{\mu}$ and $H_{\mu}$ are {\it known},
\begin{equation}\label{secondS-est}
    \wh S(\beta_\mu,\Gamma_\mu) ~ =~   \wh S_1(\beta_\mu) -  \wh S_2(\Gamma_\mu)
\end{equation}
where 
\begin{equation}\label{def_S1S2_hat} 
\wh S_1(\beta_\mu):= \beta_\mu^\T \wh\pi,\qquad 
\wh S_2(\Gamma_\mu)
:=\bigl\|\Gamma_\mu^\T\wh\pi\bigr\|_2^2
-\frac1{N-1}
\tr\!\left[
\Gamma_\mu^\T \Sigma_{\wh\pi} 
\Gamma_\mu
\right]. 
\end{equation} 
Here and throughout for any vector $v\in \Delta^{p-1}$, we write $\Sigma_v := \diag(v)-v v^\T$. 
 The second term in $\wh S_2(\Gamma_\mu)$ corrects the finite-sample bias
induced by $\|\Gamma_\mu^\T\wh\pi\|_2^2$. Indeed, by \eqref{pi-hat}, we know $\EE_{\pi}[\wh \pi]=\pi$ and $\Cov_{\pi}(\wh \pi)= \Sigma_\pi/N$, so that  
\[
\EE_{\pi}\left[\wh S_2(\Gamma_\mu)\right] 
=
\|\Gamma_\mu^\T  \pi\|_2^2
+
{1\over N}    \tr\left[ \Gamma_\mu^\T\Sigma_\pi\Gamma_\mu \right] - {1 \over N}
\tr\left[ \Gamma_\mu^\T\Sigma_\pi\Gamma_\mu\right] =  \bigl\|\Gamma_\mu^\T \pi\bigr\|_2^2.
\]
The second term in the middle is an upward bias of order $N^{-1}$ induced by the naive plug-in estimator. The correction in $\wh S_2(\Gamma_\mu)$ removes this bias and improves the finite-sample performance of   $\wh S(\beta_\mu,\Gamma_\mu)$.

Combining \eqref{firstS-est} with \eqref{secondS-est}, the first challenge mentioned in the Introduction can be addressed: although one cannot construct $h$ and $H$ such that \eqref{firstS-est} holds with equality for all instances of $\{x_1,\ldots,x_p\}$ \citep[Remark 8]{BBNW},  Theorem \ref{Riesz} provides a way to approximate $S(G)$ $\mu$-almost surely, as in \eqref{firstS-est}, which further leads to the estimator $\wh S(\beta_\mu,\Gamma_\mu)$ in (\ref{secondS-est}). Nevertheless, the expression for $\wh S(\beta_\mu,\Gamma_\mu)$ depends on those of $h_\mu$ and $H_\mu$, which, in turn, by \eqref{eq:explicit-riesz} of Theorem \ref{Riesz}, require exact knowledge of $f_\mu$. This leads to the second challenge in estimating $S(G)$ when $\mu$ is unknown.

%Consequently,  $S(G)$ is the  $\mu$-almost sure limit of the left hand side of (\ref{firstS-est}). 
%Finally, given the observed feature set $\mathcal{X}$, and access to  an i.i.d. sample $Y_1, \ldots, Y_N$ from the softmax mixture  $\pi$,  
%%Given $\mathcal X$ and the sample  $Y_1, \ldots, Y_N$, this
%and  motivates the estimator

%the  criterion  $S(G)$ can be estimated by
%\begin{equation}\label{secondS-est}
%\widetilde  S (G) := \sum_{j=1}^p   \tr\left(H_{\mu}(x_j) \right)  \wh \pi_j -  \left[ \sum_{j=1}^p  h_{\mu}(x_j) \wh \pi_j  \right ]^\T \left[  \sum_{j=1}^p  h_{\mu}(x_j) \wh  \pi_j\right].
% \end{equation}
%Although   \cref{Riesz} guarantees the existence of $H$ and $h$, 
%they do not admit closed-form expressions in general. Under the additional assumption that $\mu\in C^2(\cB)$, with  $\mu$ and $\nabla\mu$ vanishing on the boundary  of $\cB$, \cref{thm:mu-smooth} %\cite{BBNW} 
%in \cref{app_sec_expressions_h_H} provides  explicit expressions 
%\begin{equation}\label{def_h_H-1}
%		H_{\mu}(x) = {\nabla^2 \mu(x) \over \mu(x)} ,\qquad h_{\mu}(x) = -{\nabla \mu(x) \over \mu(x)},\qquad\forall~  x\in \cB.
%	\end{equation}
%    for $H$  and $h$ such that (\ref{key-for-est-1}) hold.   These closed-form expressions (\ref{def_h_H-1}) can be used to construct  an estimator (\ref{secondS-est}) whenever $\mu$ is known and satisfies \cref{ass:mu-2}.

    In principle, we could estimate $f_\mu$ from the observed sample $x_1,\ldots,x_p$ and obtain plug-in estimates of $H_\mu$ and $h_\mu$, but the resulting nonparametric estimation rates would lead to inefficient estimation of $S(G)$. In the next section, we show that the explicit construction of $H_\mu$ and $h_\mu$ can be bypassed entirely, while still yielding a fully data-driven estimator of $S(G)$ with provable guarantees and improved numerical performance.

%We can now make precise the heuristic argument from the beginning of the section, recalling the definition of the softmax component in (\ref{soft}) and softmax mixture (\ref{mod-def}).  For any given  $H_{\mu}$ and $h_{\mu}$ as in (\ref{key-for-est}),  and  from a sample $\mathcal{X}$ from $\mu$,  the strong law of large numbers yields
% \begin{equation}\label{key-for-our-est}
% \sum_{j\in [p]}  \left[  \tr\left(H_{\mu}(x_j) \right)A_{\theta}(x_j) \right]  
  % \rightarrow 
 %  \| \theta \|_2^2  \in \RR,  \quad  \quad  %\sum_{j=1}^p
 %  \sum_{j\in [p]} \left[  h_{\mu}(x_j) A_{\theta}(x_j) \right]  \rightarrow \theta \in \RR^L,  
% \end{equation}
%almost surely, as $p\to\infty$. 
%(\ref{H-func}), (\ref{thetabar})  and  thus 
%Consequently, 
%(\ref{firstS-est}) holds, with  the approximation understood as $\mu$-almost sure convergence, and   
%the  criterion  $S(G)$ can be estimated by
%\begin{equation}\label{secondS-est}
%\wh S_{\mu}(G) := \sum_{j\in[p]}  \wh \pi_j \tr [ H_{\mu}(x_j)]  -  \left[ \sum_{j \in [p]} \wh \pi_j h_{\mu}(x_j)  \right ]^\T \left[  \sum_{j\in [p]} \wh  \pi_jh_{\mu}(x_j) \right].
 %\end{equation}

%{\marten Instead of estimating of $\Sigma_Z$, perhaps we can say a little about estimation of the $\text{span}\left( \theta_1,\ldots,\theta_K \right)$.
%Recall the estimation of this span from our earlier paper in case $\mu$ is Gaussian, discuss the extension to elliptical densities. This serves as a bridge between known $\mu$ and unknown $\mu$.}

\subsection{The Trace of Latent Covariance Criterion ($\mathrm{TLC}^{2}$) for  softmax mixtures}\label{sec_score_est}

%We now develop an estimator of $S(G)$ that does not require explicit knowledge of the functions $h$ and $H$ from \cref{Riesz}.

The construction of $\mathrm{TLC}^{2}$ is guided by the identities in (\ref{key-for-est-1}), their implication (\ref{key-for-our-est})  and the estimator in (\ref{secondS-est})--\eqref{def_S1S2_hat}, but avoids the need to compute $h_{\mu}$ and $H_{\mu}$ themselves.

The key observation is that, in order to estimate $S(G)$, it suffices   to find a vector   $\beta \in \RR^p$ and a matrix $\Gamma  \in \RR^{p\times L}$
that approximate the action of $\beta_{\mu}$ and of $\Gamma_{\mu}$ on $A_{\theta}$ which led to (\ref{key-for-our-est}) above.
Formally, we seek $\beta$ and $\Gamma$ such that 
\begin{equation}\label{key-for-our-est-2}
\beta^\T A_{\theta} = \sum_{j=1}^p A_{\theta}(x_j)  \beta_j   \approx 
   \| \theta \|_2^2,  \quad  \quad  %\sum_{j=1}^p
   \Gamma^\T A_{\theta}  =\sum_{j=1}^p  A_{\theta}(x_j)\, \Gamma_{j\cdot}    \approx \theta,\qquad \forall~  \theta\in\cB.
 \end{equation} 
Under (\ref{key-for-our-est-2}), one can thus construct the estimator $\wh S(\beta,\Gamma) = \wh S_1(\beta) - \wh S_2(\Gamma)$ as in \eqref{secondS-est}. To motivate the  final estimator construction presented in this section, we focus on $\wh S_1(\beta) = \beta^\T \wh \pi$ that is intended to estimate $\sum_{k=1}^K \alpha_k \|\theta_k\|_2^2$. By \eqref{mod-def}, we can decompose the difference between $\wh S_1(\beta)$ and its estimand as
\begin{align}\label{score_err_decomp}\nonumber
    \Bigl|\wh S_1(\beta) - \sum_{k=1}^K \alpha_k \|\theta_k\|_2^2\Bigr|
    & ~=~ \Bigl| \beta^\T (\wh\pi- \pi) + \sum_{k=1}^K \alpha_k \left( \beta^\T A_{\theta_k} - \|\theta_k\|_2^2\right)\Bigr|\\
    & ~\le~ \cO_{\pi}\Bigl(\sqrt{\beta^\T \Sigma_\pi \beta \over N} + {\|\beta\|_\i\over N}\Bigr)
    + \max_{\theta\in \cB} \left| \beta^\T A_{\theta} - \|\theta\|_2^2\right|.
\end{align}
The first term represents the {\it estimation error} arising from using $\wh \pi$ to estimate $\pi$. It is easy to see that $\sqrt{\beta^\T \Sigma_\pi \beta}\le \|\beta\|_\i$ so that the estimation error is of order $\cO_\pi(\|\beta\|_\i/\sqrt N)$. The second term, on the other hand, represents the {\it approximation error} arising from the fact that we only have access to the observed support points $x_1,\ldots,x_p$ for  achieving \eqref{key-for-our-est-2}. Since we have no information about $\theta_1,\ldots,\theta_K$ beyond their membership in $\cB$, this error must be controlled uniformly over the entire set $\cB$. There is an inherent tradeoff between these two errors: reducing the approximation error may require a more irregular $\beta$ with a larger $\|\beta\|_\i$, whereas imposing more regularity on $\beta$ may result in poor control of the approximation error.

Motivated by this tradeoff, we propose the following estimator of $\beta$:
\begin{equation} \label{beta-hat}
\wh\beta
 ~ \in~ \argmin_{\beta\in\RR^p}\|\beta\|_\i 
\qquad\text{subject to}\quad
\max_{\theta\in\cN_{\epsilon}}\,\bigl|\beta^\T A_\theta-\norm\theta_2^2\bigr|\le\lambda_\beta,
\end{equation}
where $\lambda_\beta>0$ is a tuning parameter and $\cN_{\epsilon}$ is a fixed $\epsilon$-net of $\cB$ with respect to the $\ell_2$-norm for a given $\epsilon\in(0,1]$. Although the objective in \eqref{beta-hat} can be replaced by $\beta^\T \Sigma_{\wh\pi}\beta$ in view of \eqref{score_err_decomp}, the proposed formulation can be recast as a linear program and solved efficiently. Moreover, it does not require estimating $\pi$ and can be implemented using only $\cX$, which provides a significant advantage in the ensemble setting considered in \cref{sec_ensemble}. Finally, by a similar reasoning, we construct $\wh \Gamma=(\wh\gamma_1,\ldots,\wh\gamma_L)$ by solving
\begin{align}\label{gamma-hat}
\wh\gamma_\ell
&~ \in~ \argmin_{\gamma\in\RR^p}\|\gamma\|_\i 
\qquad\text{subject to}\quad
\max_{\theta\in\cN_{\epsilon}}\,\bigl|\gamma^\T A_\theta-\theta_\ell\bigr|\le\lambda_\gamma,
\qquad\ell\in[L].
\end{align}
This leads to  $\mathrm{TLC}^{2}$, given by 

\begin{equation}\label{finalS-est}
   \wh S  :=  \wh S( \wh \beta, \wh \Gamma) ~ =~   \wh S_1( \wh \beta) -  \wh S_2( \wh \Gamma),
\end{equation}
where $\wh S_1(\wh \beta)$ and   $\wh S_2(\wh \Gamma)$ follow the expressions given by (\ref{def_S1S2_hat}), relative to $\wh \beta$ and $\wh \Gamma$ given above.

%$
  %  \wh S := \wh S(\wh\beta,\wh\Gamma).
%$

A key ingredient of \eqref{beta-hat} is the choice of tuning
parameters $\lambda_\beta$ and $\lambda_\gamma$, which must be
sufficiently small while ensuring that the feasible sets in
\eqref{beta-hat}--\eqref{gamma-hat} are nonempty.
By Theorem~\ref{Riesz} and \eqref{firstS-est},
$\beta_\mu$ and
$\Gamma_\mu=(\gamma_{\mu,1},\ldots,\gamma_{\mu,L})$,
defined in \eqref{one-point}, are natural candidates for
feasible points. Their feasibility can be ensured, with high
$\mu$-probability, by choosing $\lambda_\beta$ and
$\lambda_\gamma$ according to the {\it uniform} convergence
rates of
\begin{equation}\label{aprox-error-I}
    \max_{\theta\in \cN_{\epsilon}}\,
    \bigl| \beta_\mu^\T A_{\theta} - \|\theta\|_2^2 \bigr|
    \quad \mbox{and} \quad
    \max_{\theta\in \cN_{\epsilon}}\,
    \bigl| \gamma_{\mu,\ell}^\T A_{\theta}
    - \theta_{\ell} \bigr|.
\end{equation}
These rates are established in
Proposition~\ref{prop:net-feasibility} of
Appendix~\ref{app_sec_proof_score_est}.
Obtaining such  uniform rates requires additional
tail conditions on $\mu$, along with restrictions ensuring
the existence of the necessary moments of $h_{\mu}(X)$
and $H_{\mu}(X)$, for $X\sim\mu$.
We state these conditions below.

\begin{ass} \label{ass:rate}
{\rm (a)} Assume $\EE [X] =0$ and there exists $\sigma_x<\infty$ such that $\EE[\exp(u^\T X)] \le\exp(\sigma_x^2\|u\|_2^2/2)$ for all $u\in \RR^L$. {\rm (b)} For finite constants $D_\gamma,D_\beta$ and nonnegative
integers $d,m$, that are independent of $p$, we assume that for all $x\in\RR^L$,
\begin{equation}\label{eq:poly-growth}
%\max_{\ell\in[L]}
\|h_{\mu}(x)\|_2\le D_\gamma(1+\norm{x}_2)^d,
\qquad
|\tr[H_{\mu}(x)]|\le D_\beta(1+\norm{x}_2)^m.
\end{equation}
\end{ass}
Part (a) of \cref{ass:rate}  assumes that $\mu$ has sub-Gaussian tail. Part (b) of
\cref{ass:rate} can be verified for several examples. For instance, when $\mu=\cN(0,\Sigma)$, \eqref{eq:poly-growth} holds with $d=1$ and $m=2$, where $D_\gamma$ and $D_\beta$ depend on $\Sigma$. For $\mu$ given by a finite Gaussian mixtures, we show in \cref{app:gaussian-location-mixture} that \eqref{eq:poly-growth} also holds with $d=1$ and $m=2$, where $D_\gamma$ and $D_\beta$ depend on the mean vectors and covariance matrices.  

For any $\delta\in(0,1)$, let $C_\beta$ and $C_\gamma$ denote quantities that depend on $L$, $1/\delta$, $1/\epsilon$, a higher-order term in $p$, and the fixed constants in \cref{ass:rate}. The following proposition establishes the rates of \eqref{aprox-error-I}, thereby leading to the following   order of  the tuning parameters   
\begin{equation}
\label{eq:informal-tuning}
 \lambda_\beta\asymp {C_\beta \over \sqrt{p}},\qquad  \lambda_\gamma ~ \asymp {C_\gamma \over \sqrt{p}}.
% \lambda_a
% \asymp
% C_a
% \left\{
% {1\over \sqrt p}
% +
% \frac{
% \exp\!\left[c\sigma_x\{\sqrt L+\sqrt{\log p}\}\right]
% }{p}
% \right\},
\end{equation}
%where $c\approx 3.59$ is a universal constant. 

\begin{prop}[Informal Statement]\label{prop:net-feasibility-informal}
Grant \cref{ass:theta,ass:mu,ass:rate}, and let $\lambda_\gamma$ and $\lambda_\beta$ satisfy (\ref{eq:informal-tuning}). 
Then  for any $\delta\in(0,1)$,  with $\mu$-probability at least $1-\delta$,
\begin{equation}
\label{eq:net-uniform-informal}
\sup_{\theta\in\cN_\epsilon}\abs{\beta_{\mu}^\T A_\theta-\norm\theta_2^2}
\le\lambda_\beta,\qquad
\max_{\ell\in[L]}\sup_{\theta\in\cN_\epsilon}\abs{\gamma_{\mu,\ell}^\T\, A_\theta-\theta_\ell}
\le\lambda_\gamma,
\end{equation}
whence the programs \eqref{beta-hat} and \eqref{gamma-hat} admit solutions.
\end{prop}

Proposition~\ref{prop:net-feasibility-informal} is an informal
version of Proposition~\ref{prop:net-feasibility}, stated
and proved in Appendix~\ref{app_sec_proof_score_est}.
The formal result specifies the constants in
\eqref{eq:informal-tuning}.

An immediate consequence of
\cref{prop:net-feasibility-informal} is that any solutions
$\wh\beta$ and $\wh\gamma_\ell$ to
\eqref{beta-hat}--\eqref{gamma-hat} have approximation
errors bounded by $\lambda_\beta$ and $\lambda_\gamma$,
respectively, and sup-norms no larger than those of the
feasible points $\beta_\mu$ and $\gamma_{\mu,\ell}$.
These guarantees are formalized in
Proposition~\ref{prop:net-feasibility}, stated and proved in
Appendix~\ref{app_sec_proof_score_est}, and further
motivate the choice of objectives in
\eqref{beta-hat}--\eqref{gamma-hat}.

Combining these approximation and norm bounds with the
corresponding estimation error bounds yields the main
result of this section, which establishes guarantees for  $\mathrm{TLC}^{2}$
 constructed from solutions to
\eqref{beta-hat}--\eqref{gamma-hat}, with tuning parameters
of the orders specified in \eqref{eq:informal-tuning}.
A fully explicit choice of the tuning parameters  is given in
\cref{thm_rate_score_detailed} of
\cref{app_sec_proof_score_est}, while practical choices
are discussed in \cref{sec_prac_lbd}.

% Next, we note that for {\it any other} $\beta$ and $\gamma_{\ell}$, $\ell \in [L]$ that satisfies (\ref{eq:informal-tuning}), the  difference between the  surrogate $Q$ introduced in  (\ref{aproxS}) and   $S(G)$,  for the given population level $\pi$, can be controlled in terms of the order of magnitude of $\lambda_{\ell}$ and $\lambda_{\beta}$ of order given by (\ref{eq:informal-tuning}).
% This difference reflects the approximation error of the surrogate. Ultimately, the SHARP criterion will  also  replace $\pi$ by $\wh \pi$ in the expression of $Q$, which adds estimation error to the final uncertainty in the procedure. 
% A finite sample analysis of  $|\wh S - S(G)|$ reveals that these types of  error interact, and that the overall error control of this difference requires supremum-norm control of $\beta$, $\gamma_l$. 

\begin{thm} 
\label{thm_rate_score}
Grant \cref{ass:theta,ass:mu,ass:rate}. For any $\epsilon\in (0,1)$, 
let $\cN_{\epsilon}$ be an $\epsilon$-net of $\cB$ with cardinality at most $(1+2/\epsilon)^L$ and construct $\wh S$ from
\eqref{beta-hat}--\eqref{gamma-hat} using tuning parameters of
the order in \eqref{eq:informal-tuning}.  
For  every $\delta\in(0,1)$, if $N\ge\log(Ce/\delta)$ and $p\ge C'(\sigma_x)$, then with
probability at least $1-\delta$,
\begin{equation}
\label{eq:general-hdcd-rate}
|\wh S-S(G)|
~ \le ~ 
 {C_1\over \sqrt N} + {C_2\over \sqrt p}
%+C_2\frac{\exp(c\sigma_x\sqrt L)}{p}
+C_3\,\epsilon.
\end{equation}
Here, the constants $C_1,C_2,C_3$ depend only on the fixed model constants in \cref{ass:rate}, $L$,  $1/\delta$, $1/\epsilon$ and higher order terms in $p$.
\end{thm} 

By choosing $\epsilon = 1/\sqrt{N\wedge p}$ henceforth, the bound in \eqref{eq:general-hdcd-rate} reduces to $1/\sqrt N+1/\sqrt p$, up to $C_1,C_2$ and $C_3$.  Here, the approximation error is controlled by $p$, whereas the estimation error is controlled by $N$.  Regarding the dependence on the dimension $L$, computationally, the above choice of $\epsilon$ leads to $|\cN_{\epsilon}|\le (1 + 2\sqrt{N\wedge p})^L$, which determines the number of constraints involved in solving \eqref{beta-hat}--\eqref{gamma-hat}. Statistically, \cref{thm_rate_score_detailed} in \cref{app_sec_proof_score_est} shows that $C_1$ depends polynomially on $L$, with the degree determined by $d$ and $m$ in \cref{ass:rate}, whereas the dependence of $C_2$ on $L$ is worse, with a leading term of the form $\exp(c\sigma_x\sqrt L)$ for some universal constant $c$. This leads to an unfavorable rate for large $L$ and arises from the second term in \eqref{score_err_decomp}, where the unknown $\theta_k$'s are replaced by a search over $\cB$.

This observation suggests a substantial improvement when the subspace spanned by the $\theta_k$'s has much smaller dimension than $L$, and can be estimated well. In particular, if this subspace has dimension $s\le K$ and is much smaller than $L$, then, when the subspace is known, we only need to perform the search over this subspace rather than over the $L$-dimensional $\cB$. This motivates the practical approach in \cref{sec_prac_SS}, with supporting theory given in \cref{app_sec_proof_SS}, where we first estimate the subspace and then solve for $\wh\beta$ and $\wh\Gamma$ in \eqref{beta-hat}--\eqref{gamma-hat}  within the estimated subspace. This reduces the leading exponential dependence in $C_2$ from $\exp(c\sigma_x\sqrt L)$ to $\exp(c\sigma_x\sqrt s)$ and, computationally, reduces the size of the $\epsilon$-net from order $(N\wedge p)^{L/2}$ to $(N\wedge p)^{s/2}$.

\subsection{Screening for near-purity with  $\mathrm{TLC}^{2}$}\label{sec_method_screen}
We now describe how to use  $\mathrm{TLC}^{2}$ to detect near-purity. 
Fix $\delta\in(0,1)$ and choose $t_0=t_0(\delta)$ sufficiently large,
as specified in \eqref{eq:general-hdcd-rate} of Theorem~\ref{thm_rate_score}, so that the event
$\cE_{t_0}:=\{|\wh S-S(G)|\le t_0\}$ holds with probability at least
$1-\delta$.
% $$
% t_0=C\left(\frac{1}{\sqrt p}+\frac{1}{\sqrt N}\right),
% $$
% where $C$ is sufficiently large so that the event $\cE_{t_0}:=\{|\wh S - S(G)| \le t_0\}$ holds with probability $1-\delta$, and hence may depend on $L$,
% the model parameters, and factors in $p$,
% $N$, and $1/\delta$, as specified in
% Theorem~\ref{thm_rate_score}.  
We declare a softmax mixture to be near-pure if its  $\mathrm{TLC}^{2}$   satisfies
$\wh S  \le    2 t_0.$  On the event $\cE_{t_0}$, if a mixture is declared near-pure, then  $S(G) \le \wh S + |\wh S-S(G)| \le 3 t_0$ so that Part 1 of \cref{prop_score} with $\eps = 6t_0 / \Dtmin$ further gives 
    \begin{equation}\label{typeI}
        \min_{k\in[K]} \bigl\|\alpha -e_k\bigr\|_1 ~ \le~     (12/\Dtmin)\, t_0
    \end{equation}
    provided that $t_0  < \Dt_{\min} / 12$. According to \cref{hm}, such  a mixture $\pi$ corresponds to an $\varepsilon$-pure mixture with $\eps  \le {6t_0  / \Dt_{\min}}$.  
    
    On the other hand, if a mixture is not declared near-pure, then $\wh S>2t_0$ on the event $\cE_{t_0}$, and 
    $
        S(G) \ge \wh S - |\wh S - S(G)| > t_0.
    $
    Thus, Part 2 of \cref{prop_score}, with $\eps=t_0/\Dt_{\max}$, yields  
    \begin{equation}\label{typeII}
        \min_{k\in[K]} \bigl\|\alpha -e_k\bigr\|_1   ~ > ~  (2/\Dt_{\max})\, t_0 
    \end{equation}
    meaning that the mixture $\pi$ is not $(t_0/\Dt_{\max})$-pure.

    In practice, one  may have a prespecified threshold $t_0$ reflecting the desired notion of near-purity. Whenever $t_0$ is sufficiently large so that $
    \PP\{\cE_{t_0}\}\ge 1-\delta,
    $
    the preceding argument provides finite-sample guarantees for the resulting  $\mathrm{TLC}^{2}$-based  detection criterion. In particular, \cref{thm_rate_score} provides sufficient conditions under which such a guarantee holds.\\
    
  When a fully data-driven choice of the threshold is desired, we propose the following bootstrap procedure. The basic idea is to use the empirical distribution of the bootstrap errors $\wh S^{(b)}-\wh S$ to approximate the sampling distribution of $\wh S-S(G)$, and thereby construct a confidence interval for the score $S(G)$.

Specifically, let $B\in\NN$ denote the number of bootstrap samples, and let $\wh\pi$ be as defined in \eqref{pi-hat}. For each $b\in[B]$, we resample $Y_1^{(b)},\ldots,Y_N^{(b)}$ independently from the discrete distribution $\wh\pi$ supported on $\cX$, and set
$
\wh\pi_j^{(b)}
=
N^{-1}\sum_{i=1}^N 1\{Y_i^{(b)}=x_j\} 
$
for $j\in[p]$.
We then compute the corresponding bootstrap  $\mathrm{TLC}^{2}$, denoted by  $\wh S^{(b)}$,  using $\wh\pi^{(b)}$, together with the same $\wh\beta$ and $\wh\Gamma$ used to compute $\wh S$. Let $\wh q_{1-\alpha}$ denote the empirical $(1-\alpha)$-quantile of $\{|\wh S^{(b)}-\wh S|\}_{b=1}^B$. A nominal $(1-\alpha)$ two-sided confidence interval for $S(G)$ is then given by
\begin{equation}\label{def_CI}
\bigl[
\wh S-\wh q_{1-\alpha},
\;
\wh S+\wh q_{1-\alpha}
\bigr].
\end{equation}

This bootstrap procedure is motivated by the standard central limit theorem
for $\sqrt{N}(\wh\pi-\pi)$ conditional on $\cX$. 
It captures the sampling variability of $\widehat S$ conditional on the observed support points $\mathcal X$, with $\widehat\beta$ and $\widehat\Gamma$ held fixed. However, it does not account for the approximation error that arises from the estimation of $S(G)$ using the finite set of support points and an $\epsilon$-net of the parameter space. This approximation error is controlled by the finite-sample analysis of Theorem~\ref{thm_rate_score_detailed}.
Thus, asymptotically valid bootstrap confidence intervals for $S(G)$ require additional conditions (on $N$, $p$ and $\epsilon$) ensuring that the approximation error is negligible relative to the sampling fluctuations. We do not pursue a formal bootstrap coverage result here, but investigate the finite-sample performance of the proposed intervals through simulations.

%Under suitable regularity conditions on the limiting distribution of $\wh S$, the bootstrap can be justified as a consistent approximation to its sampling distribution.
%Holding $\wh\beta$ and $\wh\Gamma$ fixed, the bootstrap procedure captures the sampling variability arising from the estimation of $\pi$. With the prescribed $\epsilon\asymp(N\wedge p)^{-1/2}$, the remaining discrepancy from  $S(G)$ is governed by the approximation error
%\begin{equation}\label{approx_bootstrap}
%\bigl|
%\wh\beta^\T\pi-\|\wh\Gamma^\top \pi\|_2^2-S(G)
%\bigr|
% = \cO(\lambda_\gamma+\lambda_\beta),
%\end{equation}
%which vanishes as $p\to\infty$. When this approximation error is negligible relative to the sampling variation, and the preceding bootstrap approximation holds in the asymptotic regime considered, the interval provides asymptotically valid inference for $S(G)$. 

Finally, the symmetric interval above is calibrated using absolute bootstrap deviations and therefore does not itself require symmetry of the limiting distribution of $\wh S$. Our simulations in \cref{sec_sim_one_doc} also suggest that it has better finite-sample performance than equal-tailed bootstrap intervals. 

\section{Estimation of the dominant atom and rates of convergence } \label{sec_rate_atom}

If the screening procedure  indicates that  the mixture  is  $\eps$-pure, we write $k^*\in[K]$ for the index of the dominant component  $A_{\theta^*}:=A_{\theta_{k^*}} \in \Delta^{p-1}$, defined in (\ref{soft}),  with weight $\alpha_{k^{*}}  := 1 -\eps$, so that the mixture can be written as
\begin{equation}\label{def_Q_eps}
\pi   =  (1 - \eps) A_{\theta^*} + \eps  ~Q_{\eps} \qquad \text{ with }\qquad  Q_\eps:=  \sum_{k \neq k^*}(\alpha_k/\eps) A_{\theta_k}.
\end{equation} 
 Our primary interest, in this section,  is to estimate the dominant mixture component $A_{\theta^*}$ and its corresponding parameter $\theta^*$, the dominant atom. 
%Since $\sum_{k \neq k^*} \alpha_k = \eps$, the contamination distribution $Q_\eps$ is itself a softmax mixture with at most $K-1$ mixture components. For such dominant mixture, our primary interest is to estimate the dominant mixture component $A_{\theta^*}$ and its corresponding parameter $\theta^*$. 

Recall that $\wh \pi$ is the empirical frequency vector based  on  $Y_1, \ldots, Y_N$ i.i.d. from $\pi$ in \eqref{def_Q_eps}. % from the $K$-component mixture $\pi$ in \eqref{mod-def},  for some finite, but unknown $K$.
We propose to estimate the dominant  atom via maximizing the log-likelihood of {\it one softmax component} or, equivalently, by computing the negative cross-entropy maximizer: 
\begin{equation}\label{def_theta_hat}
    \wh \theta ~ \in ~ \argmax_{\theta\in \cB} ~  \sum_{j=1}^p \wh \pi_j \log A_{\theta}(x_j),
\end{equation}
conditionally on the support points $\mathcal{X}$. As expected, this 
introduces a bias proportional to the $\eps$-mass assigned to the non-dominant components. 
Our main result of this section, stated in \cref{thm:softmax} below, gives non-asymptotic error bounds of  $\wh \theta$ and, in particular, characterizes the effect of various quantities such as $K$, $L$, $p$, $\eps$, on the error bounds. Our non-asymptotic analysis of this $M$-estimator under the softmax-parametrized model also covers the limit  case $\eps=0$.
%To state the result, we introduce the following notation. 
For any $\theta\in\RR^L$ and $X \sim \mu$, we let $H_{\theta}^{(\mu)}$ be the following counterpart of the  $L \times L$ Fisher information matrix of a single softmax component

\begin{equation}\label{def_H_mu}
    H_{\theta}^{(\mu)} := {\EE[XX^\T \exp(X^\T \theta)]\over \EE[\exp(X^\T \theta)]} - {\EE[X \exp(X^\T \theta)]\over \EE[\exp(X^\T \theta)]}{\EE[X  \exp(X^\T \theta)]\over \EE[\exp(X^\T \theta)]}^\T.
\end{equation} 
Denote the smallest eigenvalue of $H_{\theta^*}^{(\mu)}$ by
\begin{equation}
    \sigma_0^2 := \lambda_L( H_{\theta^*}^{(\mu)}).
\end{equation} 
Further recall $\sigma_x^2$ from \cref{ass:rate}.

 \begin{thm}\label{thm:softmax}
     Under model \eqref{def_Q_eps} with  \cref{ass:theta} and part {\rm(a)} of \cref{ass:rate}, assume $\sigma_0^2>0$. Let $ \delta, \eps \in (0,1/2)$ and assume for a sufficiently large constant $C>0$, depending only on $\sigma_x$ and $\sigma_0$ such that
    \begin{align}\label{cond_rate_atom}  
   p\ge 4K,\quad p \ge C (L+\log p)^2, \quad
        N   \ge C L(L+\log p) \log(1 /\delta), \quad 
       \eps \le   C^{-1} (L+\log p)^{-1}. 
    \end{align} 
    Then, with probability at least $1-\delta-2p^{-1}$, one has
    \[
    \max\!\left\{\|A_{\wh\theta}-A_{\theta^*}\|_1,
    \ \sigma_0\|\wh\theta-\theta^*\|_2\right\}
    \lesssim \sqrt{\frac{L\log(1/\delta)}{N}}
    +\frac{\sigma_x^2}{\sigma_0}\sum_{k\ne k^*}\alpha_k\norm{\theta_k-\theta^*}_2.
    \]
    Moreover, $\sum_{k\ne k^*}\alpha_k\norm{\theta_k-\theta^*}_2\le  \eps\sqrt{\Dtmax} \le 2\varepsilon$.
 \end{thm}  
 \begin{proof}
    See  \cref{app_sec_proof_thm_softmax}.  
\end{proof}

\begin{remark}[The MLE adapts to the geometry of the mixture]
The error bound in Theorem~\ref{thm:softmax} consists of two terms. The first is the usual parametric estimation error, of order $\sqrt{L/N}$, associated with fitting a single softmax component. The second captures the bias induced by the remaining mixture components and depends on their geometry relative to the dominant atom:
\[
\frac{\sigma_x^2}{\sigma_0}\sum_{k\ne k^*}\alpha_k\|\theta_k-\theta^*\|_2.
\]
Each contribution is small when either the corresponding mixing weight is small or the atom is close to $\theta^*$. Thus, the  MLE $\wh\theta$ adapts to the geometry of the mixture without estimating its non-dominant components.
In particular, if
\[
\max_{k\ne k^*}\|\theta_k-\theta^*\|_2^2
~ \leq~ \frac{\sigma_0^2}{ \sigma_x^4}{L\over N},
\]
the bias term is at most of parametric order, and the MLE retains the rate $\sqrt{L/N}$, provided the condition \eqref{cond_rate_atom} holds.% This includes the theorem's restriction on the impurity level $\varepsilon$.

\cref{thm:softmax} follows from the more general \cref{thm_softmax_general}, stated and proved in \cref{app_sec_rate_atom}. There, we first analyze single-component maximum likelihood estimation under
\[
\pi_Q=(1-\varepsilon)A_{\theta^*}+\varepsilon Q,
\qquad Q\in\Delta^{p-1},
\]
and then specialize to the structured contamination $Q=Q_\varepsilon$ defined in~\eqref{def_Q_eps}. The general analysis under arbitrary $Q$ yields a parametric sampling term, together with a contamination term proportional to $\varepsilon\|A_{\theta^*}-Q\|_1$ (with the constants and conditions specified in \cref{thm_softmax_general}). This formulation is related to the classical $\varepsilon$-contamination model of robust statistics \citep{huber1964robust}, although robust estimation under arbitrary contamination is not our objective.

In our setting, the contaminating distribution is itself a mixture of softmax components. This structure allows the perturbation of the population cross-entropy gradient at $\theta^*$ to be controlled by the weighted distances between the dominant and remaining atoms. Combined with the curvature of the softmax likelihood, this explains why the usual single-component MLE, without modifying the cross-entropy loss, achieves the geometrically adaptive bound of \cref{thm:softmax}.
\end{remark}

Finally, for consistency of $\wh\theta$ and $A_{\wh\theta}$, \cref{thm:softmax} indicates that one sufficient condition  is that  $\varepsilon$ vanishes as $N$ increases. Since all of our results are finite-sample bounds, this poses no difficulty: throughout the analysis, we may allow $\eps=\eps(N)\to0$ as $N\to\infty$. The assumption that $\sigma_0^2$ is bounded away from zero as $N\to \infty$ is mild and holds, for instance, when $\mu$ is a non-degenerate Gaussian distribution.

\section{Ensemble screening and atom estimation}\label{sec_ensemble}

%The methods developed in \cref{sec_sharp,sec_rate_atom}  can be easily extended to screen for purity and estimate dominant atoms when one  has access to data from multiple softmax mixtures that share a common set of mixture parameters while having different mixing weights. 

% An ensemble of softmax mixtures extends the single softmax mixture in \eqref{mod-def} as follows. Given a set of features $\cX$, for each $d\in[n]$, let
%\begin{equation}\label{model_ensemble}
%  \pi^{(d)}_j := \pi^{(d)}(x_j) :=     \sum_{k=1}^K \alpha_k^{(d)}A_{\theta_k}(x_j),\qquad \forall~ j\in [p],
%\end{equation}
%for some $K<\i$. The components $A_{\theta_k}$, and hence their atoms $\theta_k$, are common to all $n$ softmax mixtures, whereas the mixture weights $\alpha^{(d)}\in \Delta^{K-1}$ depend on $d$. For each $d\in[n]$, we observe $N_d$ i.i.d. samples $Y_1^{(d)},\ldots,Y_{N_d}^{(d)}$, taking values in $\cX$, from $\pi^{(d)}$.

%    The screening and dominant atom estimation procedures developed for one sample extend naturally to an ensemble. The extension has  two main steps.  Step 1:  compute $\mathrm{TLC}^{2}$ for each sample      $d \in [n]$ and identify which of these   samples are  generated from a near-pure  mixture. Step 2:   use  the set of the samples identified in Step 1 to estimate the corresponding dominant atoms, noting that there may be groups of near-pure mixtures in the ensemble that share the same dominant component. We expand on each of these steps below. \\ 

The methods developed in \cref{sec_sharp,sec_rate_atom} extend naturally to settings where data are available from multiple softmax mixtures sharing a common set of mixture components but having different mixing weights. In particular, these methods can be used to screen for near-pure mixtures and estimate their dominant atoms.

We consider an ensemble of softmax mixtures that generalizes the single-mixture model in \eqref{mod-def}. Given a set of features $\cX$, for each $d\in[n]$, let
\begin{equation}\label{model_ensemble}
  \pi^{(d)}_j := \pi^{(d)}(x_j) :=
    \sum_{k=1}^K \alpha_k^{(d)}A_{\theta_k}(x_j),
    \qquad \forall~ j\in [p],
\end{equation}
for some $K<\i$. The components $A_{\theta_k}$, and hence their atoms $\theta_k$, are shared across all $n$ mixtures, while the mixing weights $\alpha^{(d)}\in\Delta^{K-1}$ vary with $d$. For each $d\in[n]$, we observe $N_d$ i.i.d. samples $Y_1^{(d)},\ldots,Y_{N_d}^{(d)}$ taking values in $\cX$, drawn from $\pi^{(d)}$.

The screening and dominant atom estimation procedures developed for a single mixture extend to this ensemble setting in two steps. First, we compute $\mathrm{TLC}^{2}$ separately for each $d\in[n]$ and identify the mixtures that are near-pure. Second, we use the samples associated with these near-pure mixtures to estimate their dominant atoms, accounting for the possibility that several mixtures share the same dominant component. We expand on each of these steps below.

\noindent {\it Step 1.}  For  each mixture $d \in [n]$, let $G_d :=\sum_{k = 1}^{K}\alpha_k^{(d)}\delta_{\theta_{k}}$ be the mixing measure underlying the $d$-th mixture. Given the empirical frequency vector $\wh \pi^{(d)}$ of support points in $\cX$, and the sample size $N_d$, we estimate $S_d := S(G_d)$ using  $\mathrm{TLC}^{2}$ by  $\wh S_d := \wh S_{d,1} - \wh S_{d,2}$, where
 %   In step (1), detecting whether each document $d\in[n]$ corresponds to a dominant mixture amounts to determining whether $\min_{k\in [K]}\|\alpha^{(d)}-e_k\|_1$ is small. By \eqref{def_S} and \cref{prop_score} in \cref{criterion}, this is equivalent to evaluating the corresponding HDCD criterion
 %   $$ 
    %    S_d := S(G_d) = \tr(\Sigma_{G_d}) 
  %  $$
  %  with $G_d :=\sum_{k = 1}^{K}\alpha_k^{(d)}\delta_{\theta_{k}}$
 %   being the mixing measure of document $d$. Under \cref{ass:mu}, estimation of $S_d$ can be done by the HDCD procedure in  \eqref{def_S_hat} of \cref{sec_score_est}.  Specifically, for each $d\in[n]$, let $\wh\pi^{(d)}$ be the empirical frequencies of $Y_1^{(d)},\ldots, Y_{N_d}^{(d)}$, similar to \eqref{pi-hat}. We estimate $S_d$ by 
    \begin{equation}\label{def_S_hat_d}
          \wh S_{d,1}  =  \wh \beta^\T \wh \pi^{(d)},\qquad  \wh S_{d,2}  =   \bigl\|\wh \Gamma^\T \wh \pi^{(d)} \bigr\|_2^2   -{1\over N_d-1} \tr\Bigl[ \wh\Gamma^\T \Sigma_{\wh \pi^{(d)}} \wh \Gamma  \Bigr], 
    \end{equation}
    with $\wh\beta$ and $\wh\Gamma$ given by \eqref{beta-hat} and \eqref{gamma-hat}, respectively. 
    It is important to note  that $\wh\beta$ and $\wh\Gamma$ only need to be computed once, as they are only based on $\cX$, which are common to all the mixtures in the ensemble. Thus, evaluating   $\wh S_d$  for each sample $d\in [n]$ adds only a minor computational cost to that of evaluating  $\mathrm{TLC}^{2}$ from  only one sample.
    
%    To identify samples  with    a small  $\mathrm{TLC}^{2}$, let $t=(t_1,\ldots,t_n)^\T$ be a vector of threshold levels with 
%\begin{equation}\label{rate_delta_d}t_d = C  \sqrt{\log n}\left(1/\sqrt{p}+{1 / \sqrt{N_d}}\right) \end{equation}
%    for each $d\in [n]$. This particular  choice is justified by \cref{thm_rate_score},   
%    which implies that, for some sufficiently large constant $C \ge 1$, the event \begin{equation}\label{E} \cE := \bigcap_{d=1}^n\{ |\wh S_d -S_d|\le t_d\}\end{equation}  holds with probability at least $1 - n^{-1}$.
 %   Detailed discussion on choosing $C$ is given in  \cref{sec_prac_delta}. We then define the  set of  selected near-pure  mixtures by 
 %  \begin{equation}\label{def_S_set_hat}
 %       \wh \cS := \wh \cS(t) = \left\{d\in [n]: \wh S_d \le 2t_d\right\}.
 %  \end{equation}
 %   As reasoned in \cref{sec_method_screen}, the above screening procedure enjoys  properties \eqref{typeI} and \eqref{typeII}.\\

To identify samples with a small $\mathrm{TLC}^{2}$,
let $t=(t_1,\ldots,t_n)^\T$ be a vector of
threshold levels. By \cref{thm_rate_score},
under its stated conditions, these thresholds
can be chosen such that, for any
$\delta\in(0,1)$,
\begin{equation}\label{E}
\mathbb P(\mathcal E)\ge 1-\delta,
\qquad
\mathcal E:=
\bigcap_{d=1}^n
\left\{
|\wh S_d-S_d|\le t_d
\right\}.
\end{equation}
Such a choice follows by applying
\cref{thm_rate_score} to each ensemble member
with failure probability $\delta/n$ and
taking a union bound. We discuss the
theoretical orders of these thresholds and
their practical selection in
\cref{sec_prac_delta}.

We then define the set of selected near-pure
mixtures by
\begin{equation}\label{def_S_set_hat}
\wh\cS:=\wh\cS(t)
=\left\{
d\in[n]:\wh S_d\le 2t_d
\right\}.
\end{equation}
On the event $\mathcal E$, this screening
procedure enjoys properties
\eqref{typeI} and \eqref{typeII}, provided
that $\max_{d\in[n]}t_d<\Delta_{\min}/12$.

    \noindent{\it Step 2}. For each selected near-pure mixture $d\in \wh \cS$, we compute the cross-entropy maximizer  \eqref{def_theta_hat} of \cref{sec_rate_atom} to estimate the dominant atom: 
    \begin{equation}\label{est1} 
            \wh \theta_d  \in  \argmax_{\theta\in\cB} ~  \sum_{j=1}^{p} \wh \pi_j^{(d)} \log A_{\theta}(x_j)
    \end{equation} 
    which, by  \cref{thm:softmax}, is close to one of  the true atoms $\theta_1,\ldots,\theta_K$. Suppose the ensemble  admits the following assumption:

    % $\mu$-almost surely, with probability at least $1-\delta-n^{-1}$, there exists some $k\in [K]$ such that 
    % \[
    %      \|\wh\theta_d - \theta_k\|_2 ~ \lesssim~  {1\over \sigma_0}\sqrt{L\log(1/\delta)\over N_d} + {\sigma_x^2\sqrt{\Dt_{\max}}\over \sigma_0^2\Dt_{\min}} ~ t_d .%~ \lesssim~  {\sigma_x^2\sqrt{\Dt_{\max}}\over \sigma_0^2\Dt_{\min}} t_d.
    % \] 

    \begin{ass}\label{ass_quasi}  
        For  each $k\in [K]$, there exists at least one  $d_k\in [n]$ such that its $\pi^{(d_k)}$ is $\varepsilon$-pure, with $\theta_k$ as its dominant atom, and  with $\eps \le t_{d_k}/\Dtmax$.
    \end{ass}
    \noindent  Then, under this assumption, on the event $\cE$ and by using part 2 of \cref{prop_score}, one has $\wh S_{d_k} \le S_{d_k} + t_{d_k} \le 2t_{d_k}$ so that $d_k\in \wh\cS$ and $|\wh\cS| \ge K$. Furthermore, with probability at least $1-2n^{-1}$, there exists some permutation $\iota: [K] \to [K]$ and a partition $\wh\cS_1,\ldots,\wh\cS_K$ of $\wh \cS$ such that for each $k\in [K]$, the estimated dominant atom $\wh\theta_d $ is close to $\theta_{\iota(k)}$.

    Finally, since there could be more than one estimator per $\theta_k$, we propose to  cluster $\{\wh\theta_d\}_{d\in \wh \cS}$ into $K$ groups by solving  the $k$-means problem: 
    \begin{align}\label{kmeans}
        \wt\theta_1,\ldots,\wt \theta_K  ~ =~  \argmin_{\bar\theta_1,\ldots,\bar\theta_K}\, \sum_{d\in \wh\cS} \min_{k\in [K]} \|\wh\theta_d - \bar\theta_k\|_2^2.
    \end{align}
    The following theorem quantifies the estimation error of the above $k$-means solution. Let $N_{\min} = \min_{d\in[n]}N_d$ denote the smallest sample size over $n$ mixtures.

    \begin{thm}\label{thm_kmeans}
        Grant model \eqref{model_ensemble} with \cref{ass:mu,ass:theta,ass:rate,ass_quasi} and assume $\lambda_L(H_{\theta_k}^{(\mu)})\ge \sigma_0^2 > 0$ for all $k\in [K]$. For sufficiently large $p$ and $N_{\min}$, with probability at least $1-2p^{-1}-2n^{-1}$, 
        \begin{equation*}
            \max_{k\in[K]} ~  \bigl\|\wt \theta_k - \theta_{\iota(k)}\bigr\|_2^2 ~ \le ~   C  \left( {1 \over N_{\min} } +  {1\over p}\right).
        \end{equation*} 
    \end{thm}
    \begin{proof}
        The proof is given in \cref{app_sec_proof_thm_kmeans}. Both the requirement of $p$ and $N_{\min}$ and the quantity $C$ depend  on  $\sigma_0$, $C_1,C_2$ and $C_3$ in \cref{thm_rate_score} with $N=N_d$ and $\delta = 1/n^2$, as well as  the number of mixtures in the ensemble that are $(6t_d/\Dtmin)$-pure. Their explicit expressions  can be traced throughout.
    \end{proof} 
    
\begin{remark}
We emphasize that Step 1 of the ensemble procedure applies to any
softmax-mixture ensemble, without additional structural
assumptions such as Assumption~\ref{ass_quasi}. If the
ensemble contains no samples generated by
$\varepsilon$-pure mixtures, the procedure may return
$\wh\cS=\emptyset$. Otherwise, once dominant atoms have
been estimated as in \eqref{est1}, the next task is to
associate these estimates with the mixture parameters.
Unlike the screening and dominant-atom estimation steps, which do not require knowledge of $K$, the subsequent grouping of the estimated dominant atoms requires specifying a value of $K$. A practical choice of such $K$ is given in \cref{sec_prac_delta}. Assumption~\ref{ass_quasi} provides the conditions under which we establish theoretical guarantees for this final step.

Assumption~\ref{ass_quasi} is related to the
{\it anchor-document assumption} in topic models, which
are collections of nonparametric discrete mixtures
$\rho^{(1)},\ldots,\rho^{(n)} \in \Delta^{p-1}$ with shared components.
The anchor-document assumption corresponds to imposing
Assumption~\ref{ass_quasi} on $\rho^{(d)}$ with
$\varepsilon=0$, a substantially stronger requirement.
The existence and detection of anchor documents have
been studied in the topic-model literature
\citep{BansalBK2014,Klopp2023} and in nonnegative
matrix factorization \citep{gillis2014snpa}.
An ensemble of softmax mixtures can be viewed as a
softmax-parametrized topic model relative to dictionary that is common to the corpus \citep{dieng2020topic}, and our screening
procedure, tailored to this class, can detect
near-anchor documents with $\varepsilon>0$, when present.
A small simulation study in \cref{app_sim_anchor}
illustrates its effectiveness in this setting.

Finally, \cref{thm_kmeans} implies that consistency generally requires $ N_{\min}\wedge p\to \infty$, up to the quantities on which $C$ depends. Regarding finite-sample behavior,
Theorem~\ref{thm:softmax} establishes that the MLE
has bias proportional to $\varepsilon$.
Consider the limiting case of
Assumption~\ref{ass_quasi} with $\varepsilon=0$, and
suppose further that every non-pure mixture $d$ satisfies
\[
    \min_{k\in[K]}\|\alpha^{(d)}-e_k\|_1 ~ > ~ (12 / \Dtmin)\, t_d.
\]
Then, on the event $\cE$ defined in \eqref{E},
$\wh\cS$ coincides exactly with the index set of
pure mixtures. Consequently, each estimator
$\wh\theta_d$ in \eqref{est1} achieves the parametric
rate, and the final error bound in \cref{thm_kmeans}
improves to $C/N_{\min}$.
We include this observation for completeness, as our
analysis does not rely on the limiting assumption
$\varepsilon=0$ or any separation assumption as the one in the display above.
\end{remark}

\section{Practical considerations}\label{sec_prac}

We discuss in detail three practical aspects pertaining to screening with $\mathrm{TLC}^{2}$ for single and ensemble softmax mixtures: the choice  of $\lambda_\beta$ and $\lambda_\gamma$ in computing \eqref{beta-hat} and \eqref{gamma-hat}, the choice  of the threshold levels $t$ in \eqref{E} %\eqref{rate_delta_d}
for ensemble softmax mixtures, and the estimation of the subspace spanned by the $\theta_k$'s and its use in screening with $\mathrm{TLC}^{2}$.

\subsection{Tuning parameters  $\lambda_\beta$ and $\lambda_\gamma$ in $\mathrm{TLC}^{2}$}\label{sec_prac_lbd}

$\mathrm{TLC}^{2}$ requires solving \eqref{beta-hat}--\eqref{gamma-hat}, which depend on the tuning parameters $\lambda_\beta$ and $\lambda_\gamma$. Motivated by the rates established in \eqref{eq:net-tuning}, we set
$\lambda_\beta(c_1)=c_1\sqrt{L/p}$ and $\lambda_\gamma(c_2)=c_2\sqrt{L/p}$,
where both leading constants are selected from the candidate set $\mathcal{C}=\{0.1,0.2,\ldots,5\}$ as follows.

By fixing the support points in $\cX$, we generate validation samples to select $c_1$ and $c_2$ from $\cC$. Specifically, for each replication $r\in[R]$, with $R=1000$, we sample $K$ atoms $\theta_1^{(r)},\ldots,\theta_K^{(r)}$ independently and uniformly from $\cB$. For each $d\in[D_{\mathrm{cv}}]$, with $D_{\mathrm{cv}}=100$, we then sample the mixing weights $\alpha^{(r,d)}$ uniformly from $\Delta^{K-1}$ and generate $Y_1^{(r,d)},\ldots,Y_N^{(r,d)}$  i.i.d. according to model \eqref{mod-def}. 

The selection procedure for $(c_1,c_2)$ consists of two stages. First, for each $c_1,c_2\in\mathcal C$, we solve \eqref{beta-hat} with $\lambda_\beta(c_1)$ to obtain $\wh\beta(c_1)$ and solve \eqref{gamma-hat} with $\lambda_\gamma(c_2)$ to obtain $\wh\Gamma(c_2)$. In both programs, we construct $\cN_\epsilon$ by randomly sampling points in $\cB$: we first sample the radius uniformly from $(0,1)$ and then the direction uniformly over the sphere. Throughout the  numerical results in \cref{sec_numeric}, we use a grid of 1,000 points randomly sampled within $\cB$, uniformly over the radius, to solve \eqref{beta-hat}, and a grid of 200 such points to solve \eqref{gamma-hat}.  

Next, we evaluate the resulting $\wh\beta(c_1)$ and $\wh\Gamma(c_2)$ using the validation samples. For each $r\in[R]$ and $d\in[D_{\mathrm{cv}}]$, let $\wh\pi^{(r,d)}$ denote the empirical frequency vector of $\cX$ based on $Y_1^{(r,d)},\ldots,Y_N^{(r,d)}$. We compute $\wh S_1^{(r,d)}(c_1)$ and $\wh S_2^{(r,d)}(c_2)$ according to \eqref{def_S_hat_d}. Their corresponding true score components are $S_1^{(r,d)}=\sum_{k=1}^K\alpha_k^{(r,d)}\|\theta_k^{(r)}\|_2^2$ and $S_2^{(r,d)}=\|\sum_{k=1}^K\alpha_k^{(r,d)}\theta_k^{(r)}\|_2^2$. We then compute the validation errors
\[
\mathcal E_\beta(c_1)=\sum_{r=1}^R\sum_{d=1}^{D_{\mathrm{cv}}}
\big|\wh S_1^{(r,d)}(c_1)-S_1^{(r,d)}\big|,\qquad
\mathcal E_\Gamma(c_2)=\sum_{r=1}^R\sum_{d=1}^{D_{\mathrm{cv}}}
\big|\wh S_2^{(r,d)}(c_2)-S_2^{(r,d)}\big|.
\]
Finally, we select $\wh c_1\in\arg\min_{c_1\in\mathcal C}\mathcal E_\beta(c_1)$ and $\wh c_2\in\arg\min_{c_2\in\mathcal C}\mathcal E_\Gamma(c_2)$, and use $\wh\beta(\wh c_1)$ and $\wh\Gamma(\wh c_2)$ to compute the $\mathrm{TLC}^{2}$ in \eqref{def_S1S2_hat}.

\subsection{Tuning the screening threshold  in ensemble softmax mixtures}\label{sec_prac_delta} 
In view of \cref{thm_rate_score}, we propose to set the
threshold vector $t = (t_1,\ldots,t_n)^\T$ in \eqref{E} as  
\begin{equation}\label{rate_delta_d}t_d = C \bigl({p}^{-1/2}+N_d^{-1/2}\big),\qquad d\in [n],\end{equation}
%\eqref{rate_delta_d}. 
with $C$ selected by the 
following data-driven procedure. % and, when unknown, $K$.

We write $\wh\cS(C)$ to make the dependence of the selected set
on $C$ explicit. Before constructing these sets, we recommend
computing $\textrm{TLC}^2$ with $\wh S=\wh S_1-\wh S_2$
replaced by $\wt S=\wt S_1-\wt S_2$, where
\[
\wt S_2=\min\{1,\max\{0,\wh S_2\}\},
\qquad
\wt S_1=\min\{1,\max\{\wt S_2,\wh S_1\}\}.
\]
This truncation ensures that
$0\le\wt S_2\le\wt S_1\le1$ and hence $0\le\wt S\le1$.
We use the resulting statistic throughout the selection procedure.

As $C$ increases, the number of mixtures in $\wh\cS(C)$ is
nondecreasing. Let $\cC_{grid}$ be a predetermined finite grid
of positive values, and choose $\rho\in(0,1]$ to specify the
largest proportion of mixtures that may be selected. For a
fixed integer $K\ge2$, consider the admissible grid
\[
\cC_{grid}(K)
=
\left\{
C\in\cC_{grid}:
K+1\le|\wh\cS(C)|\le\floor{\rho n}
\right\}.
\]
We consider only values of $K$ for which $\cC_{grid}(K)$ is
nonempty and set $\rho=0.4$ in our numerical studies.

For each $C\in\cC_{grid}(K)$, we compute
$\wh\theta_t(C)$ for $t\in\wh\cS(C)$ as in \eqref{est1}
and apply the $k$-means procedure in \eqref{kmeans}.
This yields estimates of the dominant atoms and a partition
of $\wh\cS(C)$ into $K$ nonempty clusters,
$\wh\cS_1(C),\ldots,\wh\cS_K(C)$.
We evaluate this partition using its average silhouette
coefficient. Specifically, for $t\in\wh\cS_k(C)$ with $|\wh\cS_k(C)|>1$, define the
average distance to the other points in the same cluster by
\[
a_t(C)
=
\frac{1}{|\wh\cS_k(C)|-1}
\sum_{j\in\wh\cS_k(C)\setminus\{t\}}
\|\wh\theta_t(C)-\wh\theta_j(C)\|_2,
\]
and the smallest average distance to another cluster by
\[
b_t(C)
=
\min_{\ell\in[K]\setminus\{k\}}
\frac{1}{|\wh\cS_\ell(C)|}
\sum_{j\in\wh\cS_\ell(C)}
\|\wh\theta_t(C)-\wh\theta_j(C)\|_2.
\]
The silhouette coefficient of point $t$ is
\[
s_t(C)
=
\frac{b_t(C)-a_t(C)}{a_t(C)\vee b_t(C)},
\]
with $s_t(C)=0$ if $t$ belongs to a singleton cluster or
$a_t(C)=b_t(C)=0$. Averaging over the selected points gives
\[
\textsf{SC}_K(C)
=
\frac{1}{|\wh\cS(C)|}
\sum_{t\in\wh\cS(C)}s_t(C),
\]
where the subscript makes the dependence on $K$ explicit.
The coefficient ranges from $-1$ to $1$, with larger values
indicating better separation between clusters relative to
their within-cluster distances. For this fixed $K$, we select $\wh C(K)$ by maximizing  $\textsf{SC}_K(C)$ over $C\in\cC_{grid}(K)$.

When $K$ is unknown, we repeat this procedure for each $K$
in a predetermined set $\cK_{grid}$ of admissible candidate
values and select $\wh K$ by maximizing $\textsf{SC}_K(\wh C(K))$. 
The final threshold vector is determined by $\wh C(\wh K)$,
and the dominant atoms are estimated by the corresponding
$k$-means fit with $\wh K$ clusters. We use this procedure
in the real-data analysis in \cref{sec_real_data}.

\subsection{Subspace-assisted $\mathrm{TLC}^{2}$}\label{sec_prac_SS}

We describe how an estimated low-dimensional subspace can
reduce the computational cost of screening with $\mathrm{TLC}^{2}$.
The modification concerns only the coefficients $\wh\beta$
and $\wh\Gamma$ in \eqref{beta-hat} and \eqref{gamma-hat},
and applies to both single and ensemble softmax mixtures.
We first present the common construction and then describe
the subspace estimators used in the two settings. Throughout,
we consider $K\ll L$, without requiring $K$ to be known.

Let $V\in\RR^{L\times s}$, with $s\le K$, have orthonormal columns spanning
a subspace that contains the true atoms relevant to screening,
and let $\wh V\in\RR^{L\times s}$ with orthonormal columns
be an estimate of this subspace. The dimension $s$
depends on the setting and need not equal $K$.

Next, we propose to solve the finite-net programs based on $\wh V$. Specifically, let  $\cN'_\epsilon$ be an $\epsilon$-net of
$\cB_s:= \{\eta\in \RR^s: \|\eta\|_2   \le   1\}$. When  $\wh V$ estimates $V$ well,  $\{\wh V\eta: \eta\in \cN'_\epsilon\}$ is close to $\{V\eta: \eta\in \cN'_\epsilon\}$. It is easy to see that the latter forms an $\epsilon$-net of the relevant true atoms $\theta_1,\ldots,\theta_K$. In view of \eqref{beta-hat} and \eqref{gamma-hat}, since they are designed to recover $\sum_k \alpha_k\|\theta_k\|_2^2 = \sum_k \alpha_k \|\eta_k\|_2^2$ and $\|\sum_k \alpha_k \theta_k\|_2 = \|\sum_k \alpha_k \eta_k\|_2$ for some $\eta_k \in \cB_s$, respectively, we 
propose to solve
\begin{align}
\label{beta_hat_dr}
\wt\beta
&\in\argmin_{\beta \in\RR^p}\|\beta\|_\infty
\quad\text{subject to}\quad
\max_{\eta \in  \cN'_{\epsilon}} \left|\beta^\T A_{ \wh V \eta}-\|\eta\|_2^2\right|\le  \lambda_\beta',\\
\label{gamma_hat_dr}
\wt\gamma_\ell
&\in\argmin_{\gamma\in\RR^p}\norm \gamma_\infty
\quad\text{subject to}\quad
\max_{\eta\in  \cN'_{\epsilon}}\left|\gamma^\T A_{\wh V\eta}- \eta_\ell \right|\le  \lambda_\gamma',
\qquad\ell\in[s]
\end{align}
where the tuning parameters $\lambda_\beta'$ and $\lambda_\gamma'$ are given in \eqref{rate_lbd_dr}. Since $|\cN_\epsilon'| \le (1 + 2/\epsilon)^s$ and $s\ll L$,  the numbers of constraints above are greatly  reduced compared with those in \eqref{beta-hat} and \eqref{gamma-hat}.   

Let $\wt\Gamma=(\wt\gamma_1,\ldots,\wt\gamma_s)
\in\RR^{p\times s}$. We replace $\wh\beta$ and $\wh\Gamma$
in the screening procedure by $\wt\beta$ and
$\wt\Gamma\wh V^\T$, respectively.
Under the conditions of \cref{thm_rate_score_SS} in
\cref{app_sec_proof_SS}, the resulting estimator
$\wh S(\wt\beta,\wt\Gamma\wh V^\T)$ achieves essentially
the rates in \cref{thm_rate_score}, with an additional
term accounting for subspace estimation error and with
the factor $\exp(\sqrt{L})$ replaced by $\exp(\sqrt{s})$.

For a single softmax mixture, the relevant subspace is
the column space of
$\sum_{k=1}^K\alpha_k\theta_k\theta_k^\T$, which equals
the span of the atoms with positive mixing weights.
Its dimension $s$ is the rank of this matrix and satisfies
$s\le K$. We estimate this subspace using the procedure
in \cite{BBNW}. The construction and selection of $s$
are deferred to \cref{app_sec_SS_one}, where
\cref{thm_V_error_one_doc} gives an explicit convergence
rate under the additional assumption that $\mu$ is elliptical.

For an ensemble of softmax mixtures, we use a subspace
estimator tailored to the ensemble structure. Under the
assumptions in \cref{app_sec_SS_ensemble}, its population
target has column space
$\operatorname{span}\{\theta_1,\ldots,\theta_K\}$,
so the same reduced programs in \eqref{beta_hat_dr}--\eqref{gamma_hat_dr} apply with the corresponding
estimated basis and dimension $s$. The construction and
selection of $s$ are detailed in that section.
The simulation study in \cref{app_sim_SS} demonstrates the effectiveness of the resulting subspace-assisted $\mathrm{TLC}^2$.

Both procedures above show promises when the feature distribution $\mu$ is close to elliptical (see \cref{ass:elliptical}). When this condition does not hold in practice, another approach that we found numerically appealing is to fit the ensemble softmax mixture model using the EM algorithm in \cref{subsec-fit-ensem-ssm}, with an overspecified number of mixture components. The hope is that the span of the fitted atoms still approximately contains the subspace spanned by the true atoms. We use this heuristic in the real-data application in \cref{sec_real_data}.

 \subsection{Algorithm}
For the convenience of the reader, we collect all the steps above in the screening for near-pure mixtures and dominant atoms estimation under the ensemble softmax mixtures.  

\begin{algorithm}[ht]
\caption{Screening for near-pure mixtures and estimation of ensemble softmax mixtures}
\label{alg:esharp}
\begin{algorithmic}[1]
\Require Shared features $\cX=\{x_1,\ldots,x_p\}\subset\RR^L$;
empirical frequency vectors $\{\wh\pi^{(d)}\}_{d=1}^n$;
sample sizes $\{N_d\}_{d=1}^n$;
an integer $K$;
an estimated basis $\wh V\in\RR^{L\times s}$ for the atom subspace,
with orthonormal columns.
\State Compute $\wt\beta$ and
$\wt\Gamma=(\wt\gamma_1,\ldots,\wt\gamma_s)$
by solving \eqref{beta_hat_dr} and \eqref{gamma_hat_dr},
with $\lambda_\beta'$ and $\lambda_\gamma'$ selected
according to \cref{sec_prac_lbd}.
\State Compute $\wh S_1,\ldots,\wh S_n$ as in \eqref{def_S_hat_d},
using $\wt\beta$ and $\wt\Gamma\wh V^\T$ in place of
$\wh\beta$ and $\wh\Gamma$, respectively.
\State Select the set of near-pure mixtures $\wh\cS$
as in \eqref{def_S_set_hat}, with thresholds $t_1,\ldots,t_n$
selected according to \cref{sec_prac_delta}.
\State Compute the estimated dominant atoms $\wh\theta_d$
for $d\in\wh\cS$ using \eqref{est1}, and cluster them
by solving the $k$-means problem in \eqref{kmeans}
to obtain $\wt\theta_1,\ldots,\wt\theta_K$.
\State \Return $\wt\theta_1,\ldots,\wt\theta_K$.
\end{algorithmic}
\end{algorithm}

\section{Simulation studies}\label{sec_numeric}

In this section, we conduct simulation studies  to examine the performance of the proposed procedure in single and ensemble softmax mixtures. \cref{app_sim} contains additional simulation studies that illustrate the benefits of subspace assistance in $\mathrm{TLC}^2$ and compare E-$\mathrm{TLC}^2$ with existing approaches for detecting near-pure mixtures in the ensemble setting. %and real data analysis in \cref{sec_real_data} to verify the effectiveness of our proposed procedure. 

\subsection{Single softmax mixtures}\label{sec_sim_one_doc}
We focus on a single softmax mixture and investigate the factors affecting the estimation errors of $S(G)$ and, for a near-pure mixture, the dominant atom $\theta^*$. We also investigate the performance of the bootstrap procedure in \cref{sec_method_screen}. %, as well as to examine how the performance of the proposed method adapts to the geometry of the mixture.

We set $\mu=N(0,I_L)$ and generate $\theta_1$ uniformly over $\cB$. We set $K=4$ and generate $\theta_2,\ldots,\theta_K$ independently and uniformly over the sphere $\{\theta:\|\theta-\theta_1\|_2=\delta\}$, conditional on $\theta\in\cB$, via rejection sampling. Here $\delta>0$ represents the separation between $\theta_1$ and the other atoms. To generate $\eps$-pure mixtures,  we set $\alpha_1=1-\eps$ and draw $(\alpha_2,\ldots,\alpha_K)/\eps$ from $\mathrm{Dirichlet}(1_{K-1})$, so that $\theta_1$ is the dominant atom. Finally, $Y_1,\ldots,Y_N$ are generated i.i.d. from model \eqref{mod-def}.

We examine the effect of the following parameters by varying them one at a time: $N=\{1,2,3,4,5\}\times10^3$, $p=\{1,2,4,6\}\times10^2$, $L=\{4,8,12,16,20\}$, $\eps=\{0.01,0.05,0.1,0.2,0.5\}$, and $\delta=\{0.05,0.1,0.2,0.5,1\}$. The baseline setting is $L=10$, $N=2000$, $p=400$, $\eps=0.05$, and $\delta=0.5$ when the corresponding parameter is not varied. When varying $\delta$, we set $\eps=0.2$ so that the mixtures assign non-negligible weight to the nondominant components and the effect of atom separation can be more clearly assessed. 

\begin{figure}[ht]
    \centering
    \includegraphics[width=\linewidth]{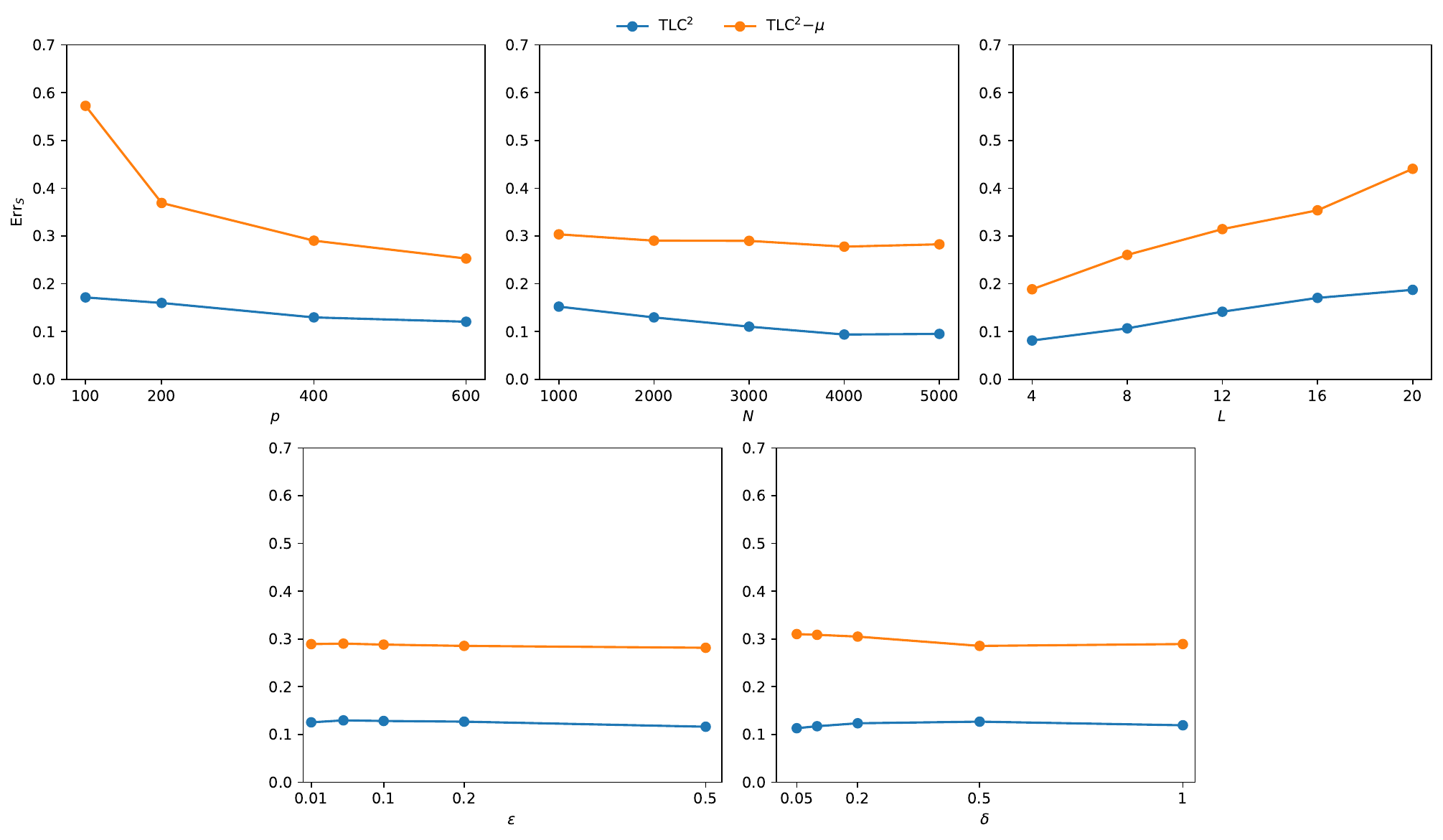}

    \caption{Dependence of  $\mathrm{Err}_S$, on $p$, $N$, $L$, $\eps$, and $\delta$. }
   
   \label{fig:one_doc_score_error}
\end{figure}

\begin{figure}[ht]
    \centering
    \includegraphics[width=\linewidth]{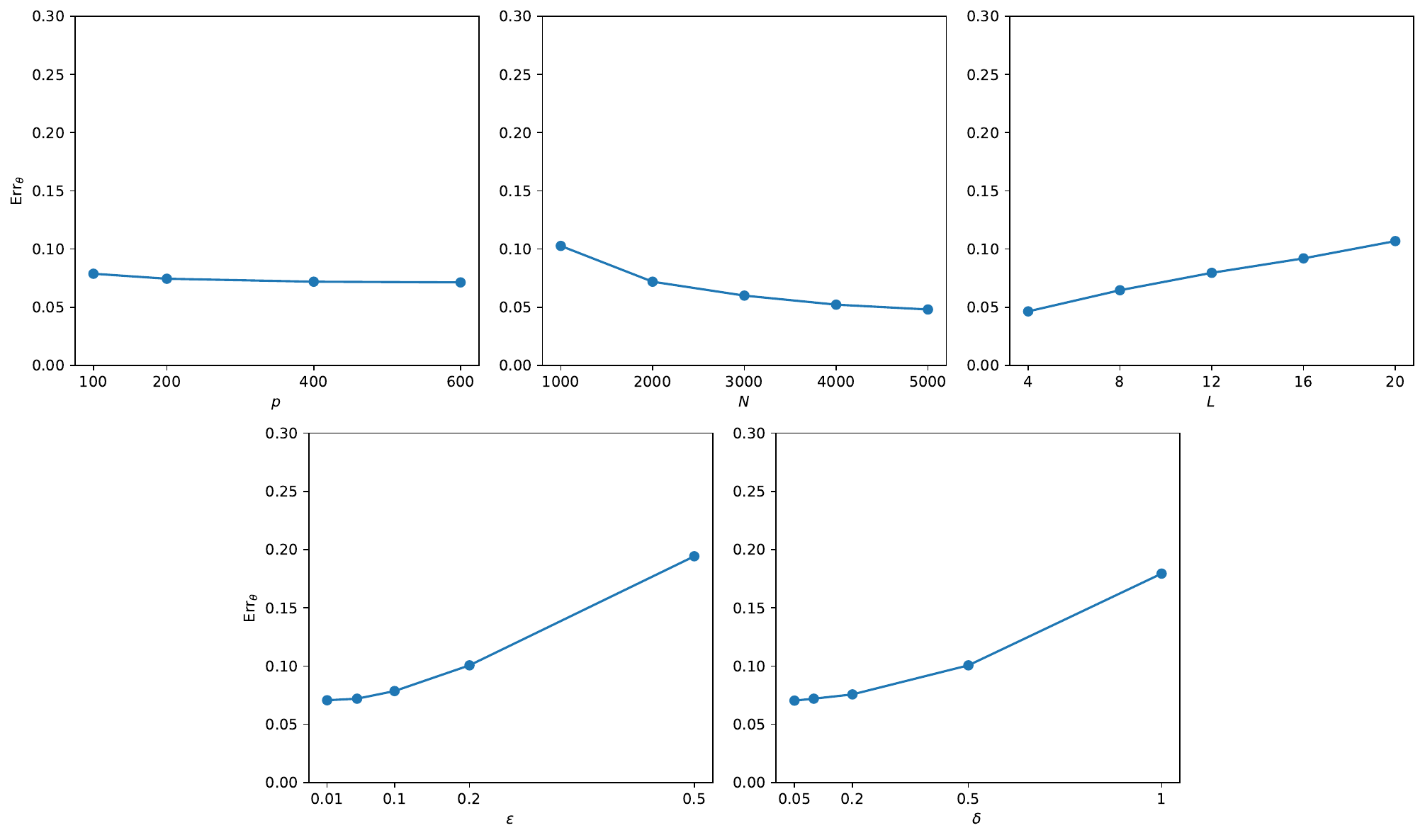}
    \caption{Dependence of   $\mathrm{Err}_\theta$ on $p$, $N$, $L$, $\eps$, and $\delta$.}
    \label{fig:one_doc_theta_error}
\end{figure}

We record the estimation error of $\mathrm{TLC}^2$, the estimator $\wh S$ in \eqref{finalS-est}, as $\mathrm{Err}_S:=|\wh S-S|$, and that of the M-estimator $\wh\theta_1$ in \eqref{def_theta_hat} as $\mathrm{Err}_\theta:=\|\wh\theta_1-\theta_1\|_2$.  The implementation details for computing $\wh S$ follow \cref{sec_prac_lbd}. For comparison, we also compute the estimation error of $\mathrm{TLC}^2$ using $\beta_\mu$ and $\Gamma_\mu$ defined in \eqref{one-point}, which we denote by $\mathrm{TLC}^2$-$\mu$. We report the average errors over 100 replications in \cref{fig:one_doc_score_error} and \cref{fig:one_doc_theta_error}.

As shown in \cref{fig:one_doc_score_error}, the error $\mathrm{Err}_S$ decreases as $N$ and $p$ increase and increases as $L$ increases, while remaining essentially stable as $\eps$ and $\delta$ vary. These patterns are consistent with the dependence on $N$, $L$, and $p$, and the absence of dependence on $\eps$ and $\delta$, predicted by \cref{thm_rate_score,thm_rate_score_detailed}. Moreover, $\mathrm{TLC}^2$ even outperforms $\mathrm{TLC}^2$-$\mu$, despite the latter using the true distribution $\mu$. This can be explained by the error decomposition in \eqref{score_err_decomp}: although $\mathrm{TLC}^2$-$\mu$ has a controlled approximation error, it can incur a larger estimation error through a larger value of $\|\beta_\mu\|_\i$, compared with $\mathrm{TLC}^2$ that uses $\wh\beta$ from \eqref{beta-hat}.

As shown in \cref{fig:one_doc_theta_error}, the atom estimation error $\mathrm{Err}_\theta$ increases with $L$, $\eps$, and $\delta$, decreases with $N$, and remains essentially stable as $p$ varies. These empirical patterns are consistent with the dependence on $N$, $L$, $\eps$, and $\delta$, and the absence of dependence on $p$, predicted by \cref{thm:softmax}.

We next investigate the finite-sample performance of the bootstrap procedure in \cref{sec_method_screen} for constructing confidence intervals for $S(G)$. Alongside the symmetric confidence intervals in \eqref{def_CI}, calibrated using absolute bootstrap deviations, we consider equal-tailed basic bootstrap intervals. Let $\wt q_{\alpha}$ denote the empirical $\alpha$-quantile of the bootstrap deviations $\{\wh S^{(b)}-\wh S\}_{b=1}^B$. A nominal $(1-\alpha)$ equal-tailed confidence interval for $S(G)$ is then given by
\[
\bigl[
\wh S-\wt q_{1-\alpha/2},
\;
\wh S-\wt q_{\alpha/2}
\bigr].
\]

% Applying the bootstrap procedure,
% For the absolute-error calibration, we instead denote $\wt q_{1-\alpha}$ as the empirical $(1-\alpha)$-quantile of $\{|\wh S^{(b)}-\wh S|\}_{b=1}^B$. This threshold defines the symmetric interval
% \[
% \bigl[
% \wh S-\wt q_{1-\alpha},
% \;
% \wh S+\wt q_{1-\alpha}
% \bigr].
% \]
% These constructions are reported separately. Both constructions use the same observed sample and the same
% bootstrap draws, allowing a paired comparison.

As mentioned towards the end of \cref{sec_method_screen}, the bootstrap directly addresses sampling fluctuations around the
finite-support surrogate
$$
Q_{\mathcal X}(\pi) := \wh{\beta}^{\top}\pi-\|\wh{\Gamma}^{\top}\pi\|_2^2.
$$
We evaluate the coverage probabilities and average lengths of the two confidence intervals for both $S(G)$ and $Q_{\mathcal X}(\pi)$ as $p$, $N$, and $\eps$ vary one at a time. The baseline setting is $N=p=1000$ and $\eps=0.2$, and we consider $N\in\{200,500,1000,1500,2000\}$, $p\in\{200,400,800,1000,1200,1600\}$ and $\eps\in\{0.01,0.05,0.1,0.2,0.5\}.$
Each setting has $R=100$ Monte Carlo replications. In each replication, we set $\mu=N(0,I_L)$ with $L=10$ and generate $\theta_1,\ldots,\theta_K$ uniformly over $\cB$, with $K=4$. For each $k\in[K]$, we generate ten $\eps$-pure mixtures and draw $N$ observations from each mixture according to \eqref{mod-def}, yielding $D=40$ documents per replication.

We also compare the theoretically motivated threshold in \eqref{rate_delta_d}, with $C=1$, to the average threshold from the absolute-error calibration. For replication $r$ and document $d$, let $\wh q_{1-\alpha,rd}$ denote the empirical $(1-\alpha)$-quantile of $\{|\wh S_{rd}^{(b)}-\wh S_{rd}|\}_{b=1}^B$. We compute
\[
\bar t^{\mathrm{boot}}
=\frac{1}{RD}\sum_{r=1}^R\sum_{d=1}^D
\wh q_{1-\alpha,rd}.
\]
The coverage probabilities, average interval lengths, and both thresholds are reported in \cref{tab:bootstrap-calibration}.

As shown in \cref{tab:bootstrap-calibration}, both interval constructions achieve coverage close to the nominal level of $0.95$ for $Q_{\mathcal X}(\pi)$ but generally undercover $S(G)$. This difference illustrates the effect of surrogate approximation error, which the bootstrap does not automatically capture. Increasing $N$ shortens the intervals and generally improves coverage for $S(G)$, while increasing $p$ also generally improves coverage for $S(G)$. The results are relatively insensitive to $\varepsilon$. Compared with the equal-tailed basic intervals, the symmetric intervals generally provide higher coverage for $S(G)$ at the cost of slightly greater lengths.

\begin{table}[ht]
\centering
\caption{Coverage probabilities, average interval lengths, and calibration thresholds for the two bootstrap intervals.}
\label{tab:bootstrap-calibration}
\small
\setlength{\tabcolsep}{4pt}
\renewcommand{\arraystretch}{1.15}
\begin{tabular}{crrr@{\hspace{14pt}}rrrrr}
\toprule
& \multicolumn{3}{c}{Equal-tailed intervals}
& \multicolumn{4}{c}{Symmetric intervals} & \\
\cmidrule(lr){2-4}\cmidrule(lr){5-8}
Parameter value & Cov.\ $S$ & Cov.\ $Q$ & Length
& Cov.\ $S$ & Cov.\ $Q$ & Length & $\overline t^{\mathrm{boot}}$ & $t_d$ \\
\midrule
\multicolumn{9}{l}{\textit{(a) Varying $p$: $N=1000$, $\varepsilon=0.2$}} \\
\addlinespace[2pt]
200  & 0.855 & 0.951 & 0.623 & 0.872 & 0.956 & 0.628 & 0.314 & 0.197 \\
400  & 0.910 & 0.954 & 0.638 & 0.922 & 0.957 & 0.642 & 0.321 & 0.157 \\
800  & 0.916 & 0.951 & 0.648 & 0.926 & 0.954 & 0.652 & 0.326 & 0.129 \\
1000 & 0.914 & 0.949 & 0.654 & 0.927 & 0.950 & 0.658 & 0.329 & 0.122 \\
1200 & 0.924 & 0.947 & 0.651 & 0.934 & 0.949 & 0.655 & 0.327 & 0.116 \\
1600 & 0.916 & 0.942 & 0.655 & 0.925 & 0.944 & 0.659 & 0.329 & 0.109 \\
\addlinespace[5pt]
\multicolumn{9}{l}{\textit{(b) Varying $N$: $p=1000$, $\varepsilon=0.2$}} \\
\addlinespace[2pt]
200  & 0.849 & 0.945 & 0.994 & 0.922 & 0.962 & 1.042 & 0.521 & 0.197 \\
500  & 0.902 & 0.945 & 0.810 & 0.926 & 0.956 & 0.821 & 0.411 & 0.147 \\
1000 & 0.914 & 0.949 & 0.654 & 0.927 & 0.950 & 0.658 & 0.329 & 0.122 \\
1500 & 0.926 & 0.953 & 0.561 & 0.930 & 0.953 & 0.563 & 0.282 & 0.110 \\
2000 & 0.926 & 0.949 & 0.500 & 0.927 & 0.951 & 0.501 & 0.251 & 0.104 \\
\addlinespace[5pt]
\multicolumn{9}{l}{\textit{(c) Varying $\varepsilon$: $N=p=1000$}} \\
\addlinespace[2pt]
0.01 & 0.915 & 0.946 & 0.661 & 0.923 & 0.952 & 0.665 & 0.333 & 0.122 \\
0.05 & 0.911 & 0.945 & 0.660 & 0.917 & 0.948 & 0.663 & 0.332 & 0.122 \\
0.10 & 0.918 & 0.951 & 0.658 & 0.927 & 0.953 & 0.662 & 0.331 & 0.122 \\
0.20 & 0.914 & 0.949 & 0.654 & 0.927 & 0.950 & 0.658 & 0.329 & 0.122 \\
0.50 & 0.915 & 0.947 & 0.643 & 0.930 & 0.952 & 0.647 & 0.323 & 0.122 \\
\bottomrule
\end{tabular}
\end{table}

\subsection{Ensemble softmax mixture}\label{sec_sim_ensemble}

We consider the ensemble setting in \eqref{model_ensemble} of \cref{sec_ensemble} and evaluate the proposed procedure for selecting near-pure softmax mixtures with $\mathrm{TLC}^2$.

To generate the data, we first set $\mu=N(0,I_L)$ and generate $\theta_1,\ldots,\theta_K$ independently and uniformly over $\cB$. We then generate $n=100$ softmax mixtures. For each $k\in[K]$, we generate $m=5$ $\eps$-pure mixtures as in \cref{sec_sim_one_doc}.  For the remaining mixtures, we generate $\alpha^{(d)}$ uniformly over the simplex $\Delta^{K-1}$. Finally, for each mixture $d\in[n]$, we generate $Y_1^{(d)},\ldots,Y_{N}^{(d)}$ i.i.d. according to model \eqref{mod-def}.

% {\mike with $R=1000,D_{cv}=100,\rho=0.4$  We use a grid of  1000 points and 200 points randomly sampled within $\cB$ uniformly over radius in place of $\cB$ for solving \eqref{beta-hat} and \eqref{gamma-hat}. }

For E-$\mathrm{TLC}^2$, screening with $\mathrm{TLC}^2$ in the ensemble setting, we follow the recommendations in \cref{sec_prac_delta,sec_prac_lbd} for selecting the tuning parameters and the threshold in \eqref{def_S_set_hat}. Let $\cS$ denote the true set of $\eps$-pure softmax mixtures and $\cS_k$ the subset whose dominant atom is $\theta_k$. For an estimated non-empty set $\wh\cS$, we evaluate its precision, false negative ratio (FN) and total coverage (TC), defined as
\begin{equation}\label{def_metrics}
\mathrm{Precision} = \frac{|\wh\cS\cap\cS|}{|\wh\cS|},
\qquad
\mathrm{FN} = \frac{|\cS\setminus\wh\cS|}{|\cS|},
\qquad
\mathrm{TC} = \frac{1}{K}\sum_{k=1}^K
1\{(\wh\cS\cap\cS_k)\neq\emptyset\}.
\end{equation}
To quantify the consequences of false positives, we also report the average true $S_d$ among the falsely selected mixtures, given by
\[
\mathsf{S}_{\mathrm{FP}}
:=
\frac{1}{|\wh\cS\setminus\cS|}
\sum_{d\in\wh\cS\setminus\cS}S_d.
\]
We set $\mathsf{S}_{\mathrm{FP}}=0$ if $\wh\cS\setminus\cS = \emptyset$.
A small $\mathsf{S}_{\mathrm{FP}}$ indicates that the falsely selected mixtures are still close to being near-pure. For comparison, we also include E-$\mathrm{TLC}^2$-$\mu$ that uses the true $\beta_\mu$ and $\Gamma_\mu$ for computing each $\wh S_d$.

\begin{figure}[ht]
    \centering
    \includegraphics[width=1\linewidth]{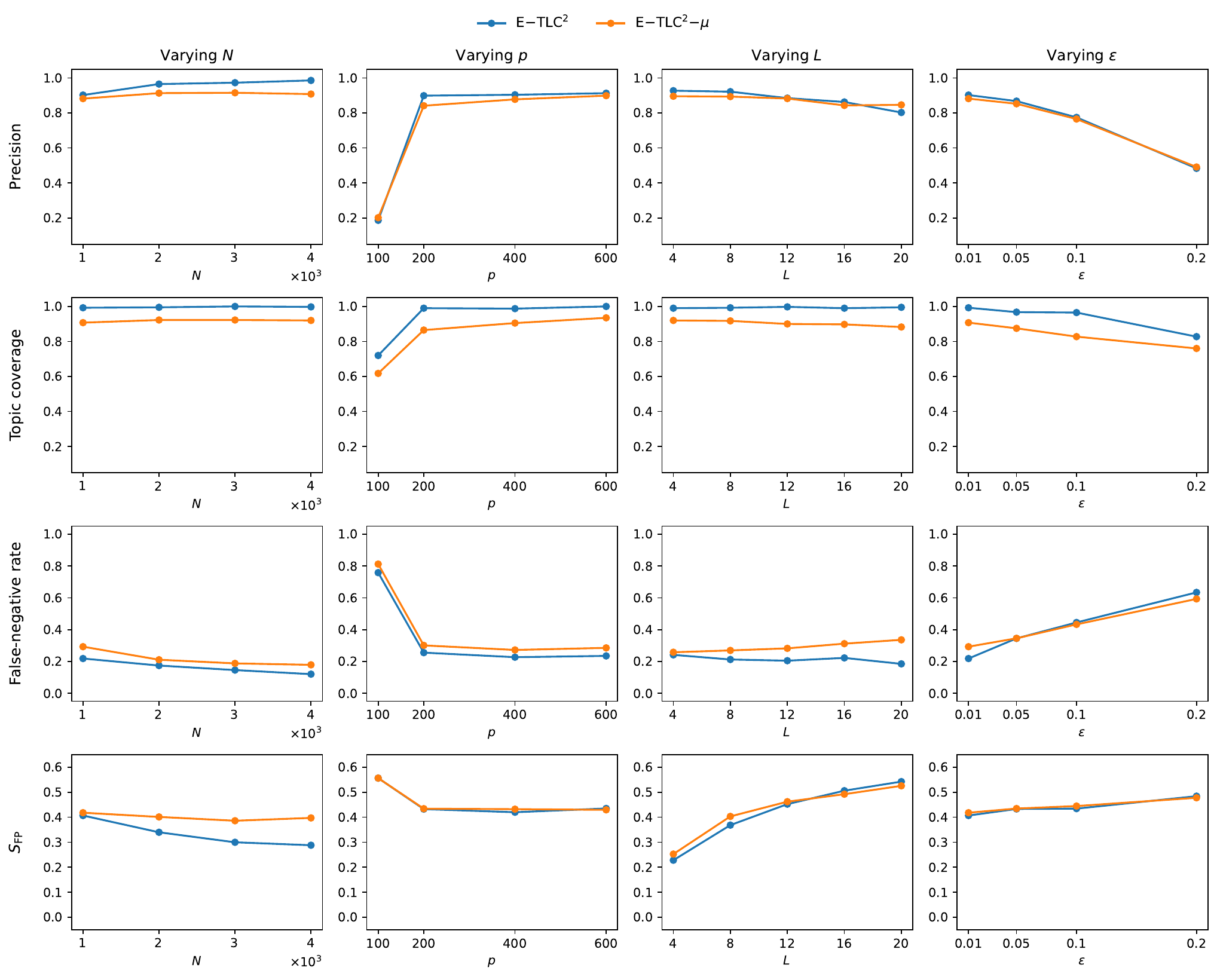}
    \caption{Dependence of estimation error on $N,L,p,\eps$. The baseline setting is $K=4,p=400,N=1000,L=10,\eps=0.01$.}%The rows report  $\mathrm{Precision}$, topic coverage $\mathrm{TC}$, the false negative proportion $\mathrm{FN}$ and averaged estimated HDCD criteria for its false positive documents $\mathsf{S}_{\text{FP}}$ respectively. Each point represents the average over 100 independent replications.}
    \label{fig:sim_vary_all_metrics}
\end{figure}

We vary $N=\{1,2,3,4\}\times10^3$, $p=\{2,4,6,8\}\times10^2$, $L=\{4,8,12,16,20\}$, and $\eps=\{0.01,0.05,0.1,0.2\}$, one at a time. The baseline setting is $K=4$, $p=400$, $N_d=1000$, $L=10$, and $\eps=0.01$ when the corresponding parameter is not varied. For each setting, we report the average metrics over 100 replications in \cref{fig:sim_vary_all_metrics}.

As shown in \cref{fig:sim_vary_all_metrics}, $\mathrm{Precision}$ increases with $N$ and $p$ and decreases with $L$ and $\eps$, indicating improved selection when the sample size and number of support points are larger, the dimension is lower, and the mixtures are closer to being pure. For $\mathrm{TC}$, the effects of $p$ and $\eps$ are most apparent: coverage improves as $p$ increases and as $\eps$ decreases, while remaining close to one across the other settings. For the two metrics where smaller values are preferable, $\mathrm{FN}$ and $\mathsf{S}_{\mathrm{FP}}$ generally decrease with $N$ and $p$ and increase with $L$ and $\eps$. These empirical patterns are consistent with the dependence on $N$, $p$, $L$, and $\eps$ established by our theoretical results. Overall, E-$\mathrm{TLC}^2$ performs consistently better than or comparably to its counterpart using the known $\mu$ across the settings considered.

\section{A real data analysis}
\label{sec_real_data}
We illustrate the proposed method by combining song-level audio
features from the Million Song Dataset (MSD) with user listening records from its associated Taste Profile
Subset\footnote{The data set is available at \url{http://millionsongdataset.com/tasteprofile/}.}. To select songs and users, we first identify users with at least 1,000 recorded plays across at least 300 distinct songs. Among these 1,886 users, we select the $p=5{,}000$ songs
with the largest numbers of distinct listeners, and with this set of songs fixed, we retain  $n=357$ users  who have listened to at least 300 distinct songs. We represent each song by $L=53$ audio features: five  track-level attributes (duration, loudness, tempo, start of fade-out, and end of fade-in), together with the means
and standard deviations of 12 pitch and 12  timbre features across
each song's segments. The pitch features form a chroma vector
representing the relative strengths of the 12 pitch classes, irrespective of octave.
Each segment's vector takes normalized values in $[0,1]$ which describe the distribution
of pitch-class content. The timbre features describe sound color and texture through
12 coefficients representing spectro-temporal patterns.
The first four dimensions are associated with average segment
loudness, brightness, sound flatness, and attack characteristics,
respectively; the remaining dimensions capture additional aspects
of spectral shape and temporal evolution.
For both pitch and timbre, the segment means summarize a song’s average acoustic profile, while the standard deviations quantify variation across segments. Together, these summaries provide interpretable acoustic contrasts. We standardize each feature to zero mean and unit variance across the selected songs.

Given the feature dimension $L=53$, we use the subspace-assisted E-TLC$^2$ procedure in \cref{sec_prac_SS}. However, an exploratory analysis suggests that the feature distribution is not elliptical, making the procedure in \cref{app_sec_SS_ensemble} inapplicable. We therefore estimate the subspace by using EM to fit an ensemble softmax mixture model with an overspecified number of components $K$, as described at the end of \cref{sec_prac_SS}. To choose $K$, we consider $K\in\{4,5,\ldots,10\}$ and find that components with very low  mixing weights, averaged over all users, emerge when $K\ge 6$. For example, when $K=6$, two fitted components have average mixing weights of $0.044$ and $0.038$, whereas the smallest average mixing weight for $K=5$ is $0.144$. We therefore select $K=5$ as the overspecified number of components for fitting the ensemble EM. In fact, additional results in \cref{subsec-msd-sensitivity} show that the subsequent data analysis  is relatively insensitive to this choice.

\begin{figure}[ht]
    \centering
    \includegraphics[width=0.5\linewidth]{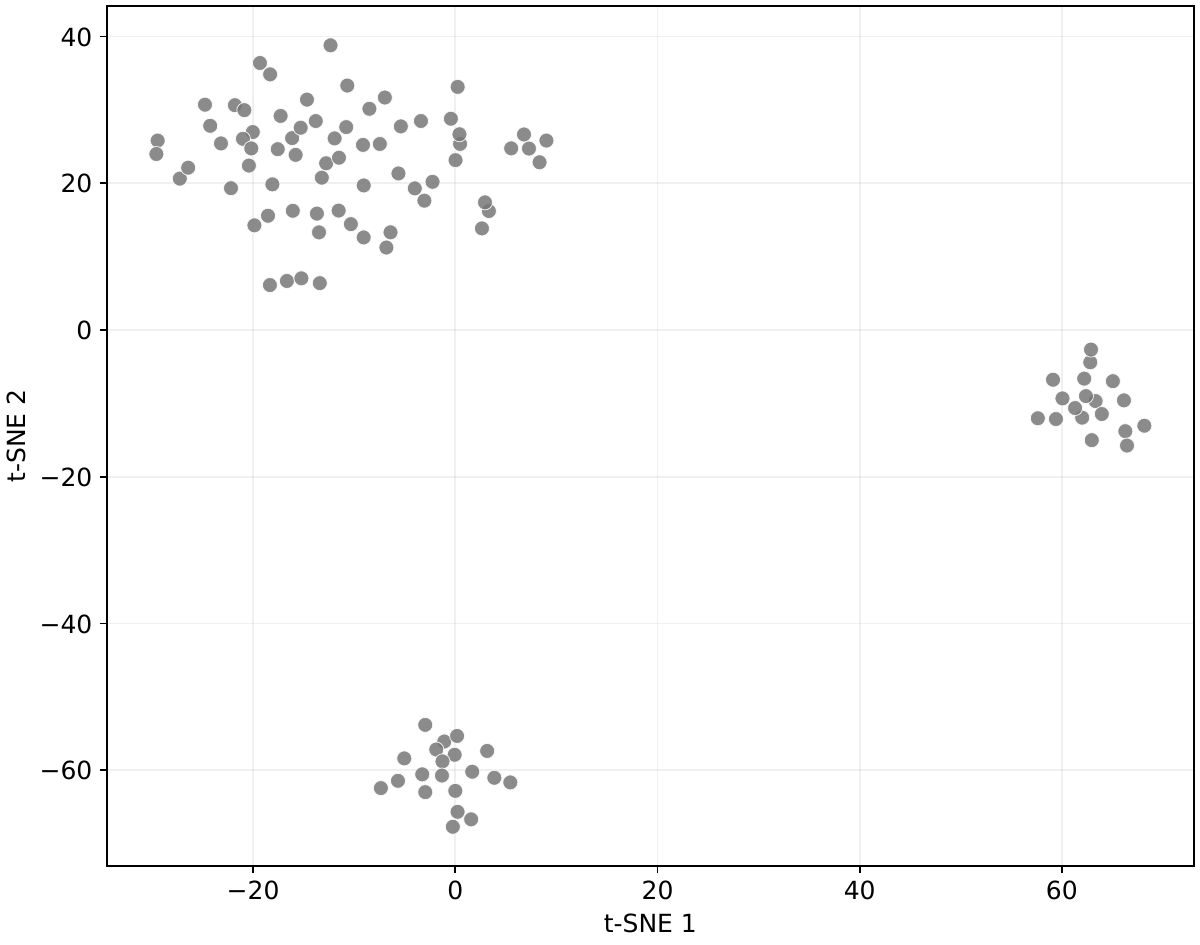}
    \caption{Two-dimensional  visualization of the dominant-atom estimates for the 107 users with the smallest TLC$^2$ scores (30\%), computed using the subspace estimated by EM with $K=5$ components.}
    \label{fig:K5-user-MLE}
\end{figure}

Using the subspace estimated by ensemble EM with $K=5$, we compute TLC$^2$ for each user as described in \cref{sec_prac_SS} and start by examining the 30\% of users with the smallest TLC$^2$. For each of these 107 users, we estimate their corresponding dominant atoms via \eqref{est1}. The two-dimensional t-SNE visualization of these estimates in \cref{fig:K5-user-MLE}, based on cosine distances, suggests three broad groups, motivating the choice of $K_0=3$ clusters in the $k$-means step in \eqref{kmeans} to merge them.
More systematically, following \cref{sec_prac_delta}, we compare the silhouette coefficients for $k$-means with $K_0\in\{3,4,5\}$ and find that the largest coefficient is indeed attained at $K_0=3$. It is worth clarifying that $K_0=3$ here serves as an estimate of the true number of components (at least the ones that have near-pure mixtures in \cref{ass_quasi}), whereas the overspecified choice $K=5$ at the beginning is used only for subspace estimation.

\begin{figure}[ht]
    \centering
    \includegraphics[width=0.5\linewidth]{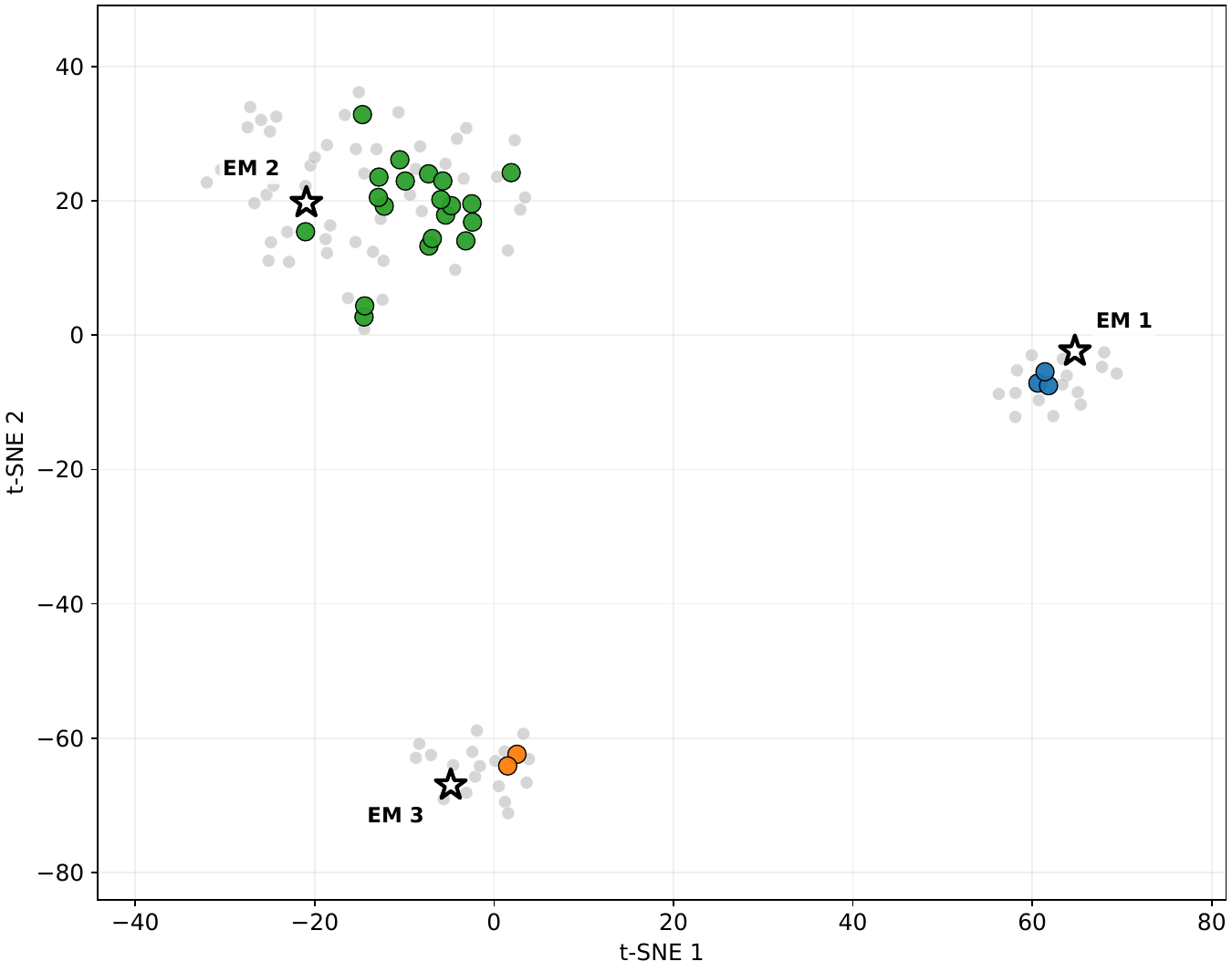}
    \caption{T-SNE visualization of the MLEs for the
107 (30\%)  users with low TLC$^2$ and the three component estimates from the
EM. Colored points represent the 25 selected users, with colors indicating
their cluster assignments; gray points represent the remaining
82 users. Stars denote the three EM estimates of $\theta$.}
    \label{fig:highlight25_EM3}
\end{figure}

With $K_0=3$, the corresponding screening threshold with E-TLC$^2$ in \cref{sec_prac_delta} selects 25 near-pure users, whose estimated dominant atoms are clustered into three groups using $k$-means. \cref{fig:highlight25_EM3} displays the t-SNE visualization of the estimated dominant atoms for the 107 lowest-TLC$^2$ users, with the 25 selected users colored by their cluster assignments and the remaining users shown in gray. For comparison, we also run ensemble EM with $K_0$ components and use the estimated atoms as reference points in \cref{fig:highlight25_EM3}. The highlighted groups lie near the corresponding EM estimates, suggesting that the selected users have listening profiles close to those identified by ensemble EM with this refined number of components. Across all 357 users, TLC$^2$ scores ($\wh S_d$) are negatively associated with the maximum mixing weights ($\|\wh\alpha^{(d)}\|_\i$) estimated by ensemble EM, with Pearson and Spearman correlations of $-0.8193$ and $-0.8518$, respectively. The 25 selected users have an average maximum mixing weight of $0.9997$, compared with $0.7727$ across all users, indicating that they are nearly pure under the EM fit. Together, these results support the use of TLC$^2$ to identify near-pure users and obtain atom estimates close to those from EM, without requiring a full model fit with a correctly specified number of components.

To interpret the centers of the three colored clusters in \cref{fig:highlight25_EM3}, we examine the 50 songs assigned the highest probabilities under the softmax model defined by each center. Representative songs and artists, along with the resulting interpretations, are summarized in \cref{tab:msd_center_interpretation}. Based on these songs, we interpret the three centers as {\sf indie/folk}, {\sf electronic/dance}, and {\sf pop/R\&B/rap} listening profiles, respectively.

\begin{table}[htbp]
\centering
\small
\setlength{\tabcolsep}{4pt}
\renewcommand{\arraystretch}{0.95}
\caption{Representative songs from each center's top 50
and the corresponding musical interpretations.}
\label{tab:msd_center_interpretation}
\begin{tabular}{@{}clll@{}}
\toprule
Center & Artist & Song & Interpretation \\
\midrule
\multirow{4}{*}{1}
 & Camera Obscura
 & ``Other Towns and Cities''
 & \multirow{4}{*}{\shortstack[l]{Indie/folk}} \\
 & Soltero
 & ``Acadian Coast'' & \\
 & Tiny Vipers
 & ``They Might Follow You'' & \\
 & The Cinematic Orchestra
 & ``That Home'' & \\
\midrule
\multirow{4}{*}{2}
 & \multirow{2}{*}{Deadmau5}
 & ``Complications''
 & \multirow{4}{*}{\shortstack[l]{Electronic/dance}} \\
 &
 & ``Not Exactly'' & \\
 & Boys Noize
 & ``The Battery'' & \\
 & Vitalic
 & ``Valletta Fanfares'' & \\
\midrule
\multirow{4}{*}{3}
 & \multirow{2}{*}{Mariah Carey}
 & ``Inseparable''
 & \multirow{4}{*}{\shortstack[l]{Pop/R\&B/rap}} \\
 &
 & ``I'll Be Lovin' U Long Time'' & \\
 & Britney Spears
 & ``3'' & \\
 & Plies
 & ``Goons Lurkin'' & \\
\bottomrule
\end{tabular}
\end{table}

\begin{figure}[htbp]
    \centering
    \includegraphics[width=1\linewidth]{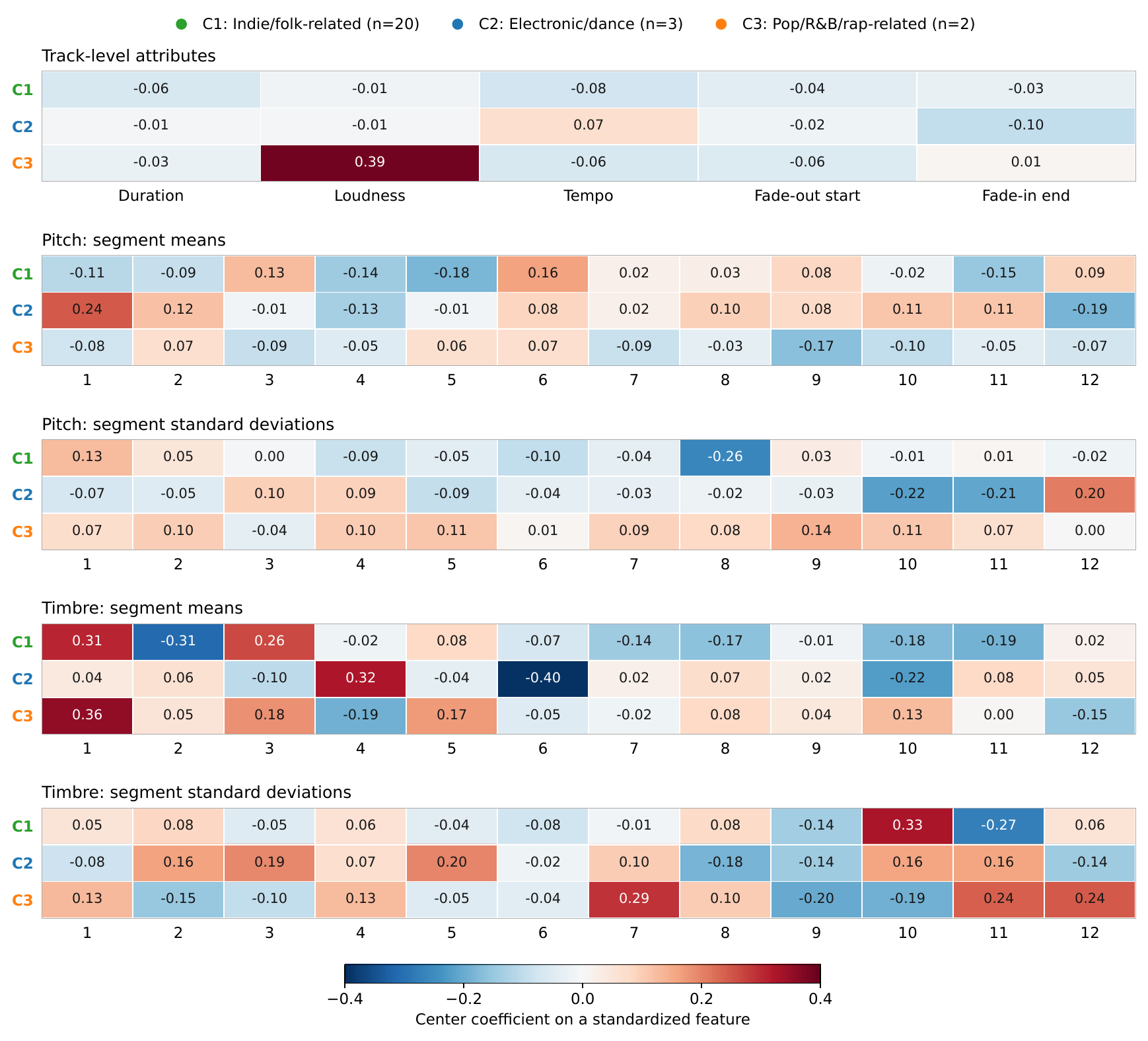}
    \caption{Feature coefficients of the three $k$-means centers obtained
from the  MLEs of the 25 selected users.
The panels display five track-level attributes and the segment
means and standard deviations of the 12 pitch and 12 timbre
features. Cell values and colors indicate center coefficients,
with a common color scale across panels.}
\label{fig:center_feature_weights}
\end{figure}

\cref{fig:center_feature_weights} displays the coefficients of the three estimated cluster centers for all 53 standardized audio features. The centers exhibit distinct feature-weight patterns. The {\sf indie/folk} center has a negative coefficient for the brightness-related timbre dimension, whereas the {\sf electronic/dance} center has a pronounced positive coefficient for the attack-related timbre dimension. The {\sf pop/R\&B/rap} center assigns a substantially larger positive coefficient to loudness. These contrasts are consistent with the interpretations suggested by the representative songs.

\newpage

 \bibliographystyle{abbrvnat}
    \bibliography{ref}

\newpage 
 
 \appendix 
 
We provide proofs section by section in \cref{app_sec_proof_score,app_sec_rate_atom,app_sec_proof_thm_kmeans,app_sec_proof_SS}, with auxiliary lemmas collected in \cref{app_sec_aux}. \cref{app_sim} contains additional simulation results.
 
 \section{Proofs of results in Section \ref{sec_sharp}} \label{app_sec_proof_score}
 \subsection{Proof of Proposition \ref{prop_score}}\label{Propscore} 
 
 \noindent The proof of this proposition follows from the next two lemmas, stated here and proved below.   
       
       \begin{lemma}\label{lem_iff}
        For any finite mixture we have 
        \begin{align*}
             {\Delta_{\min} }(
			1 - \|\alpha\|_2^2
			) ~ \le ~  2 S(G) ~ \le ~  \Delta_{\max}  (
			1 - \|\alpha\|_2^2
			).
        \end{align*}
    \end{lemma}

 \begin{lemma}\label{lem_norms}
      For any $\alpha\in \Delta^{K-1}$,  we have 
      \begin{enumerate}
          \item[(a)] $1- \|\alpha\|_2^2 \le \min_{k\in[K]} \|\alpha - e_k\|_1$.
          \item[(b)] If $2(1-\|\alpha\|_2^2) < 1$, then $\min_{k\in[K]} \|\alpha - e_k\|_1   \le 2(1- \|\alpha\|_2^2)$.
      \end{enumerate}
      % \begin{align*} 
           % 1- \|\alpha\|_2^2 \le \varepsilon \quad \Longrightarrow \quad &\min_{k\in[K]} \|\alpha - e_k\|_1 \le 2\varepsilon \quad  \iff \quad  \alpha_{(1)} \ge 1-\varepsilon,\qquad \forall ~ 0\le \varepsilon <1/2;\\
          %\alpha_{(1)} \ge 1-\varepsilon  \quad  \iff \quad &\min_{k\in[K]} \|\alpha - e_k\|_1 \le 2\varepsilon \quad  \Longrightarrow \quad  1- \|\alpha\|_2^2 \le 2\varepsilon,\qquad \forall ~ 0\le \varepsilon\le 1.
      % \end{align*}  
\end{lemma}

\begin{proof}[Proof of Proposition \ref{prop_score}]
    As a result of \cref{lem_iff,lem_norms}, if  $\Delta_{\min}>0$, then  
        $$\min_{k\in[K]} \|\alpha - e_k\|_1=0 \quad \iff \quad  \|\alpha\|_2 =1 \quad \iff \quad   S(G)=0 .$$
    This is the third claim of  \cref{prop_score}. The other claims follow by combining \cref{lem_iff,lem_norms}. 
\end{proof}

\medskip

\noindent {\it Proof of \cref{lem_iff}}\label{app_sec_proof_lem_iff}. From \eqref{def_S}, we have 
        \begin{align*}
            2S(G) &= 2\tr(\Sigma_G) =   \sum_{k\ne \ell} \alpha_k \alpha_\ell \|\theta_k-\theta_\ell\|^2 \ge \Delta_{\min} (1 - \|\alpha\|_2^2).
        \end{align*} 
        The other direction can be proved analogously. $\square$ \\

\noindent{\it Proof of \cref{lem_norms}}\label{app_sec_proof_lem_norms}. Let $\alpha_{(1)} \ge \alpha_{(2)} \ge \cdots \ge \alpha_{(K)}$ denote the  ordered entries of $\alpha \in \Delta^{K-1}$.
            To prove the first claim,  we have 
            \begin{align*}
                1-\|\alpha\|_2^2  =\sum_k \alpha_{k}(1-\alpha_k) 
                &= \alpha_{{(1)}}(1-\alpha_{{(1)}}) + \sum_{k= 2}^K \alpha_{(k)}(1-\alpha_{(k)})\\
                &\le (1-\alpha_{(1)}) + \sum_{k\ne 1} \alpha_{(k)}\\
                &= 2(1- \alpha_{(1)})\\
                &=  \min_{k\in[K]}\|\alpha - e_k\|_1.
            \end{align*}  % since $1 - \alpha_{(1)} = \sum_{k=2}^K \alpha_{(k)}$.
            To prove the second claim, note that 
            \begin{align*}
                  1- \|\alpha\|_2^2  &=  \sum_{k=1}^K\alpha_k(1-\alpha_k)  = \sum_{k=1}^K\bigl(\alpha_k \vee (1-\alpha_k)\bigr)\bigl(\alpha_k \wedge (1-\alpha_k)\bigr).
            \end{align*} 
            Using $\alpha_k \vee (1-\alpha_k) \ge 1/2$ for all $k\in [K]$ gives 
            \[
                2(1- \|\alpha\|_2^2) \ge \sum_{k=1}^K \bigl(\alpha_k \wedge (1-\alpha_k)\bigr) = \sum_{k: \alpha_k \ge 1/2} (1-\alpha_k) + \sum_{k: \alpha_k < 1/2} \alpha_k.
            \]
            Since there is at most one $k\in [K]$ such that $\alpha_k \ge 1/2$ and $2(1- \|\alpha\|_2^2)< 1$ ensures that there must be one $k\in [K]$ such that $\alpha_k \ge 1/2$ (otherwise the above display leads to contradiction), it then follows 
            \[
                2(1- \|\alpha\|_2^2) \ge \left(1 - \alpha_{(1)}\right) + \sum_{k=2}^K \alpha_{(k)}  = \min_{k\in [K]} \|\alpha - e_k\|_1
            \]
            which proves the second claim, completing the proof. $\square$

\subsection{Proof of \cref{Riesz}}
%and uniqueness of $h$ and $H$ via an RKHS argument.}
\label{proof_Riesz}

 \begin{proof}
Since $f_\mu$ is strictly positive, the formulas in \eqref{eq:explicit-riesz} define measurable mappings. By \cref{ass:mu}, for every $\theta\in\RR^L$, their entries are absolutely integrable against $e^{x^\T\theta}\mu(\dd x)$. Write
\[
M_\mu(\theta):=\int_{\RR^L}e^{x^\T\theta}f_\mu(x)\dd x.
\]
For every $\ell\in[L]$, integration by parts gives
\begin{align*}
\EE_\mu[h_{\mu,\ell}(X)e^{X^\T\theta}]
&=-\int_{\RR^L}\partial_\ell f_\mu(x)e^{x^\T\theta}\dd x\\
&=\theta_\ell\int_{\RR^L}f_\mu(x)e^{x^\T\theta}\dd x
=\theta_\ell M_\mu(\theta).
\end{align*}
Similarly, for every $i,j\in[L]$, two integrations by parts yield
\begin{align*}
\EE_\mu[H_{\mu,ij}(X)e^{X^\T\theta}]
&=\int_{\RR^L}\partial_{ij}f_\mu(x)e^{x^\T\theta}\dd x\\
&=-\theta_i\int_{\RR^L}\partial_jf_\mu(x)e^{x^\T\theta}\dd x\\
&=\theta_i\theta_j M_\mu(\theta).
\end{align*}
Since $M_\mu(\theta)>0$, the proof of \eqref{key-for-est-1} follows.
%For uniqueness, let $h,\widetilde h$ be two representers in the stated integrability class, fix $\ell$. Then
%\[\int_{\RR^L} (h_{\mu,\ell}-\widetilde h_{\mu,\ell})x)e^{x^\T\theta}\mu(\dd x)=0,
%\qquad\theta\in\cB.
%\]
%The finite signed measure $\nu(\dd x):=(h_{\mu,\ell}-\widetilde h_{\mu,\ell})(x)\mu(\dd x)$ has an absolutely convergent Laplace transform on a neighborhood of zero. As this transform vanishes on that neighborhood, uniqueness of Laplace transforms implies $\nu=0$ and hence $h_{\mu,\ell}-\widetilde h_{\mu,\ell}=0$ $\mu$-almost surely. Applying the same argument coordinatewise to $h$ and entrywise to $H$ establishes uniqueness.
\end{proof}

\subsection{Proofs of the results in \cref{sec_score_est}}\label{app_sec_proof_score_est}

For any $\beta\in\RR^p$ and $\Gamma \in \RR^{p\times L}$, recall $\wh S_1(\beta)$ and $\wh S_2(\Gamma)$ from \eqref{def_S1S2_hat}.  Further recall that 
\[
\bar \theta  = \sum_{k=1}^K \alpha_k \theta_k.
\]
The following proposition bounds the estimation errors of
$\wh S_1(\beta)$ and $\wh S_2(\Gamma)$ in terms of the approximation
errors $\beta^\T A_{\theta_k}-\norm{\theta_k}_2^2$ and
$\Gamma^\T A_{\theta_k}-\theta_k$, respectively, for $k\in[K]$.
   
\begin{prop}\label{prop:easy}
For any given $\lambda_\gamma^*,\lambda_\beta^*>0$, let $\beta$ and $\Gamma$ be any fixed quantities, conditional on $\cX$, satisfying
\begin{equation}\label{eq:atom-feasibility}
 \left\|\Gamma^\T A_{\theta_k}-\theta_k\right\|_\i \le \lambda_\gamma^*,
\qquad
\abs{\beta^\T A_{\theta_k}-\norm{\theta_k}_2^2}\le \lambda_\beta^*
\end{equation}
for every $k\in[K]$ such that $\theta_k\in\cB$.
Fix \(\eta\in(0,1)\). Provided \(N\ge\log(56e/\eta)\),  each of the following statements holds  with \(\pi\)-probability at least \(1-\eta\):
\begin{align}
&\left|\wh S_2(\Gamma)-\norm{\bar\theta}_2^2\right|
 ~ \le ~  2\|\Gamma\|_{\i,2}
\sqrt{\frac{2\log(4/\eta)}N}
+32\|\Gamma\|_{\i,2}^2\frac{\log(56e/\eta)}N  +2\sqrt{L}\, \lambda_\gamma^*  
+2L\lambda_\gamma^{*2},
\label{eq:prop-S2}\\
\label{eq:prop-S1}
&\left|\wh S_1(\beta)-\sum_{k=1}^K\alpha_k\norm{\theta_k}_2^2\right|
 ~ \le ~ 
\|\beta\|_\i\sqrt{\frac{2\log(2/\eta)}N}+\lambda_\beta^*,
\end{align}
where $\|\Gamma\|_{\i,2}\le \sqrt{L}\|\Gamma\|_\i$.
\end{prop}
\begin{proof}
    The proof is given in \cref{app_sec_proof_prop:easy}. Inspecting its proof also reveals that \eqref{eq:prop-S2}--\eqref{eq:prop-S1} remain valid when \eqref{eq:atom-feasibility} only holds for all $\theta_k$ whose corresponding $\alpha_k$ is strictly positive.
\end{proof}

\paragraph{Exact tuning parameters.}
 Under \cref{ass:rate}, the uniform second-moment bounds  
\begin{equation}\label{eq:variance-bounds}
\begin{aligned}
v_\gamma^2
&:= \sup_{\|v\|_2=1}\sup_{\theta\in\cB}\EE \left[[v^\T(h_{\mu}(X)-\theta)]^2\,e^{2X^\T\theta}\right],\\
v_\beta^2
&:=~ \sup_{\theta\in\cB}
\EE \left[(\tr[H_{\mu}(X)]-\norm\theta_2^2)^2\, e^{2X^\T\theta}\right]
\end{aligned}
\end{equation}
are finite.
For $\delta\in(0,1)$, set
\begin{equation}\label{eq:radius}
r:= \sqrt{2}\sigma_x\sqrt{L +2\log(8 p/\delta)}.
\end{equation}
Recall $D_\gamma$ and $D_\beta$ from \cref{ass:rate}. 
Define
\begin{equation}\label{eq:truncation-constants}
B_\gamma:=D_\gamma(1+r)^d,
\qquad B_\beta:=D_\beta(1+r)^m,
\end{equation}
and set
\begin{equation}\label{eq:u-rho}
u:=\log\frac{8(L+1)|\cN_{\epsilon}|}{\delta},
\qquad
\rho:=\sqrt2\,\sigma_x
\sqrt{\frac{L +2\log(8/\delta)}p}.
\end{equation}
For $a\in\{\gamma,\beta\}$, choose the tuning parameters
\begin{equation}\label{eq:net-tuning}
\lambda_a:=e^\rho\left[
v_a\left\{\sqrt{\frac{2u}{p}}
+\sqrt{\frac{\delta}{8p}}\right\}
+\frac{4e^r(B_a+1)u}{3p}
\right].
\end{equation}
Finally, define   
\begin{equation}\label{eq:effective}
\Lambda_\gamma:=\lambda_\gamma+(rB_\gamma+1)\epsilon,
\qquad
\Lambda_\beta:=\lambda_\beta+(rB_\beta+2)\epsilon.
\end{equation}

\begin{prop} 
\label{prop:net-feasibility} 
Grant \cref{ass:theta,ass:mu,ass:rate}, and let the tuning parameters
be defined by \eqref{eq:radius}--\eqref{eq:effective}.  Then, with
$\mu$-probability at least $1-\delta/2$, 
\eqref{beta-hat}--\eqref{gamma-hat} admit solutions satisfying
\begin{equation}
\label{eq:net-norm-bound}
\|\wh\Gamma\|_\i\le B_\gamma,
\qquad
\|\wh\beta\|_\i\le B_\beta.
\end{equation}
Moreover, on the same event, uniformly over $\theta\in\cB$,
\begin{equation}
\label{eq:net-uniform}
\left\|\wh\Gamma^\T A_\theta-\theta\right\|_\i 
\le\Lambda_\gamma,
\qquad
\abs{\wh\beta^\T A_\theta-\norm\theta_2^2}
\le\Lambda_\beta.
\end{equation}
\end{prop}
\begin{proof}
    \cref{prop:net-feasibility} is proved in \cref{app_sec_proof_prop:net-feasibility}.
\end{proof}

Combining \cref{prop:easy} with \cref{prop:net-feasibility} readily yields the following detailed version of  \cref{thm_rate_score} on the rate of convergence of the estimation error of  $\wh S = \wh S(\wh\beta,\wh\Gamma)$.
\begin{thm}
\label{thm_rate_score_detailed}
Grant \cref{ass:theta,ass:mu,ass:rate}.  Fix $\delta,\epsilon\in(0,1)$  and
let $\cN_{\epsilon}\subset\cB$ be a deterministic
$\epsilon$-net.  Define $r,B_\gamma,B_\beta,u,\rho$,
$\lambda_\gamma,\lambda_\beta,\Lambda_\gamma$ and $\Lambda_\beta$ by
\eqref{eq:radius}--\eqref{eq:effective}.
If, in addition,
\[
N\ge\log(224e/\delta),
\]
then, with joint probability at least $1-\delta$,
\begin{align}
\label{eq:detailed-net-hdcd-bound}
\abs{\wh S-S(G)}
&\le
\Lambda_\beta+2\sqrt L\,\Lambda_\gamma+2L\Lambda_\gamma^2\notag\\
&\quad+B_\beta\sqrt{\frac{2\log(8/\delta)}N}
+2B_\gamma\sqrt{\frac{2L\log(16/\delta)}N}
+32B_\gamma^2\frac{L\log(224e/\delta)}N.
\end{align}
\end{thm}

\begin{remark}[Simplification of the rates in \cref{thm_rate_score_detailed}]
By \eqref{eq:radius}--\eqref{eq:effective}, the net-cardinality bound
$|\cN_\epsilon|\le(1+2/\epsilon)^L$, 
%and there are functions
%\[
%C_a=C_a\!\left(L,\log p,\log(1/\delta),\log(1/\epsilon)\right),
%\qquad a\in\{\gamma,\beta\},
%\]
%such that
%\[
%\Lambda_a\leC_a\left\{{1\over \sqrt p}+\frac{\exp(c\sigma_x\sqrt L)}p+\epsilon\right\},\qquad a\in\{\gamma,\beta\},
%\]
%after changing the universal constant $c$ if necessary.  The remaining terms in \eqref{eq:detailed-net-hdcd-bound} are of orders $N^{-1/2}$ and $N^{-1}$, up to the same type of fixed-model and logarithmic factors.  Since $N^{-1}\le N^{-1/2}$, $p^{-1}\le p^{-1/2}$,
%$\epsilon^2\le\epsilon$, and the squares and cross-products of
%the displayed tolerance can be absorbed into the same form after
%increasing $c$ and the constants, we obtain
we find after some algebra,
\[
\abs{\wh S-S(G)}
\le
C_1{1\over \sqrt{N\wedge p}}
+C_2\frac{\exp(c\sigma_x\sqrt L)}p
+C_3\epsilon
\]
where the dependence of $C_1, C_2, C_3$ on $L$ is polynomial.
\end{remark}  

\begin{proof}[Proof of \cref{thm_rate_score_detailed}]
The requirements in \eqref{eq:atom-feasibility} of \cref{prop:easy}
are exactly the conclusions of \cref{prop:net-feasibility}.  It remains
to prove the exact score inequality.

We work on the event under which
\cref{prop:net-feasibility} holds.  Conditioning on $\cX$, $\wh\beta$ and $\wh\Gamma$ are fixed and satisfy the uniform bounds in
\eqref{eq:net-uniform} of \cref{prop:net-feasibility}. In particular, they satisfy \eqref{eq:atom-feasibility} of \cref{prop:easy}  with
$\lambda_\gamma^*=\Lambda_\gamma$  and
$\lambda_\beta^*=\Lambda_\beta$.

Apply the two conclusions of \cref{prop:easy} separately with
 $\eta=\delta/4$ each. They then hold
simultaneously with $\pi$-probability at least $1-\delta/2$.
The proposition's sample-size condition is precisely
\[
N\ge\log(56e/\eta)=\log(224e/\delta).
\]
Using  \eqref{eq:proof-coefficient-norms}, \cref{eq:prop-S2} of \cref{prop:easy} gives
\begin{align*}
\left|\wh S_2-\|\bar\theta\|_2^2\right|
&\le2B_\gamma\sqrt{\frac{2L\log(16/\delta)}N}
 +32B_\gamma^2\frac{L\log(224e/\delta)}N +2\sqrt L\,\Lambda_\gamma+2L\Lambda_\gamma^2.
\end{align*}
\cref{eq:prop-S1} of \cref{prop:easy} gives
\[
\left|\wh S_1-\sum_{k=1}^K\alpha_k\norm{\theta_k}_2^2\right|
\le B_\beta\sqrt{\frac{2\log(8/\delta)}N}
 +\Lambda_\beta.
\]
Therefore
\begin{align*}
\abs{\wh S-S(G)}
&\le
\Lambda_\beta+2\sqrt L\,\Lambda_\gamma+2L\Lambda_\gamma^2\\
&\quad+B_\beta\sqrt{\frac{2\log(8/\delta)}N}
+2B_\gamma\sqrt{\frac{2L\log(16/\delta)}N}
+32B_\gamma^2\frac{L\log(224e/\delta)}N,
\end{align*}
which is \eqref{eq:detailed-net-hdcd-bound}.
By \cref{prop:net-feasibility}, the  event has
$\mu$-probability at least $1-\delta/2$, and conditionally on any realization
in this event, the preceding bound holds with probability at least
$1-\delta/2$.  Hence all conclusions of the theorem hold with the
stated probabilities.
\end{proof}

\subsubsection{Proof of \cref{prop:easy}}\label{app_sec_proof_prop:easy}
\begin{proof}
We start with the first result. By the definition of  $\wh S_2(\Gamma)$ in \eqref{def_S1S2_hat} with $\Gamma = (\gamma_1,\ldots,\gamma_L)\in \RR^{p\times L}$,
\begin{align*}
\abs{\wh S_2(\Gamma)-\norm{\bar\theta}_2^2}
&\le
\abs{\sum_{\ell=1}^L(\wh\pi^\T \gamma_\ell)^2-\norm{\bar\theta}_2^2}
+\frac1{N-1}\sum_{\ell=1}^L
\gamma_\ell^\T \Sigma_{\wh \pi}\gamma_\ell,
\end{align*}
where $\Sigma_{\wh\pi}=\diag({\wh\pi})-{\wh\pi}{\wh\pi}^\T$. 
We bound the two terms separately. Observe that
\begin{align*}
\abs{\sum_{\ell=1}^L(\wh\pi^\T\gamma_\ell)^2-\norm{\bar\theta}_2^2}
&=\abs{\sum_{\ell=1}^L
(\wh\pi^\T\gamma_\ell+\bar\theta_\ell)
(\wh\pi^\T\gamma_\ell-\bar\theta_\ell)}\\
&\le2\abs{\bar\theta^\T(\Gamma^\T\wh\pi-\bar\theta)}
+\|\Gamma^\T\wh\pi-\bar\theta\|_2^2
=:\mathrm I+\rII.
\end{align*}
The first term satisfies
\begin{align*}
\rI
&\le2\abs{\bar\theta^\T\Gamma^\T(\wh\pi-\pi)}
+2\abs{\bar\theta^\T(\Gamma^\T\pi-\bar\theta)}\\
&\le2\abs{\bar\theta^\T\Gamma^\T(\wh\pi-\pi)}
+2\|\bar\theta\|_2\|\Gamma^\T\pi-\bar\theta\|_2\\
&\le2\abs{\bar\theta^\T\Gamma^\T(\wh\pi-\pi)}
+2\sqrt{L}\,\lambda_\gamma^*.
\end{align*}
Here \(\|\bar\theta\|_2\le1\), and \eqref{eq:atom-feasibility} implies
\(
\abs{\gamma_\ell^\T\pi-\bar\theta_\ell}
=\abs{\sum_k\alpha_k(\gamma_\ell^\T A_{\theta_k}-\theta_{k\ell})}
\le\lambda_\gamma^*
\).
By Hoeffding's inequality, with $\pi$-probability at least \(1-\eta/2\),
\[
\abs{\bar\theta^\T\Gamma^\T(\wh\pi-\pi)}
\le\|\Gamma\bar\theta\|_\i
\sqrt{\frac{2\log(4/\eta)}N}
\le\|\Gamma\|_{\i,2}
\sqrt{\frac{2\log(4/\eta)}N}.
\]
Consequently, on this event,
\begin{equation}\label{bd_I_prop_easy}
\rI\le2\|\Gamma\|_{\i,2}
\sqrt{\frac{2\log(4/\eta)}N}
+2\sqrt{L}\,\lambda_\gamma^*.
\end{equation}

Regarding $\rII$, use the same decomposition to obtain
\[
\rII = \norm{\Gamma^\T\wh\pi-\bar\theta}_2^2
\le2\norm{\Gamma^\T(\wh\pi-\pi)}_2^2
+2{L}\,\lambda_\gamma^{*2}.
\]
To apply the vector-valued Bernstein's inequality in \cref{lem_bernstein_vector}, note that
\[
\Gamma^\top (\wh\pi-\pi)
=\frac1N\sum_{i=1}^N\Gamma^\T(Z_i-\pi),
\]
where the  $Z_i$ for $i\in [N]$ are i.i.d. random vectors in $\RR^p$ from the categorical distribution with probabilities in $\pi$. Note
\[
\norm{\Gamma^\T(Z_i-\pi)}_2\le2\|\Gamma\|_{\i,2} =:U,
\qquad
\sum_{i=1}^N\EEpi\norm{\Gamma^\T(Z_i-\pi)}_2^2
=N\tr(\Gamma^\T\Sigma_\pi\Gamma)=:\sigma^2.
\]
Moreover,
\[
\tr(\Gamma^\T\Sigma_\pi\Gamma)
\le\sum_{j=1}^p\pi_j\|\Gamma_{j\cdot}\|_2^2\le \|\Gamma\|_{\i,2}^2.
\]
Invoking \cref{lem_bernstein_vector} with these values gives, with conditional \(\pi\)-probability at least \(1-\eta/2\),
\begin{align*}
\norm{\Gamma^\T(\wh\pi-\pi)}_2
&\le2\sqrt{\frac{\tr(\Gamma^\T\Sigma_\pi\Gamma)\log(56/\eta)}N}
+\frac{2\|\Gamma\|_{\i,2}\log(56e/\eta)}N\\
&\le2\|\Gamma\|_{\i,2}\left(\sqrt{\frac{\log(56/\eta)}N}
+\frac{\log(56e/\eta)}N\right).
\end{align*}
%Put \(a=\log(56/\eta)\) and \(b=\log(56e/\eta)=a+1\). Since \(N\ge b\), we have \(b/N\le1\). It follows that, on the same event,
%\begin{align*}
%\rII
%&\le8\|\Gamma\|_{\i,2}^2\left(\sqrt{\frac aN}+\frac bN\right)^2
%+2L\lambda_\gamma^{*2} \le\frac{(32b-16)\|\Gamma\|_{\i,2}^2}N
%+2L\lambda_\gamma^{*2}.
%\end{align*}
%The last inequality uses $b/N\le1$ and $a=b-1$:
%\[
%\left(\sqrt{\frac aN}+\frac bN\right)^2
%\le\frac{2a}{N}+\frac{2b^2}{N^2}
%\le\frac{2a}{N}+\frac{2b}{N}
%=\frac{4b-2}{N}.
%\]
It follows after some algebra that, on the same event,
\begin{align*}
\rII
\le\frac{(32 \log(56e/\eta)-16)\|\Gamma\|_{\i,2}^2}N
+2L\lambda_\gamma^{*2}.
\end{align*}
It remains to bound the correction term in \(\wh S_2\). Since  $\Sigma_{\wh\pi}$ is positive semidefinite and \(N\ge2\),
\begin{align*}
\frac1{N-1}\sum_{\ell=1}^L
\gamma_\ell^\T \Sigma_{\wh\pi} \gamma_\ell
&\le\frac1{N-1}\sum_{j=1}^p\wh\pi_j\|\Gamma_{j\cdot}\|_2^2 \le\frac{\| \Gamma\|_{\i,2} ^2}{N-1}\le\frac{2\|\Gamma\|_{\i,2}^2}N.
\end{align*}
Combining this with \eqref{bd_I_prop_easy} and the bound on \(\rII\), and using \(\| \Gamma\|_{\i,2}\le\sqrt L\|\Gamma\|_\i\), gives \eqref{eq:prop-S2}. 
 
For the second statement, write
\begin{align*}
\left|\wh S_1(\beta)-\sum_{k=1}^K\alpha_k\norm{\theta_k}_2^2\right|
&\le\abs{\beta^\T(\wh\pi-\pi)}
+\abs{\beta^\T\pi-\sum_{k=1}^K \alpha_k\norm{\theta_k}_2^2} \le\abs{\beta^\T(\wh\pi-\pi)}+\lambda_\beta^*.
\end{align*}
Hoeffding's inequality bounds the first term by \(\|\beta\|_\i\sqrt{2\log(2/\eta)/N}\) with  $\pi$-probability at least \(1-\eta\). This proves \eqref{eq:prop-S1}. 
\end{proof}

\subsubsection{Proof of \cref{prop:net-feasibility}}\label{app_sec_proof_prop:net-feasibility}

\begin{proof}
We first establish the feasibility of \eqref{beta-hat} and \eqref{gamma-hat} for $\beta_\mu$ and $\Gamma_\mu$ given in \eqref{one-point} as well as the coefficient-norm bounds using the
reference sample $X_1,\ldots,X_p$, and then extend the constraints from
the finite net to $\cB$.\\

\noindent\emph{Step 1.}
% Let $\mathcal V$ be a fixed $1/2$-net of the Euclidean unit sphere in
% $\RR^L$, with $|\mathcal V|\le 5^L$. For any $x\in\RR^L$,
% \[
% \norm{x}_2\le 2\max_{v\in\mathcal V}|v^\T x|.
% \]
Since $\mu$ is sub-Gaussian, by \cref{lem_quad}, one has
\begin{align}
\PP_\mu(\norm X_2>r)
&\le  \frac{\delta}{8p}. \label{eq:proof-tail}
\end{align}
% By the definition of the sub-Gaussian Orlicz norm in \cref{ass:rate},
% \[
% \PP_\mu(|v^\T X|>s)\le2\exp(-s^2/(2\sigma_x^2))
% \qquad(s>0)
% \]
% for every unit vector $v$. Therefore, a union bound and the explicit value of $r$ in \eqref{eq:radius} yield
% \begin{align}
% \PP_\mu(\norm X_2>r)
% &\le\sum_{v\in\mathcal V}\PP_\mu(|v^\T X|>r/2) \nonumber\\
% &\le2\cdot5^L\exp\left(-\frac{r^2}{8\sigma_x^2}\right)\nonumber\\
% &\le 2\cdot5^L
% \exp\bigl(-L\log5-\log(16p/\delta)\bigr)\nonumber\\
% &=\frac{\delta}{8p}. \label{eq:proof-tail}
% \end{align}
Moreover,  for every fixed $a>0$ and every
nonnegative integer $s$,
\[
\EE_\mu\bigl[(1+\norm X_2)^s e^{a\norm X_2}\bigr]<\infty
\]
For $\theta\in\cB$, we have  $e^{x^\T\theta}\le e^{\norm x_2}$, and  \eqref{eq:poly-growth} implies
\begin{align*}
(h_{\mu,\ell}(x)-\theta_\ell)^2e^{2x^\T\theta} \le \|h_{\mu}(x)-\theta\|_2^2\, e^{2x^\T\theta}\le2\left\{D_\gamma^2(1+\norm x_2)^{2d}+1\right\}
 e^{2\norm x_2},
\end{align*}
and
\begin{align*}
\left(\tr[H_{\mu}(x)]-\norm\theta_2^2\right)^2e^{2x^\T\theta}\le2\left\{D_\beta^2(1+\norm x_2)^{2m}+1\right\}
 e^{2\norm x_2}.
\end{align*}
We conclude that the second moments  \eqref{eq:variance-bounds} are indeed bounded.

\medskip
\noindent\emph{Step 2.}
For $\ell\in[L]$ and $\theta\in\cB$, define
\[
f_{\ell,\theta}(x)
 :=\{h_{\mu,\ell}(x)-\theta_\ell\}e^{x^\T\theta},
\qquad
g_\theta(x)
 :=\{\tr[H_{\mu}(x)]-\norm\theta_2^2\}e^{x^\T\theta}.
\]
By \cref{Riesz}, for all $\ell$ and $\theta$,
\begin{equation}\label{eq:proof-zero-means}
\EE_\mu[f_{\ell,\theta}(X)]=0,
\qquad \EE_\mu[g_\theta(X)]=0.
\end{equation}
We introduce the truncated functions
\[
\widetilde f_{\ell,\theta}(x)
 :=f_{\ell,\theta}(x)1\{ \norm x_2\le r\},
\qquad
\widetilde g_\theta(x)
 :=g_\theta(x)1\{ \norm x_2\le r\}.
\] 
By recalling $B_\gamma$ and $B_\beta$ from \eqref{eq:truncation-constants}, if $ \norm x_2\le r$,  \cref{ass:rate} implies
$|h_{\mu,\ell}(x)|\le B_\gamma$ and $|\tr[H_{\mu}(x)]|\le B_\beta$,
while $|x^\T\theta|\le r$. Hence, for every $x\in\RR^L$ and $\theta\in\cB$,
\begin{equation}\label{eq:proof-envelopes}
|\widetilde f_{\ell,\theta}(x)|\le e^r(B_\gamma+1),
\qquad |\widetilde g_\theta(x)|\le e^r(B_\beta+1).
\end{equation}
Their second moments obey
\begin{align*}
\EE_\mu[\widetilde f_{\ell,\theta}(X)^2]
 &\le\EE_\mu[f_{\ell,\theta}(X)^2]\le v_\gamma^2,\\
\EE_\mu[\widetilde g_\theta(X)^2]
 &\le\EE_\mu[g_\theta(X)^2]\le v_\beta^2.
\end{align*}
Fix $\ell\in[L]$ and $\theta\in\cN_{\epsilon}$. The variables
\[
    \widetilde f_{\ell,\theta}(X_j)
   -\EE_\mu[\widetilde f_{\ell,\theta}(X)]
\]
are independent, have mean zero, and have variance
$\operatorname{Var}_\mu(\widetilde f_{\ell,\theta}(X))
\le v_\gamma^2.
$
By \eqref{eq:proof-envelopes}, they are also bounded by
$|\widetilde f_{\ell,\theta}(X_j)
   -\EE_\mu[\widetilde f_{\ell,\theta}(X)]|\le2e^r(B_\gamma+1)$.
Bernstein's inequality gives
\begin{align}
\PP\left(
\left|\frac1p\sum_{j=1}^p
\left\{\widetilde f_{\ell,\theta}(X_j)
-\EE_\mu[\widetilde f_{\ell,\theta}(X)]\right\}\right|
>v_\gamma\sqrt{\frac{2u}{p}}+\frac{4e^r(B_\gamma+1) u}{3p}
\right)\le2e^{-u}.
\label{eq:proof-bern-gamma}
\end{align}
Applying the same argument to
$\widetilde g_\theta(X_j)-\EE_\mu[\widetilde g_\theta(X)]$
gives
\begin{align}
\PP\left(
\left|\frac1p\sum_{j=1}^p
\left\{\widetilde g_\theta(X_j)
-\EE_\mu[\widetilde g_\theta(X)]\right\}\right|
>v_\beta\sqrt{\frac{2u}{p}}+\frac{4e^r(B_\beta+1) u}{3p}
\right)\le2e^{-u}.
\label{eq:proof-bern-beta}
\end{align}
The union bound over $\ell\in [L]$ and $\theta\in \mathcal N_\epsilon$ in \eqref{eq:proof-bern-gamma}, and over $\theta\in \mathcal N_\epsilon$ in \eqref{eq:proof-bern-beta}, yields an event $\cE_B$ with
\[
\PP(\cE_B^c)
\le2(L+1)|\cN_{\epsilon}| e^{-u}=\frac\delta4
\]
taking 
  $u=\log(8(L+1)|\cN_{\epsilon}|/\delta)$.

\medskip
\noindent\emph{Step 3.}
Define
\begin{equation}\label{def_event_r}
\cE_r:=\left\{\max_{j\in[p]}\norm{X_j}_2\le r\right\}.
\end{equation}
By \eqref{eq:proof-tail} and the union bound,
\begin{equation}\label{eq:proof-sample-event}
\PP(\cE_r^c)
\le p\,\PP_\mu(\norm X_2>r)\le\frac\delta8.
\end{equation}
On $\cE_r$, the original and truncated empirical values agree
for every $j\in[p],\ell\in[L]$ and $\theta\in\cN_\epsilon$.
 By
\eqref{eq:proof-zero-means}, 
\[
\EE_\mu[\widetilde f_{\ell,\theta}(X)]
=-\EE_\mu[f_{\ell,\theta}(X)1\{ \| X\|_2 > r \} ],
\] and
the Cauchy--Schwarz inequality, \eqref{eq:variance-bounds}, and
\eqref{eq:proof-tail} imply
\begin{align*}
\left|\EE_\mu[\widetilde f_{\ell,\theta}(X)]\right|
&\le\{\EE_\mu[f_{\ell,\theta}(X)^2]\}^{1/2}
 \left\{\PP_\mu(\norm X_2>r)\right\}^{1/2} \le v_\gamma\sqrt{\frac\delta{8p}}.
\end{align*}
Similarly,
\[
\left|\EE_\mu[\widetilde g_\theta(X)]\right|
\le v_\beta\sqrt{\frac\delta{8p}}.
\]
On $\cE_B\cap\cE_r$, we therefore have, simultaneously
for $\ell\in[L]$ and $\theta\in\cN_{\epsilon}$,
\begin{align*}
\left|\frac1p\sum_{j=1}^p f_{\ell,\theta}(X_j)\right|
&=\left|\frac1p\sum_{j=1}^p\widetilde f_{\ell,\theta}(X_j)\right|\\
&\le\left|\frac1p\sum_{j=1}^p
\left\{\widetilde f_{\ell,\theta}(X_j)
-\EE_\mu[\widetilde f_{\ell,\theta}(X)]\right\}\right|
 +\left|\EE_\mu[\widetilde f_{\ell,\theta}(X)]\right|\\
&\le v_\gamma\sqrt{\frac{2u}{p}}
 +\frac{4e^r(B_\gamma+1) u}{3p}
 +v_\gamma\sqrt{\frac\delta{8p}}.
\end{align*}
Similarly,
\[
\left|\frac1p\sum_{j=1}^p g_\theta(X_j)\right|
\le v_\beta\sqrt{\frac{2u}{p}}
 +\frac{4e^r(B_\beta+1) u}{3p}
 +v_\beta\sqrt{\frac\delta{8p}}.
\]
In particular, the right-hand sides bound the original numerator
empirical averages uniformly over the required $\epsilon$-net points.

\medskip
\noindent\emph{Step 4.}
For every $\theta\in\cB$, Jensen's inequality yields
the following lower bound for  the denominator term:
\begin{align*}
Z_p(\theta)&:=\frac1p\sum_{j=1}^p e^{X_j^\T\theta}
\ge e^{ \frac1p\sum_{j=1}^pX_j^\T\theta }
 \ge e^{-\norm{\frac1p\sum_{j=1}^p X_j }_2}.
\end{align*}
Since $\frac1p\sum_{j=1}^p X_j$ is $(\sigma_x/\sqrt{p})$ sub-Gaussian, using  $\rho$ in \eqref{eq:u-rho} and \cref{lem_quad} gives
\begin{align*}
\cE_Z :=\left\{\norm{\frac1p\sum_{j=1}^p X_j}_2\le\rho \right\},\qquad  \PP(\cE_Z^c)\le \frac\delta8,
\end{align*}
and on $\cE_Z$,
\begin{equation}\label{eq:proof-denominator}
\inf_{\theta\in\cB}Z_p(\theta)\ge e^{-\rho}.
\end{equation}

\medskip
\noindent\emph{Step 5.}
Consider $\gamma_{\mu,\ell}$ and $\beta_\mu$ in \eqref{one-point}.
For every $\theta\in\cB$,
\begin{align*}
\gamma_{\mu,\ell}^\T A_\theta-\theta_\ell
&=\frac{\frac1p\sum_{j=1}^p
\{h_{\mu,\ell}(X_j)-\theta_\ell\}e^{X_j^\T\theta}}{Z_p(\theta)}
=\frac{\frac1p\sum_{j=1}^p f_{\ell,\theta}(X_j)}{Z_p(\theta)},\\
\beta_\mu^\T A_\theta-\norm\theta_2^2
&=\frac{\frac1p\sum_{j=1}^p g_\theta(X_j)}{Z_p(\theta)}.
\end{align*}
Combining Step~3 and \eqref{eq:proof-denominator}, for all
$\theta\in\cN_{\epsilon}$ and $\ell\in[L]$,
\begin{align*}
|\gamma_{\mu,\ell}^\T A_\theta-\theta_\ell|
&\le e^\rho\left[
 v_\gamma\sqrt{\frac{2u}{p}}+\frac{4e^r(B_\gamma+1) u}{3p}
 +v_\gamma\sqrt{\frac\delta{8p}}\right]
=\lambda_\gamma,\\
|\beta_\mu^\T A_\theta-\norm\theta_2^2|
&\le e^\rho\left[
 v_\beta\sqrt{\frac{2u}{p}}+\frac{4e^r(B_\beta+1) u}{3p}
 +v_\beta\sqrt{\frac\delta{8p}}\right]
=\lambda_\beta.
\end{align*}
Thus $\gamma_{\mu,\ell}$ and $\beta_\mu$ are feasible for
\eqref{gamma-hat} and \eqref{beta-hat}.
On $\cE_r$, the polynomial-growth conditions in \cref{ass:rate} give
\begin{equation}\label{bd_norm_Gamma_mu}
\norm{\gamma_{\mu,\ell}}_\infty\le \|\Gamma_{\mu}\|_{\i,2}\le  B_\gamma,
\qquad \norm{\beta_\mu}_\infty\le B_\beta.
\end{equation}
Optimality of the solutions gives 
\begin{equation}\label{eq:proof-coefficient-norms}
\norm{\wh\gamma_{\ell}}_\infty\le\norm{\gamma_{\mu,\ell}}_\infty
\le B_\gamma,
\qquad
\|\wh\beta\|_\i\le\norm{\beta_\mu}_\infty\le B_\beta.
\end{equation}
This verifies \eqref{eq:net-norm-bound}.

\medskip
\noindent\emph{Step 6.}
For $b\in\RR^p$, define
\[
m_b(\theta):=b^\T A_\theta=\sum_{j=1}^p A_{\theta,j}b_j.
\]
Direct differentiation gives
\[
\nabla A_{\theta,j}
=A_{\theta,j}\left(X_j-\sum_{\ell=1}^p A_{\theta,\ell}X_\ell\right),
\]
and consequently
\[
\nabla m_b(\theta) = (\nabla A_{\theta})^\T b 
=\sum_{j=1}^pA_{\theta,j}b_j
\left(X_j-\sum_{\ell=1}^p A_{\theta,\ell}X_\ell\right).
\]
For any unit vector $v\in \RR^L$, since 
\[
v^\T\nabla m_b(\theta)
= v^\T X^\T \Sigma_{A_\theta} b, 
\]
on the event $\cE_r$, Cauchy--Schwarz inequality gives
\begin{align*}
|v^\T\nabla m_b(\theta)|
&\le\sqrt{v^\T X^\T \Sigma_{A_\theta} X v}
\sqrt{b^\T \Sigma_{A_\theta} b} \le\norm b_\infty r,
\end{align*}
because
\[
b^\T \Sigma_{A_\theta} b 
\le\sum_{j=1}^p A_{\theta,j}b_j^2\le\norm b_\infty^2,
\qquad
v^\T X^\T \Sigma_{A_\theta} X v
\le\sum_{j=1}^p A_{\theta,j}(v^\T X_j)^2\le r^2.
\]
Taking the supremum over all unit vectors $v$ yields
$\norm{\nabla m_b(\theta)}_2\le r\norm b_\infty$.
Since $\cB$ is convex, the mean-value theorem gives
\begin{equation}\label{eq:proof-lipschitz}
|m_b(\theta)-m_b(\eta)|
\le r\norm b_\infty\norm{\theta-\eta}_2,
\qquad\theta,\eta\in\cB.
\end{equation}
Fix $\theta\in\cB$ and choose $\eta\in\cN_{\epsilon}$ with
$\norm{\theta-\eta}_2\le\epsilon$. The first net constraint,
\eqref{eq:proof-coefficient-norms}, and
$|\eta_\ell-\theta_\ell|\le\epsilon$ imply
\begin{align*}
|\wh\gamma_\ell^\T A_\theta-\theta_\ell|
&\le|m_{\wh\gamma_\ell}(\theta)
 -m_{\wh\gamma_\ell}(\eta)|
 +|m_{\wh\gamma_\ell}(\eta)-\eta_\ell|
 +|\eta_\ell-\theta_\ell|\\
&\le rB_\gamma\epsilon+\lambda_\gamma+\epsilon\\
&=\Lambda_\gamma.
\end{align*}
Likewise, since
\[
|\norm\eta_2^2-\norm\theta_2^2|
\le(\norm\eta_2+\norm\theta_2)\norm{\eta-\theta}_2
\le2\epsilon,
\]
we obtain
\begin{align*}
|\wh\beta^\T A_\theta-\norm\theta_2^2|
&\le |m_{\wh\beta}(\theta)-m_{\wh\beta}(\eta)|
 +|m_{\wh\beta}(\eta)-\norm\eta_2^2|
 +|\norm\eta_2^2-\norm\theta_2^2|\\
&\le rB_\beta\epsilon+\lambda_\beta+2\epsilon\\
&=\Lambda_\beta.
\end{align*}
Because $\theta$ was arbitrary, the inequalities hold uniformly on
$\cB$, proving \eqref{eq:net-uniform}.

\medskip
\noindent\emph{Step 7.}
The preceding arguments use the  events
$\cE_B$,   $\cE_r$, and  $\cE_Z$. Their respective failure 
probabilities are bounded by $\delta/4$, $\delta/8$, and $\delta/8$.
Consequently, all conclusions hold on their intersection 
that has probability \(
\PP(\cE_B\cap\cE_r\cap\cE_Z)\ge1-\delta/2\).
\end{proof}

\subsection{Verification of \cref{ass:rate} for Gaussian mixtures}
\label{app:gaussian-location-mixture}
An important example of $\mu$ that satisfies \cref{ass:rate} is Gaussian mixtures. 
Suppose that
\[
\mu(x)=\sum_{j=1}^J p_j\mu_j(x),
\qquad
p_j>0,
\qquad
\sum_{j=1}^J p_j=1,
\]
where $\mu_j$ is the density of $\mathcal N(m_j,\Sigma_j)$,
with $m_j\in\RR^L$ and $\Sigma_j$ positive definite. We assume that
\begin{equation}\label{eq:gaussian-mixture-centered}
\EE[X]= \sum_{j=1}^J p_jm_j=0
\end{equation}
and 
\[
\max_{j\in[J]}\norm{m_j}_2 \le M <\infty,
\qquad
0< \underline{\sigma}^2
\le
\min_{j\in[J]}\lambda_L(\Sigma_j) \le \max_{j\in[J]}\lambda_1(\Sigma_j) \le
\overline{\sigma}^2 <\infty
\]
Then \cref{ass:rate} holds, and the quantities in
\eqref{eq:variance-bounds} admit explicit finite upper bounds.

\paragraph{Verification of part (a).} We first verify the sub-Gaussianity. For any $\norm{u}_2=1$ and
$s\in\RR$,
\begin{align*}
\EE_\mu[e^{s u^\T X}]
&=
\sum_{j=1}^J
p_j
\exp\left\{
s u^\T m_j+\frac{s^2}{2}u^\T\Sigma_j u
\right\} \le
\exp\left(\frac{s^2\overline{\sigma}^2}{2}\right)
\sum_{j=1}^Jp_j e^{s u^\T m_j}.
\end{align*}
Let $V$ be the discrete random variable taking value $u^\T m_j$
with probability $p_j$. By \eqref{eq:gaussian-mixture-centered},
$\EE[V]=0$, while $|V|\le M$. Hence Hoeffding's lemma gives
\[
\sum_{j=1}^Jp_j e^{s u^\T m_j}
=
\EE[e^{sV}]
\le
\exp\left(\frac{s^2M^2}{2}\right).
\]
Therefore,
\[
\EE_\mu[e^{s u^\T X}]
\le
\exp\left\{
\frac{s^2}{2}
\left(
\overline{\sigma}^2+M^2
\right)
\right\},
\]
so one may take
\begin{equation}\label{eq:tau-gaussian-mixture}
\sigma_x^2=\overline{\sigma}^2+M^2.
\end{equation}

\paragraph{Verification of part (b).}
We next verify \eqref{eq:poly-growth}. Define 
\[
w_j(x)
:=
p_j\frac{\mu_j(x)}{\mu(x)},
\qquad j\in[J]
\]
and observe that  $w_j(x)\ge0$ and $\sum_{j=1}^J w_j(x)=1$. Since
\[
\nabla\mu_j(x)
=
-\mu_j(x)\Sigma_j^{-1}(x-m_j),
\]
we find that
\[
h_{\mu}(x)
=
-\frac{\nabla\mu(x)}{\mu(x)}
=  - \sum_{j=1}^J p_j \frac{\nabla\mu_j(x)}{\mu(x)}
=
\sum_{j=1}^J
w_j(x)\Sigma_j^{-1}(x-m_j)
\]
and it follows that
\begin{align*}
% \max_{\ell\in[L]}|h_{\mu,\ell}(x)|
%  \le
\norm{h_{\mu}(x)}_2 
&\le
\sum_{j=1}^J
w_j(x)
\norm{\Sigma_j^{-1}(x-m_j)}_2 \le
\frac{\norm{x}_2+M}{\underline{\sigma}^2} \le
\frac{1+M}{\underline{\sigma}^2}
(1+\norm{x}_2).
\end{align*}
Thus we may take
\begin{equation}\label{eq:Agamma-gaussian-mixture}
d=1,
\qquad
D_\gamma
=
\frac{1+M}{\underline{\sigma}^2}.
\end{equation}
Similarly,
\[
\nabla^2\mu_j(x)
=
\mu_j(x)
\left\{
\Sigma_j^{-1}(x-m_j)(x-m_j)^\T\Sigma_j^{-1}
-
\Sigma_j^{-1}
\right\},
\]
so
\[
H_{\mu}(x) = \sum_{j=1}^J
w_j(x)
\left\{
\Sigma_j^{-1}(x-m_j)(x-m_j)^\T\Sigma_j^{-1}
-
\Sigma_j^{-1}
\right\}
\]
and
\[
\tr[H_{\mu}(x)]
=
\sum_{j=1}^J
w_j(x)
\left\{
(x-m_j)^\T\Sigma_j^{-2}(x-m_j)
-
\tr(\Sigma_j^{-1})
\right\}.
\]
This implies
\begin{align*}
\left|\tr[H_{\mu}(x)]\right|
&\le
\frac{(\norm{x}_2+M)^2}{\underline{\sigma}^4}
+
\frac{L}{\underline{\sigma}^2} \le
\left\{
\frac{(1+M)^2}{\underline{\sigma}^4}
+
\frac{L}{\underline{\sigma}^2}
\right\}
(1+\norm{x}_2)^2.
\end{align*}
Hence we may take
\begin{equation*}%\label{eq:Abeta-gaussian-mixture}
m=2,
\qquad
D_\beta
=
\frac{(1+M)^2}{\underline{\sigma}^4}
+
\frac{L}{\underline{\sigma}^2}.
\end{equation*}

\section{Proofs of the results in \cref{sec_rate_atom}}\label{app_sec_rate_atom}

In \cref{sec_contami}, we analyze  
\begin{equation}\label{def_theta_hat_Theta}
    \wh \theta ~ \in ~ \argmax_{\theta\in \Theta} ~  \sum_{j=1}^p \wh \pi_j \log A_{\theta}(x_j)
\end{equation}
when $Y_1,\ldots,Y_N$ are generated from  a general contamination model, where the contamination distribution is arbitrary and supported on $\cX$, the \ $p\times L$ \  design matrix $\bX:= ( X_1,\ldots,X_p)^\T$ is deterministic, and the true parameter $\theta^*$   belongs to a convex set $ \Theta \subseteq \RR^L$. We assume $\wh\theta$ exists. In \cref{app_sec_proof_thm_softmax} we  specialize the result to the setting of random design $\bX$,   satisfying part (a) of \cref{ass:rate}, with $\Theta=\cB$ of Assumption  \ref{ass:theta}. In \cref{cor_softmax}, we provide a further specialization of  these results to  the softmax mixture contamination (\ref{def_Q_eps}).

\subsection{Non-asymptotic analysis of the MLE under general $\eps$-contamination}\label{sec_contami}

%We introduce first some more notation. 
%Let
%$\bX=(X_1,\ldots,X_p)^\T$.
%For any $\theta \in B_L(0,1)$, %\RR^L$, 
%we write \begin{align} A_{\theta} &:= \left( A(X_1; \theta),\ldots, A(X_p;\theta) \right)^\T := (A_1(\theta), \ldots, A_{\theta,p}),\end{align} for the $p-$dimensional softmax probability vector and
%\begin{equation}\label{def_H_theta}
%		H_{\theta} := \bX^\T
 %       \left[ \diag(A_{\theta}) - A_{\theta}A_{\theta}^\T \right] \bX  
        %\in \RR^{L\times L}.
%	\end{equation}
%	The embedding matrix $E\in \RR^{p\times L}$ satisfies $E^\T 1_p = 0_L$.  
%The following notation will be used frequently in the sequel.
%for the $L\times L$ (Hessian) matrix. 
%Let $\theta^*$ be the true parameter   of the leading component. 

We present our analysis of the MLE $\wh\theta$ in \eqref{def_theta_hat_Theta} under a  general contamination model. Specifically, suppose that $Y_1,\ldots,Y_N$ are i.i.d. random variables taking values in $\cX$ and are generated from
\begin{equation}\label{model_contamination}
    \pi_Q  ~ :=~  (1-\eps) A_{\theta^*} + \eps~  Q
\end{equation}
where $Q\in\Delta^{p-1}$ is an arbitrary distribution on $\cX$ and $\theta^*\in \Theta$. %Let $\wh \pi$ in \eqref{def_theta_hat} denote the empirical frequency of $Y_1,\ldots,Y_N$ from $\pi_Q$. 

Our main result of this section, stated in \cref{thm_softmax_general} below, gives non-asymptotic error bounds of  $\wh \theta$ and, in particular, characterizes the effect of various quantities such as $K$, $L$, $p$, $\eps$, and $Q$. Our non-asymptotic analysis of the M-estimator under the softmax-parametrized model also covers the uncontaminated case $\eps=0$.

Fitting a single-component softmax model  to data generated  from a $(1-\varepsilon)$-dominant mixture 
introduces a bias proportional to the $\eps$-mass assigned to the non-dominant components. To make this precise, we introduce the following notation. For any $\theta \in \RR^L$,  let $H_{\theta}$ be the $L \times L$ Fisher information matrix of a single softmax component by  
\begin{equation}\label{def_H_theta}
    H_{\theta} := \bX^\T
    \left[ \diag(A_{\theta}) - A_{\theta}A_{\theta}^\T \right] \bX.   
\end{equation} 
The design matrix $\bX$ in this subsection may be deterministic or random. In the latter case, our theory is understood conditionally on the observed values of $\bX$. 
As expected, the estimation error of $\wh \theta$ depends crucially on $H_{\theta^*}$. Our analysis also characterizes its impact on the bias due to contamination.  

%{\marten Why not stick to $\theta\in \cB$? This is otherwise confusing. Same for $\| \bX\|_{\i,2}\le1$ - one possibility is to state and prove stronger statement in the proofs section.}

% and requires a nontrivial two-step argument. The first step establishes a (slower) convergence rate for $\|\wh\theta-\theta^*\|_2$ that holds for any maximizer $\wh\theta$ in \eqref{def_theta_hat} once the accuracy of $\wh\pi$ as an estimator of $\pi_Q$ reaches a certain level (see, \cref{thm_theta_det}). This plays a role analogous of proving consistency in the classical analysis of M-estimation where all model parameters are treated as fixed.  The second step refines this rate through a localized analysis once $\wh\theta$ belongs to  a neighborhood of $\theta^*$ (see, \cref{thm_theta_det_fast}). Since our results explicitly characterize the size of this neighborhood, combining both steps yields a fast convergence rate for $\|\wh \theta-\theta^*\|_2$, as stated in the following theorem. 

% For any $\delta \in (0,1)$, define the sequence 
% 	\begin{equation}\label{def_omega}
% 		\omega (\delta)  := 3 \sqrt{\log(28/\delta) \over N}\sqrt{L+  \eps\|\bX H_{\theta^*}^{-1/2}\|_{\i,2}^2}.  
%   %       \sqrt{L} \left\{
% 		% \sqrt{2 %\|H_{\theta^*}\|_\i
%   %       \log(2L/\delta) \over N} + {4  %\|E\|_\i 
%   %       \log(2L/\delta)\over 3N}
% 		% \right\}.
% 	\end{equation}

 \begin{thm}\label{thm_softmax_general}
    Grant model \eqref{model_contamination} with any convex $\Theta \subseteq \RR^L$ in \eqref{def_theta_hat_Theta}. For any $ \delta\in(0,1)$, provided that
    \begin{equation}\label{cond_omega}
          {\|\bX\|_{\i,2} \over \sqrt{\lambda_L(H_{\theta^*})}}  \sqrt{L\log(28e/\delta)\over N}+   {\|\bX\|_{\i,2}^2 \over  \lambda_L(H_{\theta^*})} ~ \eps  \|A_{\theta^*}-Q\|_1   \le {1\over 12},
    \end{equation}
    we have, with probability greater than $1-\delta$, 
    % \begin{align*}
    %     \|\wh \theta - \theta^*\|_2   ~ &\le  ~ {2  \over \sqrt{\lambda_L(H_{\theta^*})}}\left(\omega(\delta) + 3\eps   \|H_{\theta^*}^{-1/2}  \bX^\T (A_{\theta^*}-Q)\|_2 \right). 
    %     %\le {8  \over \lambda_L}\left(\omega(\delta) + \eps   \right)
    % \end{align*} 
    % as well as
    \begin{align}\label{adaptive}
        \max\left\{\bigl\| A_{\wh\theta} - A_{\theta^*}\bigr\|_1, \bigl\|H_{\theta^*}^{1/2}  (\wh \theta - \theta^*)\bigr\|_2\right\}   &\le  6\sqrt{L\log(28e/\delta)\over N} + 6\eps  {\|\bX\|_{\i,2} \over  \sqrt{\lambda_L(H_{\theta^*})}}\|A_{\theta^*}-Q\|_1.
        \end{align}
    %Here $Q$ can be arbitrary distribution in $\Delta^{p-1}$ on $\cX$.
\end{thm} 
\begin{proof}
    See \cref{app_sec_proof_thm_softmax_general}.
\end{proof}

A few remarks are in order.  
The first term in the bound corresponds to the classical parametric estimation error, \(\sqrt{L/N}\), obtained under the single‑softmax parametrization. The second term captures the bias introduced by contamination. This bias depends on the contamination level \(\varepsilon\), the design matrix \(\mathbf{X}\), and the discrepancy between \(A_{\theta^*}\) and \(Q\). Since \(\|A_{\theta^*}-Q\|_1 \le 2\) for any \(Q\), the bias becomes negligible whenever  
\[ 
\varepsilon \le {\sqrt{\lambda_L(H_{\theta^*}) } \over 2\| \bX\|_{\i,2} } \, \sqrt{ L \over N}, 
\]  
in which case the MLE attains the standard parametric rate.

A notable observation is that the MLE is adaptive: if the target distribution \(A_{\theta^*}\) is sufficiently close to the contamination distribution \(Q\) in the sense that  
\[
\eps\|A_{\theta^*}-Q\|_1 \le {\sqrt{\lambda_L(H_{\theta^*})} \over \|\mathbf{X}\|_{\infty,2}}\, \sqrt{ L\over N}
\]  
then the estimation error reduces to \(\mathcal{O}_{\mathbb{P}}(\sqrt{L/N})\).

{Our analysis provides in fact a sharper characterization of the bias, which is governed by $\|\mathbf{X}^\T(A_{\theta^*}-Q)\|_2$ rather than the more coarse quantity \(\|\mathbf{X}\|_{\infty,2}\|A_{\theta^*}-Q\|_1\). For an arbitrary contamination distribution \(Q\), the former satisfies  
\[
\|\mathbf{X}^\T(A_{\theta^*}-Q)\|_2 \le \|\mathbf{X}\|_{\infty,2}\|A_{\theta^*}-Q\|_1,
\]  
and this bound is tight in the worst case. One can thus view  (\ref{adaptive}) as a conservative upper bound, corresponding to the worst-case scenario.  However, when \(Q = Q_{\varepsilon}\), the \((K-1)\)-softmax mixture defined in \eqref{def_Q_eps}, we will develop and discuss a more refined upper bound
for the bias in the next section. } 

\Cref{thm_softmax_general} applies to arbitrary design matrices
$\mathbf X$ and parameters $\theta^*\in\mathbb R^L$, provided that
condition \eqref{cond_omega} holds. For fixed $\mathbf X$, if
$\theta^*=tv$ and $x_j^\top v$ has a unique maximizer over $j\in[p]$,
then $A_{\theta^*}$ converges to the corresponding canonical basis
vector as $t\to\infty$, and $H_{\theta^*}$ converges to zero.
Thus, condition \eqref{cond_omega} becomes more restrictive in
highly concentrated regimes.\\
%In the next section, we derive simplified results for the important setting where \(\mathbf{X}\) is random and Assumptions \ref{ass:theta} and \ref{ass:mu} are met.

%\subsection{Proof of \cref{thm_softmax_general}}

Below  we provide a thorough analysis of the estimator $\wh\theta$ that is obtained from \eqref{def_theta_hat_Theta}. For any $\theta \in \RR^L$, recall that 
    $
        A_{\theta} = (A_{\theta,1},\ldots,A_{\theta,p})^\T \in \Delta^{p-1}
    $
    with 
\[ A_{\theta,j} :=  A_{\theta}(x_j) = {\exp(x_j^\T\theta) \over \sum_{i\in [p]} \exp(x_i^\T \theta)},\qquad    j\in [p]. 
\]
The next subsection contains  useful results on the distances between $A_{\wh\theta}$ and $A_{\theta}$ for any $\theta\in\RR^L$.

\subsubsection{Bounds of statistical distances between $A_{\wh\theta}$ and $A_{\theta}$}
For any $\theta \in \RR^L$, the first lemma bounds from above the KL divergence between $A_{\theta}$ and $A_{\wh\theta}$, that is,
\begin{equation}\label{def_KL}
    \KL(A_{\theta},A_{\wh\theta}) = \sum_{j=1}^p A_{\theta,j} \log {A_{\theta,j}\over A_{\wh\theta,j}}.
\end{equation}
\begin{lemma}\label{lem_KL}
Let $\wh\theta$ be the MLE in \eqref{def_theta_hat_Theta} with  $\Theta$ being any convex set. 
For every $\theta\in\Theta$,
\begin{equation}\label{ident_RHS}
\frac12\|A_{\wh\theta}-A_\theta\|_1^2
 ~ \le~  \KL(A_\theta,A_{\wh\theta})
~ \le~  (\wh\theta-\theta)^\T\bX^\T(\wh\pi-A_\theta).
\end{equation}
Furthermore, constrained first-order optimality implies
\begin{equation}\label{eq_X_pi_Atheta}
(\wh\theta-\theta)^\T\bX^\T(\wh\pi-A_{\wh\theta}) ~ \ge~ 0,
\qquad \forall ~ \theta\in\Theta.
\end{equation}
% and hence
% \begin{equation}\label{ident_RHS2}
% (\wh\theta-\theta)^\T\bX^\T(A_{\wh\theta}-A_\theta)
% \le (\wh\theta-\theta)^\T\bX^\T(\wh\pi-A_\theta).
% \end{equation}
\end{lemma}
\begin{proof}
Pinsker's inequality gives the first inequality in \eqref{ident_RHS}.
By the defining optimality of $\wh\theta$, for any $\theta\in\Theta$,
\[
\sum_{j=1}^p \wh\pi_j\log A_{\wh\theta,j}
\ge \sum_{j=1}^p \wh\pi_j\log A_{\theta,j}.
\]
Adding and subtracting $\wh\pi_j$ in the definition of
the KL divergence therefore gives
\[
\begin{aligned}
\KL(A_\theta,A_{\wh\theta})
&=-\sum_{j=1}^p A_{\theta,j}
  \log\frac{A_{\wh\theta,j}}{A_{\theta,j}}\\
&=\sum_{j=1}^p(\wh\pi_j-A_{\theta,j})
  \log\frac{A_{\wh\theta,j}}{A_{\theta,j}}
  -\sum_{j=1}^p\wh\pi_j
  \log\frac{A_{\wh\theta,j}}{A_{\theta,j}}\\
&\le\sum_{j=1}^p(\wh\pi_j-A_{\theta,j})
  \log\frac{A_{\wh\theta,j}}{A_{\theta,j}}.
\end{aligned}
\]
The softmax parametrization implies
\[
\log\frac{A_{\wh\theta,j}}{A_{\theta,j}}
=x_j^\T(\wh\theta-\theta)
-\log\frac{\sum_{\ell=1}^p\exp(x_\ell^\T\wh\theta)}
           {\sum_{\ell=1}^p\exp(x_\ell^\T\theta)}.
\]
By
$\sum_{j=1}^p(\wh\pi_j-A_{\theta,j})=0$, we obtain
\[ 
\KL(A_\theta,A_{\wh\theta}) \le\sum_{j=1}^p(\wh\pi_j-A_{\theta,j})
       x_j^\T(\wh\theta-\theta) =(\wh\theta-\theta)^\T\bX^\T(\wh\pi-A_\theta). 
\]
Since the loss in \eqref{def_theta_hat_Theta} is concave, the
first-order optimality  is
\[
(\theta-\wh\theta)^\T\bX^\T
(A_{\wh\theta}-\wh\pi)\ge0
\quad\text{for all }\theta\in\Theta,
\]
which proves \eqref{eq_X_pi_Atheta}, completing the proof.
\end{proof}

   %  For future reference, since 
   %  \begin{equation}\label{taylor_log_A}
   %      \log A_{\wh\theta,j} = \log A_{\theta,j} + { A_{\wh\theta,j} - A_{\theta,j} \over A_{\theta,j}}- { (A_{\wh\theta,j} - A_{\theta,j})^2 \over 2\bar A_j^2}
   %  \end{equation}
   %  for some $\bar A_j := u_j A_{\wh\theta,j} + (1-u_j)  A_{\theta,j}$ with $u_j \in [0,1]$, the LHS of \eqref{eq_start} equals to
   %  \[
   %       \sum_{j=1}^{p}  A_{\theta,j}  \left[ - { A_{\wh\theta,j} - A_{\theta,j} \over A_{\theta,j}} + { (A_{\wh\theta,j} - A_{\theta,j})^2 \over 2\bar A_j^2}\right] = {1\over 2}\sum_{j=1}^p {A_{\theta,j}  \over \bar A_j^2} (A_{\wh\theta,j} - A_{\theta,j})^2
   %  \]
   %  so that, by the above proof, 
   %  \begin{equation}\label{bd_chi_sq_A}
   %  {1\over 2}\sum_{j=1}^p {A_{\theta,j}  \over \bar A_j^2} (A_{\wh\theta,j} - A_{\theta,j})^2 ~ \le~   (
			% \wh \theta - \theta
			% )^\T \bX^\T \left(
			% \wh \pi- A_{\theta}
			% \right).
   %  \end{equation}

\cref{lem_KL} together with Cauchy-Schwarz inequality gives the following  bounds for  $\| A_{\wh\theta}- A_{\theta}\|_1$. It can be extended to a result on $\|A_{\theta}-A_{\theta'}\|_1$ for any $\theta,\theta'\in \Theta$. We write $D\succ 0$ to indicate that the matrix $D$ is positive definite.

\begin{lemma}\label{lem_TV}
		For any $\theta \in \Theta$ with $H_\theta\succ 0$, we have 
		\begin{align}\label{rate_A_diff}
			\bigl\| A_{\wh\theta} - A_{\theta}\bigr\|_1^2 &~ \le~ 2  \|H_\theta^{1/2}(\wh \theta - \theta)\|_2 \bigl\|H_\theta^{-1/2}\bX^\T (\wh \pi - A_{\theta})\bigr\|_2.
		\end{align}
    Furthermore, for any $\theta,\theta' \in \Theta$, one has 
    \begin{align}\label{rate_TV_L2} 
        \|A_{\theta} - A_{\theta'}\|_1 \le 2\|\bX\|_{\i,2} \|\theta - \theta'\|_2,
    \end{align}
    and if  $H_\theta\succ 0$, then 
    \begin{align}\label{rate_A_diff_2}
        \bigl\| A_{\theta} - A_{\theta'}\bigr\|_1^2 \le  2\|H_\theta^{1/2}(\theta - \theta') \|_2 \bigl\|H_\theta^{-1/2}\bX^\T (A_{\theta} - A_{\theta'})\bigr\|_2.
    \end{align}
	\end{lemma} 
\begin{proof}
    For any $\theta, \theta' \in \RR^L$, note that
	\[
	\sum_{j=1}^p A_{\theta,j} \log A_{\theta',j} \le \sum_{j=1}^p A_{\theta,j} \log A_{\theta,j},
	\]
	as this is simply a rewriting of the fact that the Kullback-Leibler divergence between the two probability measures $A_{\theta}$ and $A_{\theta'}$ is non-negative.
    This inequality is all we need to be able to repeat 
    the  arguments of proving \cref{lem_KL} %can be used 
    to obtain 
    \[
         \|A_{\theta} - A_{\theta'}\|_1^2\le 2|(\theta'-\theta)^\T\bX^\T (A_{\theta} - A_{\theta'})| \le 2\|\theta'-\theta\|_2  \|\bX\|_{\i,2} \|A_{\theta} - A_{\theta'}\|_1
    \]
    implying \eqref{rate_TV_L2}. The result of \eqref{rate_A_diff_2} is then trivial to obtain from the preceding display.
\end{proof}

\subsubsection{Upper bounds of the estimation error $\|\wh\theta-\theta^*\|_2$}
 
% Bounding $\|\wh\theta-\theta^*\|_2$ from above is more challenging.  
% \cref{thm_theta_det} and \cref{thm_theta_det_fast} both give bounds for $\|\wh \theta-\theta^*\|_2$. The bound in \cref{thm_theta_det_fast} is faster, but requires $\wh \theta$ lies in a specified neighborhood  of $\theta^*$. 
% In turn, \cref{thm_theta_det} is used to ensure this property.
% The price to pay for obtaining a faster rate of $\|\wh \theta - \theta^*\|_2$ is a stronger condition \eqref{cond_E_det} in \cref{cor_theta_det} than  that in \eqref{cond_E_det_init} in \cref{thm_theta_det}. Condition \eqref{cond_E_det} ensures that \eqref{cond_r_det} in \cref{thm_theta_det_fast} holds by using the bound in \eqref{rate_theta_slow}.

%	The main difficulty of  analyzing $\|\wh\theta - \theta^*\|_2$ is to establish an explicit rate of the Lipschitz ``constant'' of the inverse function of the softmax function $A: \RR^L \to \Delta_p$, which translates the bound of $\|\wh\theta - \theta^*\|_2$ to that of  $\|A_{\wh\theta} - A_{\theta^*}\|_1$.  
	
The next result shows that the crux to control $\|\wh \theta-\theta^*\|_2$ is the quantity
$\|H_{\theta^*}^{-1/2}\bX^\T(\wh \pi-A_{\theta^*}) \|_2$.

	\begin{lemma}\label{thm_theta_det}
		Let $\wh \theta$ be defined in \eqref{def_theta_hat_Theta}.
        %by using any $\wh A\in \Delta_p$.
        On the event 
     \begin{equation}\label{cond_E_det_init}
			 4 \|\bX\|_{\i, 2}  \|H_{\theta^*}^{-1/2}\bX^\T (
				\wh \pi-   A_{\theta^*}
                )\|_2  \le  \sqrt{\lambda_L(H_{\theta^*})},
		\end{equation}   
		we have 
		\begin{equation}\label{rate_theta_slow}
			\bigl\|H_{\theta^*}^{1/2}(\wh \theta - \theta^*)\bigr\|_2  \le   2\|H_{\theta^*}^{-1/2}\bX^\T (\wh \pi-A_{\theta^*})\|_2.
		\end{equation}
	\end{lemma}
    \begin{proof}
    Write  
    $r=\|H_{\theta^*}^{1/2}(\wh\theta-\theta^*)\|_2$.
    If $r=0$, the conclusion is immediate, so assume $r>0$.
    
    We first bound the Hessian along the segment
    $\theta_t=\theta^*+t(\wh\theta-\theta^*)$, $t\in[0,1]$, which lies in
    $\Theta$ by convexity. Since
    $\|\wh\theta-\theta^*\|_2\le r/\sqrt{\lambda_L(H_{\theta^*})}$,
    for every $j\in [p]$,
    \[
    \begin{aligned}
    \frac{A_{\theta_t,j}}{A_{\theta^*,j}}
    &=\frac{e^{t x_j^\T(\wh\theta-\theta^*)}}
    {\sum_{\ell=1}^p A_{\theta^*,\ell}\, e^{t x_\ell^\T(\wh\theta-\theta^*)}} \ge e^{-2t\|\bX\|_{\i,2}\|\wh\theta-\theta^*\|_2}
    \ge \exp\left(
    -\frac{2t\|\bX\|_{\i,2}r}{\sqrt{\lambda_L(H_{\theta^*})}}
    \right).
    \end{aligned}
    \]
    The variational characterization of weighted covariance
    therefore gives
    \[
    \begin{aligned}
    (\wh\theta-\theta^*)^\T H_{\theta_t}(\wh\theta-\theta^*)
    &=\min_{c\in\RR}
      \sum_{j=1}^p A_{\theta_t,j}\left(x_j^\T(\wh\theta-\theta^*)-c\right)^2\\
    &\ge \exp\left(
    -\frac{2t\|\bX\|_{\i,2}r}{\sqrt{\lambda_L(H_{\theta^*})}}
    \right)
    \min_{c\in\RR}
    \sum_{j=1}^p A_{\theta^*,j}\left(x_j^\T(\wh\theta-\theta^*)-c\right)^2\\
    &=r^2\exp\left(
    -\frac{2t\|\bX\|_{\i,2}r}{\sqrt{\lambda_L(H_{\theta^*})}}
    \right).
    \end{aligned}
    \]
    By 
    Cauchy--Schwarz inequality,
    \begin{align*}
    r\|H_{\theta^*}^{-1/2}\bX^\T(\wh\pi-A_{\theta^*})\|_2
    &\ge (\wh\theta-\theta^*)^\T\bX^\T(\wh\pi-A_{\theta^*})\\
    &\ge (\wh\theta-\theta^*)^\T\bX^\T(A_{\wh\theta}-A_{\theta^*}) &&\text{by \eqref{eq_X_pi_Atheta}}\\
    &=\int_0^1(\wh\theta-\theta^*)^\T H_{\theta_t}(\wh\theta-\theta^*)\,\dd t\\
    &\ge r^2\int_0^1\exp\left(
    -\frac{2t\|\bX\|_{\i,2}r}{\sqrt{\lambda_L(H_{\theta^*})}}
    \right)\,\dd t.
    \end{align*}
    Since $H_{\theta^*}\succ0$ implies $\|\bX\|_{\i,2}>0$,
    evaluating the integral and dividing by $r>0$ yields
    \[
    \begin{aligned}
    1-\exp\left(
    -\frac{2\|\bX\|_{\i,2}r}{\sqrt{\lambda_L(H_{\theta^*})}}
    \right)
    &\le
    \frac{2\|\bX\|_{\i,2}}{\sqrt{\lambda_L(H_{\theta^*})}}
    \|H_{\theta^*}^{-1/2}\bX^\T(\wh\pi-A_{\theta^*})\|_2 \overset{ \eqref{cond_E_det_init}}{\le} \frac12.
    \end{aligned}
    \] Hence the integrand in the preceding
    bound is at least $1/2$ throughout $[0,1]$. Substituting this
    bound and dividing by $r$ gives
    \[
    r\le
    2\|H_{\theta^*}^{-1/2}\bX^\T(\wh\pi-A_{\theta^*})\|_2,
    \]
    which proves \eqref{rate_theta_slow}.
    \end{proof}

	\subsubsection{Proof of Theorem \ref{thm_softmax_general}: the  rate of convergence of $\|\wh\theta-\theta^*\|_2$ for fixed design}\label{app_sec_proof_thm_softmax_general}

    \begin{proof}  
    We will invoke \cref{lem_TV} and \cref{thm_theta_det}.
    	In view of 
    	\cref{thm_theta_det} and the identity $A_{\theta^*} = \pi + \eps ( A_{\theta^*} -Q)$, both the condition \eqref{cond_E_det_init} and the rate in \eqref{rate_theta_slow} depend on 
        %$\|H_{\theta^*}^{-1/2}\bX^\T (\wh \pi - A_{\theta^*}  ) \|_2$ which is bounded as
        \begin{align}
               \|H_{\theta^*}^{-1/2}\bX^\T (\wh \pi - A_{\theta^*}  ) \|_2
               &\le \|H_{\theta^*}^{-1/2} \bX^\T (\wh \pi-\pi)\|_2 + \eps \| H_{\theta^*}^{-1/2}\bX^\T ( A_{\theta^*}- Q) \|_2
      \label{bd_X_A_Q}\\
      &\le \|H_{\theta^*}^{-1/2} \bX^\T (\wh \pi-\pi)\|_2 + 
       \eps \| \bX H_{\theta^*}^{-1/2}\|_{\i,2} \|A_{\theta^*}-Q\|_1 \nonumber
     \end{align}
    %     &\le \| \bX^\T (\wh \pi-\pi)\|_2 + \eps  
    % \| A_{\theta^*}-Q\|_1	 &&\text{since $\|X_j\|_2\le 1$}\\
    %  &\le \sqrt{L} \max_{k\in [L]} | \bX_{\cdot k}^\T (\wh \pi-\pi) | + \eps  \| A_{\theta^*}-Q\|_1
    We bound $\|H_{\theta^*}^{-1/2} \bX^\T (\wh \pi-\pi)\|_2$ by an application of Bernstein's inequality  \cref{lem_bernstein_vector} with $U=2\|\bX H_{\theta^*}^{-1/2}\|_{\i,2}$ and $\sigma^2 = N\tr(H_{\theta^*}^{-1/2} \bX^\T \Sigma_\pi \bX H_{\theta^*}^{-1/2} )$. Here $\Sigma_\pi:=  \diag(\pi)  - \pi \pi^\T$.
    We find,
for any $0<\delta<1$, with probability at least $1-\delta$, 
    		% \[
    		% \PP\left\{	
    		% \|\bX^\T (\wh \pi-\pi)\|_2 \ge   
      %       \sqrt{t \bX_{\cdot k}^\T \Sigma  \bX_{\cdot k} \over N} + {2t \|\bX_{\cdot k}\|_\i \over 3N} ~ \Big| ~ \bX
    		% \right\} \le 2e^{-t/2},
    		% \]
    \begin{align}\label{een}
    		\|H_{\theta^*}^{-1/2} \bX^\T (\wh \pi-\pi)\|_2 &\le  
            2\sqrt{\tr(H_{\theta^*}^{-1/2} \bX^\T \Sigma_\pi \bX H_{\theta^*}^{-1/2})  (\log(1/\delta)+\log 28)  \over N}\nonumber\\
            &\qquad + { 2\|\bX H_{\theta^*}^{-1/2}\|_{\i,2}~ (\log(1/\delta)+\log (28e))  \over N} 
    \end{align}
   % Integrate out over $\bX$, and we find  that the above also holds unconditionally. 
  %  {\marten This doesn't make sense: 1) the design is fixed; 2) even if it weren't, the RHS bound still depends on X}
   % 
Next, we bound  the variance term as follows. First, we recall
    that $\pi = A_{\theta^*} + \eps(Q - A_{\theta^*})$ and $H_{\theta^*} =\bX^\T (\diag(A_{\theta^*}) -A_{\theta^*}A_{\theta^*}^\T)\bX $, and we find that  
\begin{align*}
    \Sigma_\pi&= \diag(\pi) - \pi\pi^\T\\
    &= \diag (A_{\theta^*}) - {A_{\theta^*}} {A_{\theta^*}}^\T + \eps \diag( Q-A_{\theta^*}) - \eps A_{\theta^*} (Q- A_{\theta^*})^\T - \eps (Q- A_{\theta^*}) A_{\theta^*}^\T\\ &\qquad
    - \eps^2 (Q- A_{\theta^*})(Q- A_{\theta^*})^\T
\end{align*}
By the linearity and cyclic property of the trace, we find    
    \begin{align*}
        &\tr(H_{\theta^*}^{-1/2} \bX^\T \Sigma_\pi \bX H_{\theta^*}^{-1/2})\\ 
        &\le \tr(H_{\theta^*}^{-1/2} H_{\theta^*} H_{\theta^*}^{-1/2}) + \eps  |\tr(H_{\theta^*}^{-1/2} \bX^\T \diag(A_{\theta^*}-Q) \bX H_{\theta^*}^{-1/2})| \\ &\quad+ 2\eps |\tr(H_{\theta^*}^{-1/2} \bX^\T A_{\theta^*}(A_{\theta^*}-Q)^\T \bX H_{\theta^*}^{-1/2})|\\
   %     & \le L + \eps  |\tr(H_{\theta^*}^{-1/2} \bX^\T \diag(A_{\theta^*}-Q) \bX H_{\theta^*}^{-1/2})| + 2\eps |\tr(H_{\theta^*}^{-1/2} \bX^\T A_{\theta^*}(A_{\theta^*}-Q)^\T \bX H_{\theta^*}^{-1/2})|\\
        &\le L + \eps   \|\bX H_{\theta^*}^{-1/2}\|_{\i,2}^2 \|A_{\theta^*}-Q\|_1 + 2\eps \|H_{\theta^*}^{-1/2} \bX^\T A_{\theta^*}\|_2 \|H_{\theta^*}^{-1/2} \bX^\T (A_{\theta^*}-Q)\|_2\\
        &\le L + \eps   \|\bX H_{\theta^*}^{-1/2}\|_{\i,2}^2 \|A_{\theta^*}-Q\|_1 + 2 \eps \|\bX H_{\theta^*}^{-1/2}\|_{\i,2} \|H_{\theta^*}^{-1/2} \bX^\T(A_{\theta^*}-Q)\|_2.
    \end{align*}
    We used the equality  $\|A_{\theta^*}\|_1=1$ and the fact that $\tr( A \diag(v) A^\T) = \tr(\diag(v) A^\T A)= \sum_i v_i \| a_i\|_2^2 \le \| A^\T\|_{\infty,2}^2 \| v\|_1$ for any  $A\in \RR^{m\times n}$ and  $v\in \RR^n$.
    Hence, using the above variance bound in (\ref{een}), we find that,  with probability at least $1-\delta$,
           \begin{align}\label{bd_X_diff_init}\nonumber
               &\| H_{\theta^*}^{-1/2} \bX ^\T (\wh \pi-\pi)\|_2\\\nonumber 
               &\le   2 \sqrt{\log(28/\delta) \over N}\sqrt{L + \eps   \|\bX H_{\theta^*}^{-1/2}\|_{\i,2}^2 \|A_{\theta^*}-Q\|_1}+  { 2 \| \bX H_{\theta^*}^{-1/2}\|_{\i,2} \log(28e/\delta)\over N} \\\nonumber
               & \quad + 2 \sqrt{\log(28/\delta) \over N}\sqrt{  2 \|\bX H_{\theta^*}^{-1/2}\|_{\i,2} ~ \eps \|H_{\theta^*}^{-1/2} \bX^\T(A_{\theta^*}-Q)\|_2} \\\nonumber
               &\le  2 \sqrt{\log(28/\delta) \over N}\sqrt{L + \eps   \|\bX H_{\theta^*}^{-1/2}\|_{\i,2}^2 \|A_{\theta^*}-Q\|_1}+  { 4 \| \bX H_{\theta^*}^{-1/2}\|_{\i,2} \log(28e/\delta)\over N}\nonumber \\
               & \quad +   \eps \|H_{\theta^*}^{-1/2} \bX^\T(A_{\theta^*}-Q)\|_2 %&&\text{using $2\sqrt{ab}\le  {a}+ {b}$}
               \nonumber\\
 %          \end{align}
%    Using \eqref{bd_X_A_Q} and   our assumption \eqref{cond_omega} further bounds from above  the right hand side by 
%    \begin{align*}
        &\le 2 \sqrt{L\log(28/\delta) \over N} +  5 \| \bX H_{\theta^*}^{-1/2}\|_{\i,2} { \log(28e/\delta)\over N}+2\eps \| \bX H_{\theta^*}^{-1/2}\|_{\i,2} \|A_{\theta^*}-Q\|_1\\
        & \le ~  3\sqrt{L\log(28 e/\delta) \over N} + 2\eps \| \bX H_{\theta^*}^{-1/2}\|_{\i,2} \|A_{\theta^*}-Q\|_1 &&\text{by \eqref{cond_omega}} .\nonumber
    \end{align} 
Here, we used the inequalities $\sqrt{a+b}\le \sqrt{a} + \sqrt{b}$ and $2\sqrt{ab}\le {a}+{b}$ for any $a,b\ge0$, and the last inequality used \eqref{cond_omega} to deduce
\[ 5 \| \bX H_{\theta^*}^{-1/2} \|_{\i,2} { \log(28e/\delta)\over N} 
\le
 {5\| \bX  \|_{\i,2}\over{\sqrt{ \lambda_L( H_{\theta^*}) }}}  { \log(28e/\delta)\over N} \le
{5\over12} \sqrt { \log(28e/\delta)\over N}
\]
We can now easily verify that condition \eqref{cond_E_det_init} in \cref{thm_theta_det} is met using display \eqref{bd_X_diff_init} and  our assumption \eqref{cond_omega}. Hence,
\eqref{rate_theta_slow} of \cref{thm_theta_det} implies that with probability at least $1-\delta$, 
            \begin{align*}
    			\|H_{\theta^*}^{1/2}(\wh \theta - \theta^*)\|_2   ~ &\le  2\|H_{\theta^*}^{-1/2}\bX^\T (\wh \pi - A_{\theta^*}  ) \|_2\\   ~&\le 6\sqrt{L\log(28 e/\delta) \over N}  + 6\eps \| \bX H_{\theta^*}^{-1/2}\|_{\i,2} \|A_{\theta^*}-Q\|_1   .
    		\end{align*}
         \cref{lem_TV} with $\theta = \theta^*$ further	implies  
    \begin{align*}
        \| A_{\wh\theta} - A_{\theta^*}\|_1   ~ &\le~ 2\|H_{\theta^*}^{-1/2}\bX^\T (\wh \pi - A_{\theta^*}  ) \|_2 \\ 
    ~&\le ~ 6\sqrt{L\log(28 e/\delta) \over N}  + 6\eps \| \bX H_{\theta^*}^{-1/2}\|_{\i,2} \|A_{\theta^*}-Q\|_1.
    \end{align*} 
            This completes the proof of \cref{thm_softmax_general}.  
    \end{proof}

\subsection{Proof of \cref{thm:softmax}: the  rate of convergence of $\|\wh\theta-\theta^*\|_2$ for random design}\label{app_sec_proof_thm_softmax}

We  specialize \cref{thm_softmax_general} to the setting where we take $\Theta=\cB$ and $\bX$  is random satisfying part (a) of \cref{ass:rate}.  Recall $H_{\theta^*}^{(\mu)} $ from \eqref{def_H_mu} which has the smallest eigenvalue $\sigma_0^2>0$.  Further recall \eqref{def_event_r} that  
\begin{equation}\label{def_event_X_inf_2}
\PP\left\{ \|\bX\|_{\i,2}\le \sqrt{2}\sigma_x\sqrt{L+6\log(p)}\right\} \ge 1-p^{-2}.
\end{equation} 
For any $u\in [0,1]$ and $k\in [K]$, with $\theta_{k,u} = u\theta^* + (1-u)\theta_k$,  the largest eigenvalue of $H_{\theta_{k,u}}^{(\mu)}$ satisfies (see, \citet[Lemma 15]{BBNW})
\[ 
     \lambda_1(H_{\theta_{k,u}}^{(\mu)}) \le \left\|{\EE[XX^\T \exp(X^\T \theta_{k,u})]  \over \EE[\exp(X^\T \theta_{k,u})]}\right\|_\op \le C(1+\sigma_x^2)\sigma_x^2  
\]  
for some universal constant $C>0$. 
Using \citet[Lemma 16]{BBNW} with $\delta=1/2$, $s=5$, $\epsilon = 1/2$, $B=1$ therein also gives:  there exists some constants $C,C'>0$, depending on $\sigma_x$ and $\sigma_0$, such that for $p\ge \max\{4K, C'(L+\log p)^2\}$, one has 
\begin{equation}\label{def_event_H_eig}
    \PP\left\{    \lambda_L(H_{\theta^*})  \ge {\sigma_0^2}/2, ~~ \max_{k\in [K]}\lambda_1(H_{\theta_{k,u}})  \le C \sigma_x^2  \right\}\ge 1-4Kp^{-4}.
\end{equation}

The following lemma strengthens \eqref{def_event_H_eig} to a high-probability statement that holds for $\lambda_1(H_{\theta_{k,u}})$ uniformly over $u\in [0,1]$. 
\begin{lemma}\label{lem:uniform-hessian-path}
%Fix $\theta^*,\theta_k\in\cB$, and let $\theta_{k,u}=u\theta^*+(1-u)\theta_k$ for $u\in[0,1]$.
Assume $\min_{k\in [K]}\lambda_L(H_{\theta_k}^{(\mu)}) \ge \sigma_0^2$, $p\ge\max\{\exp(24\sigma_x^2),4K\}$ and that
the probability bounds in \eqref{def_event_X_inf_2}
and \eqref{def_event_H_eig} hold, the latter for every
fixed $u\in[0,1]$. Then 
\[
\PP\left\{
\min_{k\in [K]}\lambda_L(H_{\theta_k})\ge\frac{\sigma_0^2}{2},
\quad
\max_{k\in [K]}\sup_{u\in[0,1]}\lambda_1(H_{\theta_{k,u}})
\le e^2C\sigma_x^2
\right\}
\ge 1-p^{-1}-p^{-2},
\]
where $C$ is the constant in \eqref{def_event_H_eig}.
\end{lemma}

\begin{proof}
We first apply the pointwise bound on a deterministic grid.
Let $u_j=(j-1)/(p-1)$ for $j\in[p]$.
A union bound in \eqref{def_event_H_eig} gives
\[
\PP_{\cX}\left\{
\min_{k\in [K]}\lambda_L(H_{\theta_k})\ge\frac{\sigma_0^2}{2},
\quad
\max_{j\in[p]}\max_{k\in [K]}\lambda_1(H_{\theta_{k,u_j}})
\le C\sigma_x^2
\right\}\ge1-4K^2p^{-3}\ge 1-p^{-1}.
\]

We next control the change in the Hessian between grid
points. Since $\operatorname{rank}(H_{\theta^*})\le p-1$, $H_{\theta^*}\succ 0$  requires $p\ge L+1$.
Together with $p\ge\exp(24\sigma_x^2)$ and
$\log p\le p-1$, this implies
\[
2\sigma_x^2(L+6\log p)
\le \frac{\log p}{12}(L+6\log p)
\le \frac{7}{12}(p-1)^2.
\]
Thus, on the event in \eqref{def_event_X_inf_2},
$\|\bX\|_{\i,2}\le p-1$.

For every $u\in[0,1]$, choose $j\in[p]$ such that
$|u-u_j|\le1/\{2(p-1)\}$. Since
$\|\theta^*-\theta_k\|_2\le2$,
\[
\|\theta_{k,u}-\theta_{k,u_j}\|_2
=|u-u_j|\|\theta^*-\theta_k\|_2
\le\frac{1}{p-1}.
\]
The softmax formula therefore gives, for every $i\in[p]$,
\[
\begin{aligned}
\frac{A_{\theta_{k,u},i}}{A_{\theta_{k,u_j},i}}
&=
\frac{
\exp\{X_i^\T(\theta_{k,u}-\theta_{k,u_j})\}
}{
\sum_{\ell=1}^p A_{\theta_{k,u_j},\ell}
\exp\{X_\ell^\T(\theta_{k,u}-\theta_{k,u_j})\}
} \le
\exp\left(
2\|\bX\|_{\i,2}
\|\theta_{k,u}-\theta_{k,u_j}\|_2
\right)
\le e^2.
\end{aligned}
\]
To transfer this comparison to the Hessians, for every $v\in\RR^L$, observe that 
\[
\begin{aligned}
v^\T H_{\theta_{k,u}}v
&=
\min_{a\in\RR}
\sum_{i=1}^p A_{\theta_{k,u},i}(v^\T X_i-a)^2\\
&\le
e^2\min_{a\in\RR}
\sum_{i=1}^p A_{\theta_{k,u_j},i}(v^\T X_i-a)^2\\
&=e^2v^\T H_{\theta_{k,u_j}}v.
\end{aligned}
\]
Consequently, on the intersection of the grid event
and the feature-radius event,
\[
\sup_{u\in[0,1]}\lambda_1(H_{\theta_{k,u}})
\le
e^2\max_{j\in[p]}\lambda_1(H_{\theta_{k,u_j}})
\le e^2C\sigma_x^2.
\]
The intersection has probability at least
$1-4Kp^{-2}-p^{-2}$, proving the claim.
\end{proof}

\cref{thm:softmax} corresponds to case 2 of the following more general result. 

 \begin{cor}\label{cor_softmax}
     Consider model \eqref{model_contamination} under Assumptions \ref{ass:theta} and part (a) of  \cref{ass:rate}. Assume $\sigma_0^2>0$.
        Let $\delta, \eps \in (0,1/2)$. 
Assume
\begin{equation}\label{conditions}
\begin{aligned}
    p &~ \ge~ \max\{4K,\exp(24\sigma_x^2), C(\sigma_x^4/\sigma_0^4)(L+\log p)^2\},\\ 
  N   &~ \ge~ C (\sigma_x^2/\sigma_0^2) L(L+\log p) \log(1 /\delta),\\
  \eps &~ \le~ C^{-1} (\sigma_0^2/\sigma_x^2) (L+\log p)^{-1}
\end{aligned} 
\end{equation}
for some sufficiently large constant $C>0$. Then the following holds with  probability at least $1-\delta-2p^{-1}$.
\begin{enumerate}[noitemsep]
    \item For any $Q\in \Delta^{p-1}$,  we have
    \[
        \max\left\{\|A_{\wh\theta} - A_{\theta^*}\|_1,~ \sigma_0 \|\wh\theta-\theta^*\|_2\right\}    \lesssim    \sqrt{L\log(1 /\delta) \over N}    + { \sigma_x  \over \sigma_0 }\eps \sqrt{L+\log p}\, \|A_{\theta^*}-Q\|_1.
    \]
    \item For any $Q = Q_\eps$ given by \eqref{def_Q_eps}, we have
    \begin{align}\label{risk-sm}
        \max\left\{\|A_{\wh\theta} - A_{\theta^*}\|_1,~ \sigma_0 \|\wh\theta-\theta^*\|_2\right\}    \lesssim    \sqrt{L\log(1/\delta) \over N}+  { \sigma_x^2 \over \sigma_0} \sum_{k\ne k^*}{\alpha_k }\|\theta^*-\theta_k\|_2.
    \end{align} 
\end{enumerate}
\end{cor} 

\begin{proof}
On the event in \eqref{def_event_X_inf_2} and the event
$\lambda_L(H_{\theta^*})\ge \sigma_0^2/2$,
condition \eqref{conditions} implies \eqref{cond_omega},
provided that the constant in \eqref{conditions} is sufficiently
large. The first claim therefore follows from
\cref{thm_softmax_general}. For the second claim, work on the event in
\eqref{def_event_X_inf_2} on which
\[
    \lambda_L(H_{\theta^*})\ge \frac{\sigma_0^2}{2},
    \qquad
    \max_{k\in[K]\setminus\{k^*\}}
    \sup_{u\in[0,1]}\lambda_1(H_{\theta_{k,u}})
    \lesssim \sigma_x^2,
\]
where $\theta_{k,u}=u\theta^*+(1-u)\theta_k$.  
\cref{lem:uniform-hessian-path}  shows that this event has
probability at least $1-2p^{-1}$.
Condition on such a realization of $\cX$.
The following bounds hold on the same conditional event of
probability at least $1-\delta$ supplied by
\eqref{bd_X_diff_init}.

We first verify the condition needed to apply
\cref{thm_theta_det}.
Since $\pi-A_{\theta^*}=\eps(Q-A_{\theta^*})$,
the last bound in \eqref{bd_X_diff_init} and the triangle
inequality give
\[
\begin{aligned}
    \|H_{\theta^*}^{-1/2}\bX^\T(\wh\pi-A_{\theta^*})\|_2
    &\le
    3\sqrt{\frac{L\log(28e/\delta)}{N}}
    +3\eps
    \frac{\|\bX\|_{\i,2}}{\sqrt{\lambda_L(H_{\theta^*})}}
    \|A_{\theta^*}-Q\|_1.
\end{aligned}
\]
Consequently, \eqref{cond_omega} implies
\[
\begin{aligned}
    &\frac{4\|\bX\|_{\i,2}}
    {\sqrt{\lambda_L(H_{\theta^*})}}
    \|H_{\theta^*}^{-1/2}\bX^\T(\wh\pi-A_{\theta^*})\|_2\\
    &\quad\le
    12\left\{
    \frac{\|\bX\|_{\i,2}}
    {\sqrt{\lambda_L(H_{\theta^*})}}
    \sqrt{\frac{L\log(28e/\delta)}{N}}
    +\eps\frac{\|\bX\|_{\i,2}^2}
    {\lambda_L(H_{\theta^*})}\|A_{\theta^*}-Q\|_1
    \right\}
    \le 1.
\end{aligned}
\]
Thus, \eqref{cond_E_det_init} holds.

We next sharpen the bound on
$\|H_{\theta^*}^{-1/2}\bX^\T(\wh\pi-A_{\theta^*})\|_2$.
Using $\|A_{\theta^*}-Q\|_1\le 2$ in
\eqref{bd_X_diff_init} and adding the contamination term gives
\[
\begin{aligned}
     \|H_{\theta^*}^{-1/2}\bX^\T(\wh\pi-A_{\theta^*})\|_2 
    & \le
    2\sqrt{\frac{\log(28/\delta)}{N}}
    \sqrt{L+2\eps\|\bX H_{\theta^*}^{-1/2}\|_{\i,2}^2}
    +\frac{4\|\bX H_{\theta^*}^{-1/2}\|_{\i,2}
    \log(28e/\delta)}{N}\\
    &\qquad
    +2\eps\|H_{\theta^*}^{-1/2}
    \bX^\T(A_{\theta^*}-Q)\|_2.
\end{aligned}
\]
On the design event under consideration,
\[
    \|\bX H_{\theta^*}^{-1/2}\|_{\i,2}^2
    \le
    \frac{\|\bX\|_{\i,2}^2}{\lambda_L(H_{\theta^*})}
    \le
    \frac{4\sigma_x^2(L+6\log p)}{\sigma_0^2}.
\]
The bounds on $\eps$ and $N$ in \eqref{conditions} therefore imply
\[
    \eps\|\bX H_{\theta^*}^{-1/2}\|_{\i,2}^2\lesssim 1,
    \qquad
    \frac{\|\bX H_{\theta^*}^{-1/2}\|_{\i,2}
    \log(28e/\delta)}{N}
    \lesssim
    \sqrt{\frac{L\log(1/\delta)}{N}}.
\]
It follows that
\[
    \|H_{\theta^*}^{-1/2}\bX^\T(\wh\pi-A_{\theta^*})\|_2
    \lesssim
    \sqrt{\frac{L\log(1/\delta)}{N}}
    +\eps\|H_{\theta^*}^{-1/2}
    \bX^\T(A_{\theta^*}-Q)\|_2.
\]

To control the contamination term, differentiate
$\bX^\T A_{\theta_{k,u}}$ with respect to $u$ and integrate.
For every $k\in[K]\setminus\{k^*\}$, this gives
\begin{equation}\label{eq_taylor_XA}
    \bX^\T(A_{\theta^*}-A_{\theta_k})
    =
    \int_0^1 H_{\theta_{k,u}}(\theta^*-\theta_k)\,du.
\end{equation}
Since
$\eps(A_{\theta^*}-Q)
=\sum_{k\in[K]\setminus\{k^*\}}
\alpha_k(A_{\theta^*}-A_{\theta_k})$,
we obtain
\[
\begin{aligned}
     \eps\|H_{\theta^*}^{-1/2}
    \bX^\T(A_{\theta^*}-Q)\|_2 
    &\quad\le
    \sum_{k\in[K]\setminus\{k^*\}}\alpha_k
    \int_0^1
    \|H_{\theta^*}^{-1/2}H_{\theta_{k,u}}
    (\theta^*-\theta_k)\|_2\,du\\
    &\quad\lesssim~~
    \frac{\sigma_x^2}{\sigma_0}
    \sum_{k\in[K]\setminus\{k^*\}}
    \alpha_k\|\theta^*-\theta_k\|_2.
\end{aligned}
\]

Finally, since \eqref{cond_E_det_init} holds,
\cref{thm_theta_det,lem_TV} and
$\lambda_L(H_{\theta^*})\ge\sigma_0^2/2$ imply
\[
\begin{aligned}
    \max\left\{
    \|A_{\wh\theta}-A_{\theta^*}\|_1,\,
    \sigma_0\|\wh\theta-\theta^*\|_2
    \right\}
    &\lesssim
    \|H_{\theta^*}^{-1/2}
    \bX^\T(\wh\pi-A_{\theta^*})\|_2\\
    &\lesssim
    \sqrt{\frac{L\log(1/\delta)}{N}}
    +\frac{\sigma_x^2}{\sigma_0}
    \sum_{k\in[K]\setminus\{k^*\}}
    \alpha_k\|\theta^*-\theta_k\|_2.
\end{aligned}
\]
Combining the conditional sampling probability with the
probability of the design event proves the second claim
with probability at least
$1-\delta- p^{-1}-p^{-2}$.
\end{proof}

    \section{Proof of \cref{thm_kmeans} in \cref{sec_ensemble}}\label{app_sec_proof_thm_kmeans}
    \begin{proof}
    For $d\in [n]$, let 
    $$
        t_d = {C_1 \over \sqrt{N_d}} + {C_2\over \sqrt p} +{C_3\over \sqrt{N_{\min}\wedge p}}
    $$
    where $C_1,C_2$ and $C_3$ are specified in \cref{thm_rate_score} with the original $\delta$ replaced by $1/n^2$, so that by a union bound argument
    $\PP(\cE) \ge 1-n^{-1}$ with $\cE$ given in \eqref{E}.  Recall the connection between a given thresholding level and $\varepsilon$ in \eqref{typeI}--\eqref{typeII} of Section \ref{sec_method_screen}. Let
    \[
        \eps_{-}^{(d)} = {t_d \over \Dtmax},\qquad \eps_+^{(d)} = 6{t_d \over \Dtmin}.
    \]
    Further let 
    $$
        \cS_- = \left\{d\in [n]: \pi^{(d)} \text{ is } \eps_-^{(d)}\text{-pure}\right\}
    $$ 
    and let $\cS_{-,k}$ denote the subset of mixtures that have dominant atoms in the $k$-th component. \cref{ass_quasi} ensures $|\cS_{-,k}|\ge 1$ for all $k\in [K]$ so that $|\cS_-| \ge K$. Similarly define $\cS_+$ using $\eps_+^{(d)}$ and $\cS_{+,k}$. Note that $|\cS_+| \ge |\cS_-|\ge K$. Finally, we take $p$ and $N_{\min}$ sufficiently large in \cref{thm_kmeans} to ensure that
    \begin{equation}\label{cond_Nmin_p}
        \max_{d\in [n]}  \, t_d ~ \le~ c \sqrt{\Dtmin \over \Dtmax} {\Dtmin \over \sqrt{|\cS_+|}} {\sigma_0^2 \over \sigma_x^2}
    \end{equation}
    for a sufficiently small  constant $c>0$, as well as \eqref{cond_rate_atom} of \cref{thm:softmax} holds with $N = N_d$, $\eps = \eps^{(d)}_+$ and $\delta = 1/n^2$. 

    Since the above implies $t_d <  \Dtmin/12$, by \eqref{typeI} and its subsequent argument, we find that on the event $\cE$,  
    for any $d\in \wh\cS$, there exists some $k\in [K]$ such that 
    $
        d\in \cS_{+,k}.
    $
    Collect such $d$ in $\wh\cS_k$ and we also have
    \begin{equation}\label{inclusion}
    \wh\cS \subseteq \cS_+,\qquad \text{on $\cE$}.
    \end{equation} 
    Invoking  \cref{thm:softmax} correspondingly for each $d\in \cS_{+,k^*}$ and $k^*\in [K]$,   together with a union bound, gives that on $\cE$, 
    \begin{equation}\label{required_dist}
        \|\wh\theta_d - \theta_{k}\|_2  ~ \lesssim~  {1\over \sigma_0}\sqrt{L\log(n)\over N_d} + {\sigma_x^2 \over \sigma_0^2} {12\sqrt{\Dtmax} \over \Dtmin}\,t_d   ,\qquad \forall~  d\in \wh\cS_k,~ k\in [K]
    \end{equation}     
    with   probability at least $1- |\cS_+|/n^{2} - 2 / p \ge 1-n^{-1} - 2/p$. Note that the union bound argument only affects the $\pi^{(d)}$-probability conditioning on $\cX$ and the random event that has failure probability of $2/p$ depends on $\cX$ which is shared by all mixtures in the ensemble. The first term in \eqref{required_dist} can be absorbed into  the second one by enlarging $C_1,C_2,C_3$ defining $t_d$, if necessary.

    It remains to analyze the $k$-means solution. This follows a classical deterministic analysis (see, for instance, \cite{AwasthiSheffet2012}): on the event that \eqref{required_dist}
        and 
        \begin{equation}\label{total_error}
               \sum_{k=1}^K\sum_{d\in\wh\cS_k}
        \|\wh\theta_d-\theta_k\|_2^2 <  {\Dtmin \over 8}  
        \end{equation}
        hold, then there exists some group permutation $\iota:[K]\to[K]$ such that  
        \begin{align*}
        \max_{k\in [K]}\bigl\|\wt \theta_k - \theta_{\iota(k)}\bigr\|_2^2 &\le  \max_{k\in [K]}  {8\over |\wh\cS_k|}\sum_{k=1}^K \sum_{d\in \wh\cS_k}  \|\wh\theta_d - \theta_{k}\|_2 ^2\\ 
        &\lesssim {\sigma_x^4 \Dt_{\max}\over \sigma_0^4\Dt_{\min}^2}{ |\wh\cS|\over \min_{k\in[K]}|\wh \cS_k| }\max_{d\in[n]}t_d^2 &&\text{by \eqref{required_dist}}.
        \end{align*} 
    Now observe that on $\cE$,   \eqref{total_error}   is ensured by \eqref{required_dist}  and 
    \[
     \frac{\sigma_x^4\Dt_{\max}}
        {\sigma_0^4\Dt_{\min}^2}
        |\wh\cS| \,\max_{d\in[n]}t_d^2 
        ~ \overset{\eqref{inclusion}}{\le}~    \frac{\sigma_x^4\Dt_{\max}}
        {\sigma_0^4\Dt_{\min}^2}
        |\cS_+| \,\max_{d\in[n]}t_d^2   ~ \le ~  c\Dt_{\min} 
    \]
     using \eqref{cond_Nmin_p} for $c$ sufficiently small. To bound from below $|\wh\cS_k|$, using \eqref{typeII} gives $|\wh\cS_k| \ge |\cS_{-,k}| \ge 1$ on $\cE$ so that 
    \[
        \max_{k\in [K]}\bigl\|\wt \theta_k - \theta_{\iota(k)}\bigr\|_2^2   ~ \lesssim ~ {\sigma_x^4 \Dt_{\max}\over \sigma_0^4\Dt_{\min}^2}{ |\cS_+|\over \min_{k\in[K]}|\cS_{-,k}| }\max_{d\in[n]}t_d^2
    \]
    with probability at least $1-2n^{-1}-2p^{-1}$. This completes  the proof. 
    \end{proof}

    \section{Proof of \cref{sec_prac_SS}}\label{app_sec_proof_SS}

	% To bound $\|\wh P - P\|_\op$,  we first note that condition   \eqref{cond_eig_min}  and \eqref{bd_dk} ensure that 
	% $
	% 	\lambda_r(M) \ge c.
	% $
	% On the event $\cE_{\lambda}:=\{\lambda_r(M) \ge 2\|\wh M -M\|_\op\}$, Weyl's inequality thus gives that $  \lambda_r(\wh M) \ge \lambda_r(M)/2$.  
 % 	The Davis-Kahan theorem  further  implies 
	% \begin{equation}\label{bd_P_diff}
	% 	\|\wh P - P\|_\op \le 2\min_{Q\in \OO_{r}} \|\wh V - V Q\|_\op \lesssim {\|\wh M - M\|_\op \over \lambda_r(M)} \lesssim \|\wh M - M\|_\op.
	% \end{equation}
	% As a result, on the event $\cE_r\cap \cE_{\lambda}$, one has  
	% \[
	% 	\rI  \le 2B \| \wt \gamma_\ell   \|_\i \max_{j\in[p]} \|x_j\|_2 \|\wh P -P\|_\op \lesssim  B \| \wt \gamma_\ell   \|_\i\sqrt{L+\log p}~  \|\wh M - M\|_\op
	% \]
	% as well as 

 In this section, we state a general result that is valid for any generic estimator $\wh V$ of $V$. The matrix $V\in \RR^{L\times s}$ has orthonormal columns and needs to satisfy $VV^\T \theta_k = \theta_k$ for all $\theta_k$ with its corresponding weight $\alpha_k>0$. For the single softmax mixture setting in \cref{app_sec_SS_one}, $V$ corresponds to the $s$ leading  eigenvectors of $\sum_k \alpha_k \theta_k \theta_k^\T$ which has rank $s$. For the ensemble setting in \cref{app_sec_SS_ensemble}, we take the span of $V$ as the span of $\theta_1,\ldots,\theta_K$.
 
 Define the estimation error of $\wh V$ as 
 \begin{equation}\label{def_error_V}
    E(\wh V) := \min_{Q\in \bbO_s} \|\wh V - V Q\|_\op := \|\wh V - V\wh Q\|_\op
 \end{equation}
where  $\bbO_s$ denotes the set of all $s\times s$ orthogonal matrices. 
    % It is easy to see that 
    % \begin{equation}\label{bd_P_diff}
    % \|\wh P - P\|_\op \le  2 E(\wh V).
    % \end{equation}
For technical reasons, we split the data $Y_1,\ldots,Y_N$ into two halves of equal size and assume that $\wh V$ is obtained from the first half and that $\mathrm{TLC}^2$ is constructed using the second half. Note that both halves share the same support points in $\cX$. For simplicity, we keep the sample size of each half as $N$ in this section by treating the original sample size as $2N$.

The choice of tuning parameters in \eqref{beta_hat_dr} and
\eqref{gamma_hat_dr} naturally depends on this error.
Fix $\delta\in(0,1)$, and let
$\cN_\epsilon'\subset\cB_s$ be a deterministic $\epsilon$-net.
Recall that $B_a$, $\rho$, $v_a$, and $r$ are given in
\eqref{eq:variance-bounds}--\eqref{eq:u-rho}.
For the subspace result, set
\[
    u:=\log\frac{16(s+1)|\cN_\epsilon'|}{\delta}.
\]
Define
\begin{equation}\label{rate_lbd_dr}
\begin{aligned}
    \lambda_\gamma'
    &:=
    (2rB_\gamma+1)\bigl(E(\wh V)+\epsilon\bigr)
    +\sqrt{s}\,e^\rho\left[
    v_\gamma\left\{
    \sqrt{\frac{2u}{p}}+\sqrt{\frac{\delta}{4p}}
    \right\}
    +\frac{4e^{r_s}(B_\gamma+1)u}{3p}
    \right],\\
    \lambda_\beta'
    &:=
    2rB_\beta E(\wh V)+2(rB_\beta+1)\epsilon
    +e^\rho\left[
    v_\beta\left\{
    \sqrt{\frac{2u}{p}}+\sqrt{\frac{\delta}{4p}}
    \right\}
    +\frac{4e^{r_s}(B_\beta+1)u}{3p}
    \right],
\end{aligned}
\end{equation}
where
\[
    r_s:=\sqrt{2}\sigma_x\sqrt{s+2\log(8p/\delta)}
\]
is chosen such that
\[
    \PP_\mu\left\{
    \max_{j\in[p]}\|V^\T X_j\|_2>r_s
    \right\}\le\frac{\delta}{8}.
\]
This choice replaces the truncation factor $e^r$  in \eqref{eq:net-tuning} by $e^{r_s}$,
reducing the dimension entering this exponential factor from
$L$ to $s$.
Further define
\begin{equation}\label{rate_Lbd_dr}
\begin{aligned}
    \Lambda_\gamma'
    &:=
    (2rB_\gamma+1)\bigl(E(\wh V)+\epsilon\bigr)
    +\lambda_\gamma',\\
    \Lambda_\beta'
    &:=
    2(rB_\beta+1)\bigl(E(\wh V)+\epsilon\bigr)
    +\lambda_\beta'.
\end{aligned}
\end{equation}

  % The choice of tuning parameters in \eqref{beta_hat_dr} and \eqref{gamma_hat_dr} naturally depends on this error.  
  %   Fix $\delta \in (0,1)$.  Recall that $B_a$, $\rho$, $v_a$ and $r$ are given in \eqref{eq:variance-bounds}--\eqref{eq:u-rho}. Define
  %   \begin{equation}\label{rate_lbd_dr}
  %   \begin{aligned}
  %       \lambda_\gamma'&:=   (2 r  B_\gamma +1)(E(\wh V) +\epsilon)+ e^\rho\left[
  %       v_\gamma\left\{\sqrt{\frac{2u}{p}}
  %       +\sqrt{\frac{\delta}{8p}}\right\}
  %       +\frac{4e^{r_s}( B_\gamma+1)u}{3p}
  %       \right]\\
  %       \lambda_\beta'&:=    2 r B_\beta    E(\wh V) + e^\rho\left[
  %       v_\beta\left\{\sqrt{\frac{2u}{p}}
  %       +\sqrt{\frac{\delta}{8p}}\right\}
  %       +\frac{4e^{r_s}(B_\beta+1)u}{3p}
  %       \right]
  %   \end{aligned}
  %   \end{equation}
  %   where
  %   \[
  %   r_s =  \sqrt{2}\sigma_x\sqrt{s +2\log(8 p/\delta)}
  %   \]
  %   is  chosen such that 
  %   \[
  %       \PP_\mu \left\{
  %           \max_{j\in [p]} \|V^\T x_j\|_2 > r_s
  %       \right\}  \le {\delta \over 8}.
  %   \] 
  %   It is this choice of $r_s$ that makes   \eqref{rate_lbd_dr} depend  on $\exp(\sqrt s)$ rather than $\exp(\sqrt L)$. 
  %   Further define 
  %   \begin{equation}\label{rate_Lbd_dr}
  %       \Lambda_\gamma':=   (2r  B_\gamma +1) \left(E (\wh V) +  \epsilon\right) + \lambda_\gamma',\qquad \Lambda_\beta' :=2(r B_\beta + 1)\left(E (\wh V) +  \epsilon\right) + \lambda_\beta'
  %   \end{equation}

  The following theorem states the error rates of TLC$^2$ using the estimated subspace. 

\begin{thm}\label{thm_rate_score_SS}
	Grant \cref{ass:theta,ass:mu,ass:rate}. Fix $\delta,\epsilon\in(0,1)$  and
    let $\cN_{\epsilon}'\subset\cB_s$ be a deterministic
    $\epsilon$-net.  Define  
    $\lambda_a'$ and $\Lambda_a'$ for $a\in \{\beta,\gamma\}$  as in \eqref{rate_lbd_dr}--\eqref{rate_Lbd_dr}. 
    If $
    N\ge 2\log(224e/\delta),$
    then, with joint probability at least $1-\delta$, \eqref{eq:detailed-net-hdcd-bound} of \cref{thm_rate_score_detailed} holds for $\wh S(\wt \beta, \wt\Gamma \wh V^\T)$ with $\Lambda_\beta'$, $\Lambda_\gamma'$, $\sqrt{s} B_\gamma$ in lieu of $\Lambda_\beta$, $\Lambda_\gamma$ and $B_\gamma$, respectively. 
	% \begin{align*}
	% 	\left| \wh S - S(G)\right| \lesssim \sqrt{L\log(p)\log(\sum_t N_t) \over N_t} 
	% 	+   \sqrt{(L\log p)^4 \over p}.
	% \end{align*}
    \end{thm}
 
\begin{proof} 
	We follow the same argument  of proving \cref{thm_rate_score_detailed}. Recall from  \eqref{def_S1S2_hat} that  
    \begin{align*}
    \wh S(\wt\beta,\wt\Gamma\wh V^\T) &= \wh S_1(\wt\beta) - \wh S_2(\wt\Gamma \wh V^\T).
    \end{align*}
    By inspecting the proof of \cref{prop:easy} and the conditional independence between $\wh V$ and the responses $Y_i$'s in the second half of the data that is used to compute TLC$^2$, it is easy to see that the claim in \eqref{eq:prop-S2} continues to hold for 
    $\wh S_2(\Gamma'\wh V^\T)$ with $\Gamma'\in \RR^{p\times s}$ and  $\wh S_1(\beta)$ with $\beta\in \RR^p$ that satisfy   
    \begin{equation}\label{eq:atom-feasibility-subspace}
     \left\| \Gamma^{'\T} A_{\theta_k}-\wh Q^\T V^\T \theta_k\right\|_\i \le \lambda_\gamma^*, \qquad \abs{\beta^\T A_{\theta_k}-\norm{\theta_k}_2^2}\le \lambda_\beta^*
    \end{equation}  
    for each $k\in [K]$ with $\alpha_k > 0$.  Here $\wh Q\in \bbO_s$ is defined in \eqref{def_error_V}.
    
    Since \cref{prop_beta_est_dr} ensures that \eqref{eq:atom-feasibility-subspace}  holds for $\beta = \wt\beta$ and $\Gamma' = \wt \Gamma $ with   $\lambda_\gamma^* = \Lambda_\gamma'$ and $\lambda_\beta^* = \Lambda_\beta'$, the result follows by invoking \cref{prop:easy} with $\beta = \wt\beta$ and $\Gamma = \wt\Gamma \wh V^\T$ together with $\|\wt\Gamma\wh V^\T\|_{\i,2} \le \|\wt\Gamma\|_{\i,2} \le \sqrt{s} \|\wt\Gamma\|_\i \le \sqrt{s} B_\gamma$.
\end{proof}

The following result is the analogue of \cref{prop:net-feasibility} confined in the subspace.

 \begin{prop}\label{prop_beta_est_dr}
 	Grant \cref{ass:mu,ass:theta,ass:rate}.   Choose $\lambda_\gamma'$ and $\lambda_\beta'$ as in \eqref{rate_lbd_dr}. 
 	 Then, with $\mu$-probability at least $1-\delta/2$, 
        \eqref{beta_hat_dr}--\eqref{gamma_hat_dr} admit solutions satisfying
        \begin{equation}
        \label{eq:net-norm-bound-dr}
        \|\wt\Gamma\|_\i\le  B_\gamma,
        \qquad
        \|\wt\beta\|_\i\le B_\beta.
        \end{equation}
        Moreover, on the same event, the following holds for every $k\in[K]$ with $\alpha_k>0$,
        \begin{align}\label{eq:atom-feasibility_dr_gamma}
         \left\|\wt \Gamma^\T A_{\theta_k}-\wh Q^\T V^\T\theta_k\right\|_\i &~ \le~   \Lambda_\gamma' ,\\\label{eq:atom-feasibility_dr_beta}
        \abs{\wt \beta^\T A_{\theta_k}-\norm{\theta_k}_2^2}&~ \le~ \Lambda_\beta'  .
        \end{align} 
 \end{prop}

    \begin{proof}
Regarding the claims for $\wt\Gamma$, we first prove that
$\Gamma_\mu V\wh Q\in\RR^{p\times s}$ is feasible for
\eqref{gamma_hat_dr}, where $\wh Q$ is the aligning
orthogonal matrix in \eqref{def_error_V}.
It suffices to prove
\begin{equation}\label{feasibility_Gamma_mu}
    \max_{\eta\in\cN_\epsilon'}
    \|\wh Q^\T V^\T\Gamma_\mu^\T A_{\wh V\eta}-\eta\|_\i
    \le\lambda_\gamma'.
\end{equation}
On $\cE_r$ in \eqref{def_event_r},
\eqref{bd_norm_Gamma_mu} gives
$\|\Gamma_\mu V\wh Q\|_\i
\le\|\Gamma_\mu\|_{\i,2}\le B_\gamma$.
Comparing $A_{\wh V\eta}$ with $A_{V\wh Q\eta}$, we obtain
\[
\begin{aligned}
    \max_{\eta\in\cN_\epsilon'}
    \|\wh Q^\T V^\T\Gamma_\mu^\T A_{\wh V\eta}-\eta\|_\i 
    & \le
    \|\Gamma_\mu V\wh Q\|_\i
    \max_{\eta\in\cN_\epsilon'}
    \|A_{\wh V\eta}-A_{V\wh Q\eta}\|_1 +\sup_{\zeta\in\cB_s}
    \|\wh Q^\T
    (V^\T\Gamma_\mu^\T A_{V\zeta}-\zeta)\|_\i\\
    &\le
    2rB_\gamma E(\wh V)
    +\sup_{\zeta\in\cB_s}
    \|V^\T\Gamma_\mu^\T A_{V\zeta}-\zeta\|_2.
\end{aligned}
\]
The last step uses \eqref{rate_TV_L2} and
$\|\wh Q^\T z\|_\i\le\|z\|_2$.

To bound the supremum, repeat the truncated Bernstein argument
in the proof of \cref{prop:net-feasibility} for the fixed basis
$V$ and the deterministic net $\cN_\epsilon'$.
For $\eta\in\cN_\epsilon'$ and $\ell\in[s]$, apply that argument to
\[
    \bigl\{(V^\T h_\mu(X))_\ell-\eta_\ell\bigr\}
    e^{X^\T V\eta},
    \qquad
    \bigl\{\tr[H_\mu(X)]-\|\eta\|_2^2\bigr\}
    e^{X^\T V\eta},
\]
truncating both quantities to
$\{\|X\|_2\le r,\ \|V^\T X\|_2\le r_s\}$.
Their means are zero by \cref{Riesz}, their second moments
are bounded by $v_\gamma^2$ and $v_\beta^2$, respectively,
and their truncated absolute values are bounded by
$e^{r_s}(B_\gamma+1)$ and $e^{r_s}(B_\beta+1)$.

The probability of truncation for one observation is at most
$\delta/(4p)$, so the corresponding truncation biases are at most
$v_a\sqrt{\delta/(4p)}$ for $a\in\{\gamma,\beta\}$.
With the stated choice of $u$, the Bernstein union bound over
the $(s+1)|\cN_\epsilon'|$ quantities has failure probability
at most $\delta/8$.
The two sample-radius events together have failure probability
at most $\delta/4$, and the denominator event in
\eqref{eq:proof-denominator} has failure probability at most
$\delta/8$.
Thus, with probability at least $1-\delta/2$, these events hold
simultaneously, and
\[
\begin{aligned}
    \max_{\eta\in\cN_\epsilon'}
    \|V^\T\Gamma_\mu^\T A_{V\eta}-\eta\|_\i 
    &\le
    e^\rho\left[
    v_\gamma\left\{
    \sqrt{\frac{2u}{p}}+\sqrt{\frac{\delta}{4p}}
    \right\}
    +\frac{4e^{r_s}(B_\gamma+1)u}{3p}
    \right],\\
    \max_{\eta\in\cN_\epsilon'}
    |\beta_\mu^\T A_{V\eta}-\|\eta\|_2^2|
    &\le
    e^\rho\left[
    v_\beta\left\{
    \sqrt{\frac{2u}{p}}+\sqrt{\frac{\delta}{4p}}
    \right\}
    +\frac{4e^{r_s}(B_\beta+1)u}{3p}
    \right].
\end{aligned}
\]
We work on this common event for the remainder of the proof.

We next extend the first bound to a uniform result over $\eta\in \cB_s$ in the $\ell_2$-norm.
Since $\|\Gamma_\mu V\|_{\i,2}\le B_\gamma$,
\eqref{rate_TV_L2} implies, for any $\eta,\zeta\in\cB_s$,
\[
    \|V^\T\Gamma_\mu^\T(A_{V\eta}-A_{V\zeta})\|_2
    \le 2rB_\gamma\|\eta-\zeta\|_2.
\]
Choosing a point of $\cN_\epsilon'$ within distance $\epsilon$
of each $\eta\in\cB_s$ therefore gives
\[
\begin{aligned}
    \sup_{\eta\in\cB_s}
    \|V^\T\Gamma_\mu^\T A_{V\eta}-\eta\|_2
    &\le
    \sqrt{s}\max_{\eta\in\cN_\epsilon'}
    \|V^\T\Gamma_\mu^\T A_{V\eta}-\eta\|_\i
    +(2rB_\gamma+1)\epsilon.
\end{aligned}
\]
Combining these bounds with \eqref{rate_lbd_dr} proves
\eqref{feasibility_Gamma_mu}.
Optimality in \eqref{gamma_hat_dr} then gives
\[
    \|\wt\Gamma\|_\i
    \le\|\Gamma_\mu V\wh Q\|_\i\le B_\gamma
\]
which is the first claim in \eqref{eq:net-norm-bound-dr}.

For $\wt\beta$, the bound $\|\beta_\mu\|_\i\le B_\beta$,
\eqref{rate_TV_L2}, and
$|\|\eta\|_2^2-\|\zeta\|_2^2|
\le2\|\eta-\zeta\|_2$
give
\[
\begin{aligned}
    \sup_{\eta\in\cB_s}
    |\beta_\mu^\T A_{V\eta}-\|\eta\|_2^2|
    &\le
    \max_{\eta\in\cN_\epsilon'}
    |\beta_\mu^\T A_{V\eta}-\|\eta\|_2^2|
    +2(rB_\beta+1)\epsilon.
\end{aligned}
\]
Since $\|\wh Q\eta\|_2=\|\eta\|_2$, it follows that
\begin{equation}\label{feasibility_beta_mu}
     \max_{\eta\in\cN_\epsilon'}
    |\beta_\mu^\T A_{\wh V\eta}-\|\eta\|_2^2| 
     \le
    2rB_\beta E(\wh V)
    +\sup_{\zeta\in\cB_s}
    |\beta_\mu^\T A_{V\zeta}-\|\zeta\|_2^2|
    \le\lambda_\beta'.
\end{equation}     
Thus, $\beta_\mu$ is feasible for \eqref{beta_hat_dr}, and
optimality gives $\|\wt\beta\|_\i\le B_\beta$.
This proves \eqref{eq:net-norm-bound-dr}.

To prove   \eqref{eq:atom-feasibility_dr_gamma}--\eqref{eq:atom-feasibility_dr_beta}, we first  extend the constraints in \eqref{beta_hat_dr}--\eqref{gamma_hat_dr} to the uniform result in $\eta\in \cB_s$.
The same discretization argument, using the coefficient-norm
bounds just established, gives
\[
\begin{aligned}
    \sup_{\eta\in\cB_s}
    \|\wt\Gamma^\T A_{\wh V\eta}-\eta\|_\i
    &\le
    (2rB_\gamma+1)\epsilon+\lambda_\gamma',\\
    \sup_{\eta\in\cB_s}
    \left|\wt\beta^\T A_{\wh V\eta}-\|\eta\|_2^2\right|
    &\le
    2(rB_\beta+1)\epsilon+\lambda_\beta'.
\end{aligned}
\]
By writing 
$\theta_k=V\eta_k$ with $\eta_k=V^\T\theta_k\in\cB_s$, one has
\[
    \|\wh V\wh Q^\T\eta_k-\theta_k\|_2
    =
    \|(\wh V-V\wh Q)\wh Q^\T\eta_k\|_2
    \le E(\wh V).
\]
Consequently, \eqref{rate_TV_L2} and the preceding uniform
bounds in $\cB_s$ imply
\[
\begin{aligned}
    \|\wt\Gamma^\T A_{\theta_k}
    -\wh Q^\T V^\T\theta_k\|_\i
    &\le
    \|\wt\Gamma^\T
    (A_{\theta_k}-A_{\wh V\wh Q^\T\eta_k})\|_\i 
    +\|\wt\Gamma^\T A_{\wh V\wh Q^\T\eta_k}
    -\wh Q^\T\eta_k\|_\i\\
    &\le
    2rB_\gamma E(\wh V)
    +(2rB_\gamma+1)\epsilon+\lambda_\gamma'\\
    &\le\Lambda_\gamma'.
\end{aligned}
\]
Likewise, using
$\|\wh Q^\T\eta_k\|_2^2=\|\theta_k\|_2^2$, we obtain
\[
\begin{aligned}
    \left|\wt\beta^\T A_{\theta_k}-\|\theta_k\|_2^2\right|
    &\le
    2rB_\beta E(\wh V)
    +2(rB_\beta+1)\epsilon+\lambda_\beta' \le\Lambda_\beta'.
\end{aligned}
\]
Both inequalities hold for every $k\in[K]$ on the same event,
completing the proof.
\end{proof}

    \subsection{Subspace estimation under one softmax mixture}\label{app_sec_SS_one}
  
    We let $V$ contain the $s$ leading eigenvectors of $\sum_k \alpha_k \theta_k \theta_k^\T$ which has rank $s\le K$. 
    In view of \cref{thm_rate_score_SS}, it suffices to derive an explicit rate for $E(\wh V)$ in \eqref{def_error_V} with $\wh V$ constructed as follows. 
    
    Define
    \[
    \wh\Sigma_X:=\frac1p\sum_{j=1}^pX_jX_j^\T,\qquad
    \wh W:=\wh\Sigma_X^{-1/2}\left(\frac1N\sum_{i=1}^N Y_iY_i^\T\right)\wh\Sigma_X^{-1/2},
    \]
    where $\wh\Sigma_X^{-1/2}$ denotes the matrix square root of $\wh\Sigma_X^{-1}$. 
   For a given $s\le L$, let $\wh U$ contain the $s$ leading eigenvectors of $\wh W$, and let $\wh V$ contain the $s$ leading eigenvectors  of $\wh\Sigma_X^{-1/2}\wh U\wh U^\T\wh\Sigma_X^{-1/2}$.   

     When $s$ is unknown, one can estimate $s$ based on the following eigen-ratio criterion 
    \[
        \wh s ~ \in~   \argmax_{1\le r \le \bar r} {\lambda_r(\wh W) - \lambda_{r+1}(\wh W)\over \lambda_{r+1}(\wh W)-\lambda_{r+2}(\wh W)}
    \]
    for some pre-specified $\bar r < L-1$. We use the convention $a/0 = \infty$ for $a>0$ and $0/0=1$. This criterion is motivated by the spiked eigenvalue structure of the population-level counterpart of $\wh W$ given in \eqref{def_W_decomp}.
   
    The estimator $\wh V$   is proposed in \cite{BBNW} under 
    the following elliptical assumption on $\mu$ \begin{ass}
    \label{ass:elliptical}
        The density of  $\mu$  has the form
        $$
        f_\mu(x)=c\,|\Sigma|^{-1/2}g(x^\T\Sigma^{-1}x),\qquad \text{for }x\in\RR^L,$$
        where $\Sigma\succ0$ and $g\ge0$.
    \end{ass}
    The following theorem quantifies the estimation error of $\wh V$ as well as the estimation error of its corresponding  $\mathrm{TLC}^2$.

    \begin{thm}\label{thm_V_error_one_doc}
        Grant \cref{ass:theta,ass:mu,ass:rate,ass:elliptical} and 
         \begin{equation}\label{rank_signal}
            \lambda_s\left(\sum_{k=1}^K \alpha_k \theta_k\theta_k^\T\right) > c_0
        \end{equation}
        for some constant $c_0>0$.
        Fix any $\delta,\delta_0\in (0,1)$. Then with probability at least $1-\delta$, the following holds for sufficiently large $p$ and $N$,
        \[ 
       E(\wh V) =  \min_{Q\in \bbO_s}\|\wh V - VQ\|_\op ~\le~ 
        {
        C\{ L+\log(ep/\delta)\} \sqrt{\log(eL/\delta)} \over \sqrt{p^{1-\delta_0}\wedge N}}
        := e(L,p,N,\delta)
           \]
        where $C>0$ depends on $\Sigma$ and $g$ in \cref{ass:elliptical}, $\sigma_x$,   $c_0$ in \eqref{rank_signal}, $\log(1/\delta)$ and  $L$  polynomially. 

        Consequently, by choosing the $\epsilon =
        %= E(\wh V) \asymp 1/\sqrt{p^{1-\delta_0}\wedge N}$ 
        e(L,p,N,\delta)$
        in \cref{thm_rate_score_SS} with $|\cN_\epsilon'|\le (1+2/\epsilon)^s$, $\mathrm{TLC}^2$ based on  this $\wh V$ with $\lambda_\beta'$ and $\lambda_\gamma'$ given in \eqref{rate_lbd_dr} satisfies: with probability $1-\delta$,
        \[
            | \wh S(\wt \beta,\wt\Gamma) - S(G) | \lesssim {C_1\over \sqrt{p^{1-\delta_0}\wedge N}}
        \]
        where $C_1$ depends  polynomially on $L$, $\log(ep)$ and $\log(e/\delta)$ and  exponentially
        on $\sqrt s$, in addition to the fixed model parameters and $\delta_0$. 
    \end{thm}
    \begin{proof}
        Under \cref{ass:mu,ass:rate,ass:elliptical},  we have the following
standard facts \cite[Chapter~2]{FangKotzNg1990}  
\begin{align}\label{basic-ellip}
\Sigma_X :=\Cov(X)=\kappa \Sigma \quad \text{ 
 and } \quad
\EE\!\left[\exp({\theta^\T X})\right]=
\varphi(\theta^\T \Sigma \theta),
\end{align}
for some $\kappa>0$ and 
$\varphi:\RR\to\RR$ depending on $g$, that satisfies
\begin{align}
    \label{diag-ellip}
\text{$\varphi(q)>0$ and  $\varphi''(q)>0$ for all $q\ge 0$}
\end{align}
The key identity, proved in \cite{BBNW},% the appendix [REFERENCE],
\begin{align}
\label{KEY-ellip}
\frac{\EE[XX^\T \exp({X^\T\theta})]}
{\EE[ \exp({X^\T\theta})]}
&= 2\frac{\varphi'(q)}{\varphi(q)}\Sigma +
4\frac{\varphi''(q)}{\varphi(q)} \Sigma \theta\theta^\T \Sigma 
\end{align} with $q:=\theta^\T \Sigma \theta$,
shows that 
\begin{align}\label{algebra-ellip}
    \Lambda:= \sum_{k=1}^K\alpha_k\frac{\EE[XX^\T\exp(X^\T\theta_k)]}{\EE[\exp(X^\T\theta_k)]} =a\Sigma +\Sigma \sum_{k=1}^K B_k \theta_k \theta_k^\T \Sigma 
\end{align}
where we write \[
q_k:=\theta_k^\T \Sigma \theta_k,
\qquad
a:=2\sum_{k=1}^K\alpha_k\frac{\varphi'(q_k)}{\varphi(q_k)},
\qquad
B_k= 4\alpha_k\frac{\varphi''(q_k)}{\varphi(q_k)}.
\]
Then the whitened version takes the form
\begin{equation}\label{def_W_decomp}
W:=\Sigma_X^{-1/2}\Lambda \Sigma_X^{-1/2}=\frac{a}{\kappa}I_L+\frac{1}{\kappa^2}\Sigma_X^{1/2} \sum_{k=1}^K B_k \theta_k\theta_k^\T  \Sigma_X^{1/2} =: \frac{a}{\kappa}I_L + W^*.
\end{equation}
The first term in the decomposition of $W$ is a scalar multiple of the identity and therefore does not affect the eigenvectors.  In view of \eqref{diag-ellip} and $\kappa>0$, the second term $W^*$ is positive semidefinite of rank $s$.
% $$
%     s = \rank(\sum_{k=1}^K \alpha_k \theta_k\theta_k^\T).
% $$
Therefore the $s$ leading eigenvectors
$U = (u_1,\ldots,u_s)$ of $W$ are precisely those of $W^*$. Consequently, 
%their span is \[\textrm{span}(u_1,\ldots,u_K)=\textrm{span}(C^{1/2}\theta_1,\ldots,C^{1/2}\theta_K).\] Equivalently,
\begin{equation}\label{eq_span}
\textrm{span}(\Sigma_X^{-1/2}u_1,\ldots,\Sigma_X^{-1/2}u_s) =  \textrm{col}(\sum_k B_k \theta_k\theta_k^\T) =\textrm{col}(\sum_k \alpha_k \theta_k\theta_k^\T)
\end{equation}
where the last step uses \eqref{diag-ellip} again, and  we use $\textrm{col}(M)$ to denote the column space of any given matrix $M$.   This provides justification for the procedure of estimating the subspace in one-sample setting in \cref{sec_prac_SS}.
% We propose the following procedure:
% \begin{itemize}
% \item
% Estimate $\Lambda$, $C$ and $W$ by
% \[
% \wh\Lambda=\frac1n\sum_{i=1}^nY_iY_i^\T,\qquad\wh C=\frac1p\sum_{j=1}^pX_jX_j^\T,\qquad
% \wh W=\wh C^{-1/2}\wh\Lambda\,\wh C^{-1/2},
% \] respectively.
% \item Compute the $K$ leading eigenvectors $\wh u_1,\ldots,\wh u_K$ of
% $\wh W$.
%  \item Estimate 
%  the parameter subspace  by
% \(
% \textrm{span}(\wh C^{-1/2}\wh u_1,\ldots,\wh C^{-1/2}\wh u_K).
% \)
% \end{itemize}
% This  strategy recovers the
% span of $\theta_1,\ldots,\theta_K$ without requiring knowledge of the
% generator $g$, the radial function $\varphi$, or the proportionality
% constant $\kappa$.  These unknown quantities affect the eigenvalues of
% $W$, but not its informative eigenspace.

To analyze the estimation error, by writing $\wh\Sigma_Y = N^{-1} \sum_{i=1}^N Y_iY_i^\T$ for short, we observe that  
\begin{align*}
   &\wh\Sigma_Y - \Lambda  =\frac1N\sum_{i=1}^N Y_iY_i^\T - \Lambda \\ &= \sum_{j=1}^p (\wh\pi_j-\pi_j) X_jX_j^\T + \sum_{j=1}^p  \pi_j  X_jX_j^\T - \Lambda\\
    &= \sum_{j=1}^p (\wh\pi_j-\pi_j) X_jX_j^\T +\sum_{k=1}^K \alpha_k \left\{\frac{ \sum_{j=1}^p X_jX_j^\T \exp(X_j^\T \theta_k) }{ \sum_{j=1}^p \exp(X_j^\T \theta_k)} -  \frac{\EE[XX^\T\exp(X^\T\theta_k)]}{\EE[\exp(X^\T\theta_k)]}\right\}.
\end{align*}  
Using \cref{lem_pi_diff_XX} with $\cE_r$ in \eqref{def_event_r} and $r$ in \eqref{eq:radius} gives   
\begin{align}\label{bd_pi_diff_XX}
\PP_{\pi}\left\{
\left\|\sum_{j=1}^p(\wh\pi_j-\pi_j)X_jX_j^\T\right\|_\op
>
C  r^2\sqrt{\frac{\log(eL/\delta)}{N}}
\right\}
\le {\delta \over 4},
\end{align} 
For the second term in the above decomposition,  by invoking \citet[Lemmas 9, 11 \& 16]{BBNW} for some $\delta_0>0$ that can be taken arbitrarily small, we obtain that for all $p\ge p_0$ with $p_0$ depending on $\delta_0$ and $\sigma_x^2$,  with $\mu$-probability at least $1-\delta/4$, 
\begin{align}\label{bd_mu_pointwise}
 \max_{k\in[K]}\left\|\frac{ \sum_{j=1}^p X_jX_j^\T \exp(X_j^\T \theta_k) }{ \sum_{j=1}^p \exp(X_j^\T \theta_k)} -  \frac{\EE[XX^\T\exp(X^\T\theta_k)]}{\EE[\exp(X^\T\theta_k)]}\right\|_\op \lesssim \sqrt{L+\log(p/\delta) \over p^{1-\delta_0}}.
\end{align} 
Combining \eqref{bd_pi_diff_XX} and \eqref{bd_mu_pointwise} gives with probability $1-3\delta/4$,
\[
    \|\wh \Sigma_Y - \Lambda\|_\op ~ \lesssim  ~ \left[L+\log(ep/\delta)\right]\sqrt{\log(L/\delta)\over N} + \sqrt{L+\log(p/\delta) \over p^{1-\delta_0}} 
    %:=  {C_\Lambda\over \sqrt{N\wedge p^{1-\delta_0}}}.
\] 
Regarding $\wh \Sigma_X$, 
a standard  concentration result of the sample covariance matrix under the operator norm (see, \citet[Theorem 4.6.1]{vershynin2018high}) gives that  with probability at least $1-\delta/4$, 
\[
     \|\wh \Sigma_X -\Sigma_X\|_\op  ~ \lesssim  ~  \sqrt{L+\log(1/\delta) \over p} + {L+\log(1/\delta)\over p} .
\]
Let $\cE_X$ denote the event on which the above holds. Thus, when $C[L+\log(1/\delta)]\le p$ for some large $C>0$ and $\lambda_L(\Sigma_X) = \kappa \lambda_L(\Sigma) \ge c$, Weyl's inequality gives
 \[
   c/2 \le \lambda_L(\Sigma_X)/2\le \lambda_L(\wh\Sigma_X) \le  \lambda_1(\wh\Sigma_X) \le  2 \lambda_1(\Sigma_X) \le  2\sigma_x^2.
 \] 
 Furthermore, by using the fact
 \[ 
    \|A^{1/2}-B^{1/2}\|_{\op} \le
\frac{\|A-B\|_{\op}}
{\sqrt{\lambda_{\min}(A)}+\sqrt{\lambda_{\min}(B)}},
\qquad A,B\succ  0, 
 \]
 one has 
 \[
    \|\wh \Sigma_X^{1/2} - \Sigma_X^{1/2}\|_\op \le {\|\wh \Sigma_X-\Sigma_X\|_\op \over \sqrt{\lambda_L(\Sigma_X)}} \lesssim \sqrt{L+\log(1/\delta)\over p}
 \]
 as well as 
 \[
    \|\wh \Sigma_X^{-1/2} - \Sigma_X^{-1/2}\|_\op\le {\|\wh \Sigma_X^{-1}-\Sigma_X^{-1}\|_\op \over \sqrt{\lambda_L(\Sigma_X^{-1})}} \le \sigma_x  \|\wh \Sigma_X^{-1}(\wh\Sigma_X - \Sigma_X)\Sigma_X^{-1}\|_\op\lesssim \sqrt{L+\log(1/\delta)\over p}.
 \]
 Consequently, adding and subtracting terms gives 
 \begin{align}\label{bd_scaled_Y_W}\nonumber
    & \left\|\wh\Sigma_X^{-1/2} \wh\Sigma_Y  \wh\Sigma_X^{-1/2} -   W\right\|_\op \\\nonumber
    &\le  \|\wh\Sigma_X^{-1/2}-\Sigma_X^{-1/2}\|_\op  \|\wh\Sigma_Y\|_\op \left( \|\wh\Sigma_X^{-1/2}\|_\op + \|\Sigma_X^{-1/2}\|_\op\right) + \|\Sigma_X^{-1}\|_\op \|\wh \Sigma_Y - \Lambda\|_\op\\
    &\lesssim 
    { 
    \{L+\log(ep/\delta) \} \sqrt{\log(eL/\delta)} \over
    \sqrt{N\wedge p^{1-\delta_0}}
     }
    %{C_\Lambda\over \sqrt{N\wedge p^{1-\delta_0}}}.
 \end{align}
    Finally, since $\wh V$ and $V$ consist of, respectively, the eigenvectors of $\wh\Sigma_X^{-1/2}\wh U\wh U^\T \wh \Sigma_X^{-1/2} $ and $\Sigma_X^{-1/2} UU^\T \Sigma_X^{-1/2}$ from \eqref{eq_span}, by the variant of Davis-Kahan's theorem \citep{yu2015useful}, 
    \begin{align*}
        &\min_{Q\in \bbO_s}\|\wh V - VQ\|_\op \\ &\lesssim {\|\wh\Sigma_X^{-1/2}\wh U\wh U^\T \wh \Sigma_X^{-1/2} - \Sigma_X^{-1/2} UU^\T \Sigma_X^{-1/2}\|_\op \over \lambda_s(\Sigma_X^{-1/2} UU^\T \Sigma_X^{-1/2})}\\
        &\le \sigma_x^2  \|\wh\Sigma_X^{-1/2}-\Sigma_X^{-1/2}\|_\op \left( \|\wh\Sigma_X^{-1/2}\|_\op + \|\Sigma_X^{-1/2}\|_\op\right) +  \sigma_x^2 \|\Sigma_X^{-1}\|_\op \|\wh U\wh U^\T -UU^\T\|_\op\\
        &\lesssim \sqrt{L+\log(1/\delta)\over p} + \min_{Q\in \bbO_s}\|\wh U - UQ\|_\op. 
    \end{align*} 
    By another application of Davis-Kahan's theorem,  
    \[
        \min_{Q\in \bbO_s}\|\wh U - UQ\|_\op \lesssim {\|\wh\Sigma_X^{-1/2} \wh\Sigma_Y  \wh\Sigma_X^{-1/2} -   W\|_\op \over \lambda_s(W) - \lambda_{s+1}(W)} 
    \]
    and the fact that 
    \[
        \lambda_s(W) - \lambda_{s+1}(W) = \lambda_s(W^*) \ge {\lambda_L(\Sigma_X)\over \kappa^2}\lambda_s\left(\sum_k B_k \theta_k \theta_k^\T\right)\ge c
    \]
    for some $c$ depending on  $\kappa$, $\Sigma$, $\varphi$ and $c_0$ in \eqref{rank_signal}, invoking \eqref{bd_scaled_Y_W} concludes the proof.  
    \end{proof}

    The following lemma bounds the operator norm of $\sum_{j=1}^p (\wh\pi_j-\pi_j) X_j X_j^\T$.
    
    \begin{lemma}\label{lem_pi_diff_XX}
    Suppose $N\ge C_0 \log(14L /\delta)$ for some $C_0>0$. 
    With $\wh\pi$ given in \eqref{pi-hat}, for some universal constant $C>0$,
        \begin{align}\label{bd_pi_diff_XX_simple}
        \PP_\pi\left\{
        \left\|
        \sum_{j=1}^p(\wh\pi_j-\pi_j)X_jX_j^\T
        \right\|_\op
        >
        C \|\bX\|_{\i,2}^2\sqrt{\frac{\log(e L/\delta)}{N}}
        \right\}
        \le \delta.
        \end{align}
    \end{lemma}
    \begin{proof}
    Since $\wh\pi_j=N^{-1}\sum_{i=1}^N 1\{Y_i=X_j\}$ for $j\in [p]$, 
        we have
        \[
        \sum_{j=1}^p(\wh\pi_j-\pi_j)X_jX_j^\T =
        \frac{1}{N}\sum_{i=1}^N
        \left\{
        Y_iY_i^\T-\sum_{j=1}^p\pi_jX_jX_j^\T
        \right\}.
        \]
        Define
        \[
        Z_i := \frac{1}{N}\left\{
        Y_iY_i^\T-\sum_{j=1}^p\pi_jX_jX_j^\T
        \right\}.
        \]
        Then $Z_1,\ldots,Z_N$ are independent, symmetric, mean-zero
        random matrices. Since 
        %$\max_{j\in[p]}\|X_j\|_2\le R_X := \|\bX\|_{\i,2}$,
        \[
        \|Z_i\|_\op
        \le
        \frac{1}{N}\left\{
        \|Y_iY_i^\T\|_\op+
        \left\|\sum_{j=1}^p\pi_jX_jX_j^\T\right\|_\op
        \right\}
        \le \frac{2 \|\bX\|_{\i,2}^2}{N}
        \]
       and  moreover,
        \begin{align*}
        \left\|\sum_{i=1}^N\EE_\pi[Z_i^2]\right\|_\op
        &=
        \frac{1}{N}
        \left\|
        \EE_\pi\left[
        \left\{
        Y_iY_i^\T-\EE_\pi[Y_iY_i^\T]
        \right\}^2
        \right]
        \right\|_\op\\
        &\le
        \frac{1}{N}
        \left\|
        \EE_\pi[(Y_iY_i^\T)^2]
        \right\|_\op\\
        &=
        \frac{1}{N}
        \left\|
        \sum_{j=1}^p\pi_j\|X_j\|_2^2X_jX_j^\T
        \right\|_\op\\
&        \le \frac{\|\bX\|_{\i,2}^4}{N}
        \end{align*}
        we can 
        apply \cref{lem_bernstein_mat} to get
        \begin{align}
        \PP_\pi\left\{
        \left\|
        \sum_{j=1}^p(\wh\pi_j-\pi_j)X_jX_j^\T
        \right\|_\op>t
        \right\}
        \le
        14\exp\left\{
        -\frac{Nt^2/2}{\|\bX\|_{\i,2}^4+2\|\bX\|_{\i,2}^2t/3}
        +\log L
        \right\}.
        \label{bd_pi_diff_XX_raw}
        \end{align}
        The choice of $t$ below satisfies the lower bound required in
        \cref{lem_bernstein_mat}, since $U\le 2\|\bX\|_{\i,2}^2/N$ and
        $\sigma^2\le \|\bX\|_{\i,2}^4/N$.
        Taking
        \[
        t =
        \|\bX\|_{\i,2}^2\left\{
        \sqrt{\frac{2\log(14L/\delta)}{N}}
        +
        \frac{4\log(14L/\delta)}{3N}
        \right\}
        \]
        gives
        \begin{align*}
        \PP_\pi\left\{
        \left\|
        \sum_{j=1}^p(\wh\pi_j-\pi_j)X_jX_j^\T
        \right\|_\op
        >
        \|\bX\|_{\i,2}^2\left\{
        \sqrt{\frac{2\log(14L/\delta)}{N}}
        +
        \frac{4\log(14L/\delta)}{3N}
        \right\}
        \right\}
        \le \delta.
        \end{align*}
        Since $\log(L/\delta)/N$ is sufficiently small by the condition on $N$,  we obtain \eqref{bd_pi_diff_XX_simple}.
    \end{proof}

    \subsection{Subspace estimation under the ensemble  softmax mixtures}\label{app_sec_SS_ensemble}

    For the ensemble setting, let $V\in\RR^{L\times s}$ have orthonormal
    columns spanning $\operatorname{span}\{\theta_1,\ldots,\theta_K\}$,
    where $s\le K$ denotes the dimension of this subspace.  We propose to estimate $V$ by $\wh V \in \RR^{L\times s}$ which contains the $s$ leading eigenvectors of the following   $L\times L$ matrix 
    \begin{equation}\label{def_M_hat}
       \wh M := \sum_{d=1}^n {N_d\over N_d-1}\wh\Sigma_X^{-1} \bX^\T  \left(N_d\,  \wh\pi^{(d)}\wh\pi^{(d)\T} -  \diag(\wh\pi^{(d)})\right) \bX \wh\Sigma_X^{-1}.
    \end{equation} 
    This procedure differs from the one given in \cref{app_sec_SS_one}. We discuss its intuition below under the same \cref{ass:elliptical}.

    %We provide intuition for the estimator $\wh V$ based on the leading $r$ eigenvalues of the matrix in \eqref{def_M_hat} under \cref{ass:elliptical}.

    The matrix $\wh M \in \RR^{L\times L}$ in  \eqref{def_M_hat}  is an unbiased estimator (conditioning on $\bX$) of 
    \[
         \sum_{d=1}^n    \wh\Sigma_X^{-1} \bX^\T   N_d\,  \pi^{(d)}\pi^{(d)\T}   \bX \wh\Sigma_X^{-1}.
    \]
   Recall $\pi^{(d)}$ from \eqref{model_ensemble}. 
    It follows that, by the law of large numbers, $\mu$-almost surely as $p\to\infty$,
    \[ 
    \bX^\T\pi^{(d)} =
    \sum_{k=1}^K\alpha_k^{(d)}
    \frac{\sum_{j=1}^pX_j\exp(X_j^\T\theta_k)}
    {\sum_{j=1}^p\exp(X_j^\T\theta_k)} \to 
    \sum_{k=1}^K\alpha_k^{(d)}
    \frac{\EE[X\exp(X^\T\theta_k)]}
    {\EE[\exp(X^\T\theta_k)]}.
    \]
    For $\mu$ satisfying \cref{ass:elliptical} with covariance
    matrix $\Sigma_X=\kappa\Sigma$ and moment-generating function
    $
    \EE[\exp(X^\T\theta)]
    =
    \phi(\theta^\T\Sigma\theta).
    $
    Differentiating with respect to $\theta$ gives
    $
    \EE[X\exp(X^\T\theta)]
    =
    2\phi'(\theta^\T\Sigma\theta)\Sigma\theta.
    $
    It then follows that
    \[
    \Sigma_X^{-1}\bX^\T\pi^{(d)}
    \to 
    \sum_{k=1}^K\alpha_k^{(d)}c_k\,\theta_k,
    \qquad
    c_k
    :=
    {2\over \kappa}\frac{\phi'(\theta_k^\T\Sigma\theta_k)}
    {\phi(\theta_k^\T\Sigma\theta_k)}>0.
    \]
    As a result, by letting
    \[
    {\bm \Theta}=(\theta_1,\ldots,\theta_K),
    \qquad
    D_c=\diag(c_1,\ldots,c_K),
    \]  
    $\mu$-almost surely as $p\to\infty$,
    \begin{align*}
      \wh\Sigma_X^{-1}\bX^\T
    \left\{ 
    \sum_{d=1}^nN_d\,
    \pi^{(d)}\pi^{(d)\T}
    \right\}
    \bX\wh\Sigma_X^{-1} 
    &~ \to ~ 
    \sum_{d=1}^nN_d
    \left(
    \sum_{k=1}^K\alpha_k^{(d)}c_k\theta_k
    \right)
    \left(
    \sum_{k=1}^K\alpha_k^{(d)}c_k\theta_k
    \right)^\T\\
    & ~ =~  {\bm \Theta} D_c
    \left\{ 
    \sum_{d=1}^nN_d\,
    \alpha^{(d)}\alpha^{(d)\T}
    \right\}
    D_c {\bm \Theta}^\T.
    \end{align*}
    In particular, since $c_k>0$ for every $k\in[K]$, if 
    \[ 
    \sum_{d=1}^nN_d\,
    \alpha^{(d)}\alpha^{(d)\T}
    \succ0,
    \]
    then the $\mu$-almost sure limit of $\EE[\wh M\mid\bX]$ as
    $p\to\infty$ has column space
    $\operatorname{span}\{\theta_1,\ldots,\theta_K\}$, an
    $s$-dimensional subspace of $\RR^L$.

    When $s$ is unknown, we propose to select it by maximizing the ratio
    of consecutive eigenvalues of $\wh M$ truncated at zero. Set
    $\wh s=0$ if $\lambda_1(\wh M)\le0$; otherwise, choose
    \[
        \wh s\in\argmax_{1\le i<L}
        \frac{\lambda_i(\wh M)\vee0}
             {\lambda_{i+1}(\wh M)\vee0},
    \]
    with the conventions $a/0=+\infty$ for $a>0$ and $0/0=1$.
    In particular, if exactly $r<L$ eigenvalues of $\wh M$ are positive, then
    $\wh s=r$.

\section{Auxiliary lemmas}\label{app_sec_aux}

  The following lemma states upper bounds of the quadratic form of a sub-Gaussian random vector \citep{Hsu2012}. 
	
	\begin{lemma} \label{lem_quad}
		Let $\xi\in \RR^d$ be a  sub-Gaussian random vector with parameter $\gamma_\xi$. Then, for all symmetric positive semi-definite matrices $H$, and all $t\ge 0$, 
		\[
		\PP\left\{
		\xi^\T H\xi > \gamma_\xi^2\left(
		\sqrt{{\rm tr}(H)}+ \sqrt{2 t \|H\|_{\rm op} }
		\right)^2
		\right\} \le e^{-t}.
		\] 
	\end{lemma}

The following lemmas state the  vector-valued and matrix-valued Bernstein inequality; see, for instance, \citet[Theorem 3.1 and Corollary 4.1]{Minsker2017}. 
\begin{lemma}\label{lem_bernstein_vector}
    Let $X_1,\ldots, X_n$ be independent random vectors with zero mean and $\max_{i\in [n]}\|X_i\|_2 \le U$ almost surely. Denote $\sigma^2 := \sum_{i=1}^n \EE[\|X_i\|_2^2]$. Then for all $t \ge {1\over 6}(U + \sqrt{U^2 + 36\sigma^2})$, 
    \[
    \PP\left(
    \left\|\sum_{i=1}^n  X_i\right\|_2 > t
    \right) \le 28\exp\left(
    - {t^2 / 2\over \sigma^2 + Ut/3}
    \right).
    \]
\end{lemma}

\begin{lemma}[Matrix-valued Bernstein inequality]\label{lem_bernstein_mat}
    Let $\bX_1,\ldots, \bX_n \in \RR^{d\times d}$ be independent, symmetric random matrices with zero mean and $\max_{i\in [n]}\|\bX_i\|_\op \le U$ almost surely. Denote $\sigma^2 := \|\sum_{i=1}^n \EE[\bX_i^2]\|_\op$. Then for all $t \ge {1\over 6}(U + \sqrt{U^2 + 36\sigma^2})$, 
    \[
    \PP\left(
    \left\|\sum_{i=1}^n \bX_i\right\|_\op > t
    \right) \le 14   \exp\left(
    - {t^2 / 2\over \sigma^2 + Ut/3} + \log(d)
    \right).
    \]
\end{lemma}

\section{Additional simulation results}\label{app_sim}

We present additional simulation studies on subspace-assisted $\mathrm{TLC}^2$ in \cref{app_sim_SS} and comparison of E-TLC$^2$ with existing approaches for selecting near-pure mixtures in \cref{app_sim_anchor}.

\subsection{Effectiveness of subspace estimation assistance}\label{app_sim_SS}

We verify the effectiveness of using   subspace estimation in estimating $S(G)$ for large $L$ under both the
single and ensemble softmax mixture settings.

\begin{figure}[ht]
    \centering
    \includegraphics[width=.8\linewidth]{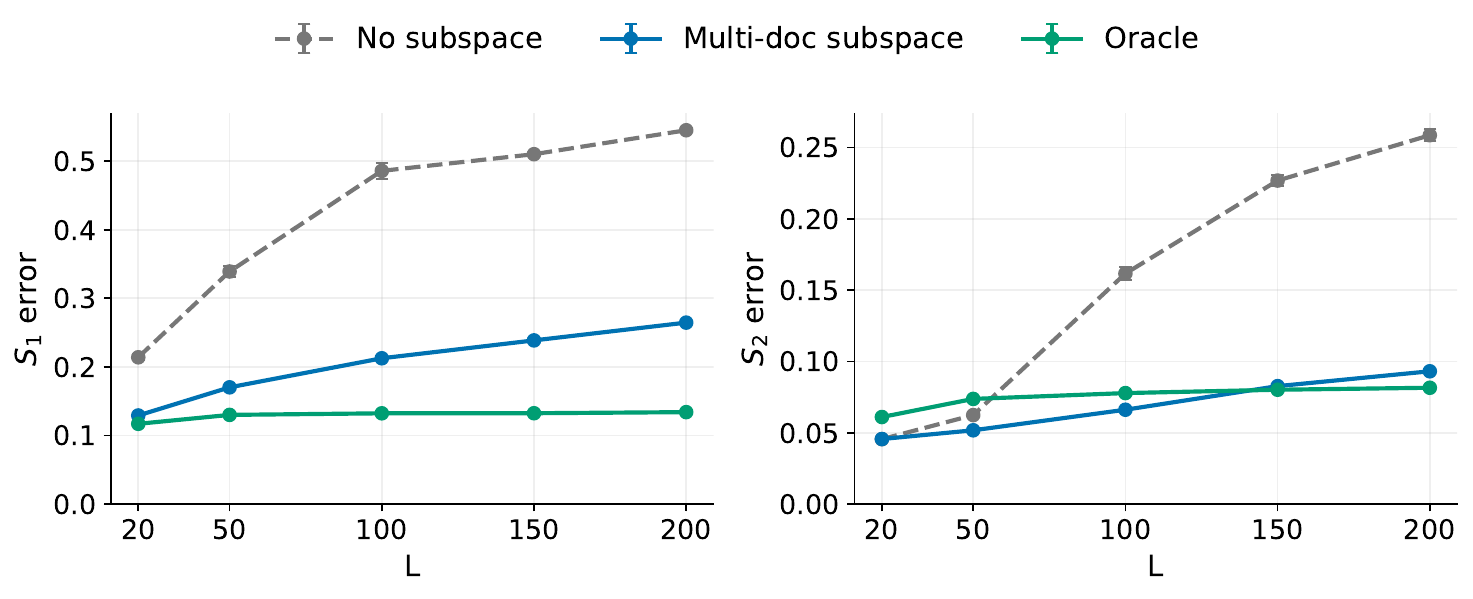}
    \caption{Estimation errors of $\mathrm{TLC}^2$ in the ensemble setting as $L$ varies, with $K=4$, $p=2000$, $D=100$, and $N=1000$ observations per document. The left and right panels report $\mathrm{Err}_{S_1}$ and $\mathrm{Err}_{S_2}$, respectively, comparing no subspace reduction, the estimated subspace, and the oracle subspace.}
    \label{fig:multi-subspace}
\end{figure}

For the ensemble setting, we set $K=4$ and draw
$\theta_1,\ldots,\theta_K$ independently and uniformly from $\cB$.
We generate $D=100$ mixtures according to \eqref{mod-def} using the $\eps$-pure weighting
scheme with $\eps=0.01$ and five $\eps$-pure mixtures per component.
We draw $N=1000$ independent observations from each mixture.
We compare three methods: no subspace reduction (called No subspace), the estimated
subspace from \cref{app_sec_SS_ensemble} (called Multi-doc subspace), and the oracle subspace
$\operatorname{span}\{\theta_1,\ldots,\theta_K\}$  (called Oracle).
The estimated subspace is spanned by the leading $K$ eigenvectors
of $\wh M$ in \eqref{def_M_hat} of \cref{app_sec_SS_ensemble}. For both subspace methods,
we compute $\wt\beta$ and $\wt\Gamma$ using
\eqref{beta_hat_dr} and \eqref{gamma_hat_dr} with the corresponding
orthonormal basis. 
To illustrate the effect of subspace reduction on the estimation
of the two terms in $S(G_d)$, we compute the errors
\[
\mathrm{Err}_{S_1}
:=\left|\wh S_{d,1}-\sum_{k=1}^K\alpha_k^{(d)}\|\theta_k\|_2^2\right|,
\qquad
\mathrm{Err}_{S_2}
:=\left|\wh S_{d,2}-\Bigl\|\sum_{k=1}^K \alpha_k^{(d)} \theta_k\Bigr\|_2^2\right|,
\]
where $\wh S_{d,1}$ and $\wh S_{d,2}$ are computed as in
\eqref{def_S_hat_d} using $\wt\beta$ and $\wt\Gamma$ in lieu of $\wh \beta$ and $\wh \Gamma$.
With $p=2000$, we vary $L\in\{20,50,100,150,200\}$ and report
these errors averaged over the $D$ documents in \cref{fig:multi-subspace}. The results show that the
estimated subspace substantially reduces the estimation errors
for both terms compared with no subspace reduction. It also
mitigates the effect of increasing $L$: $\mathrm{TLC}^2$ estimation is
carried out in the $K$-dimensional subspace, while the ambient
dimension $L$ affects the accuracy of subspace estimation.

\begin{figure}[ht]
    \centering
    \includegraphics[width=0.5\linewidth]{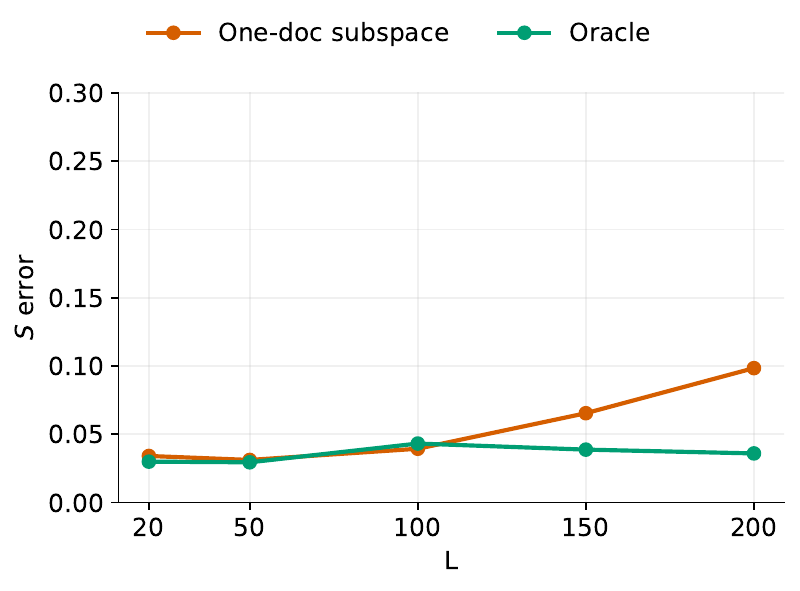}
    \caption{Overall score estimation error $\mathrm{Err}_{S}$ of $\mathrm{TLC}^2$ in the single mixture setting as $L$ varies, with $K=4$, $p=N=10000$, and uniform mixture weights. The estimated subspace is compared with the oracle subspace.}
    \label{fig:single-sub}
\end{figure}

For the single mixture setting, we use the same construction for
$\theta_1,\dots,\theta_K$ and set the mixture weights to
$\alpha_1=\cdots=\alpha_K=1/K$. We generate
$Y_1,\dots,Y_N$ independently from model \eqref{mod-def} with
$N=10000$. We compare the  estimated subspace described
in \cref{app_sec_SS_one} with the oracle subspace
$\operatorname{span}\{\theta_1,\dots,\theta_K\}$.
With $p=10000$, we vary $L\in\{20,50,100,150,200\}$ and report
only the overall $\mathrm{TLC}^2$ estimation error,
$\mathrm{Err}_{S}$, over 50 repetitions in \cref{fig:single-sub}. The  subspace estimator remains  effective as $L$ grows as long as  both $p$ and $N$ are large. It is worth mentioning that for such size of $p$ and $L$, computing $\mathrm{TLC}^2$ without using subspace estimation  assistance  is essentially infeasible.

\subsection{Comparison with existing approaches for selecting near-pure mixtures}\label{app_sim_anchor}

We compare E-$\mathrm{TLC}^2$ in \cref{alg:esharp} with   the existing procedure in \cite{Klopp2023} for selecting near-pure documents in the topic model literature, which is referred to as the Successive Projection Algorithm (SPA). It is based on the method in \cite{gillis2014snpa} developed in the context of non-negative matrix factorization.  Unlike E-$\mathrm{TLC}^2$, which aims to identify a collection of near-pure documents, SPA selects $K$ representative documents, ideally one for each topic. We report $\mathrm{Precision}$ and total coverage ($\mathrm{TC}$), defined in \eqref{def_metrics}, to evaluate selection performance. We also compare the accuracy of the recovered atoms for both methods, which in this case is defined as
\[
\mathrm{Err}_{\theta}
=
\min_{\iota}
\frac{1}{K}\sum_{k=1}^{K}
\|\widehat{\theta}_k-\theta_{\iota(k)}\|_2^2
\] 
over all permutations $\iota:  [K]\to [K].$
It is worth pointing out that since the two methods have different selection objectives, $\mathrm{TC}$ and $\mathrm{Err}_\theta$ provide more direct measures of their relative effectiveness for atom recovery.

We set $K=8$, $p=400$, $N=1000$, $L=10$, and $\eps=0.1$, and generate the data as described in \cref{sec_sim_ensemble}, except that we introduce a group of closely spaced atoms. Specifically, we first generate $\theta_1,\ldots,\theta_4$ independently and uniformly within $\cB$, and then construct $\theta_5,\ldots,\theta_8$ such that
$\|\theta_k-\theta_4\|_2\leq\delta$ for $k=5,\ldots,8$.
We vary the  parameter $\delta$ to examine the effect of atom separation on the two procedures.

The results of both methods are reported in \cref{tab:comparison_results}. Although SPA achieves higher $\mathrm{Precision}$ throughout, E-$\mathrm{TLC}^2$ consistently achieves higher $\mathrm{TC}$ and substantially smaller $\mathrm{Err}_\theta$, particularly when $\delta$ is small. For example, at $\delta=0.05$, E-$\mathrm{TLC}^2$ achieves $\mathrm{TC}=0.941$ and $\mathrm{Err}_\theta=0.110$, compared with $0.618$ and $0.408$, respectively, for SPA. This indicates that SPA can select nearly-pure documents while failing to recover all underlying topics when several atoms are closely spaced. In contrast, E-$\mathrm{TLC}^2$ remains effective at recovering the underlying atoms even under weak separation with small $\delta$.

We also observe that the precision of E-$\mathrm{TLC}^2$ generally improves as $\delta$ increases. This is consistent with the behavior of  
$
S(G)=(1/2)\sum_{k,\ell}\alpha_k\alpha_\ell
\|\theta_k-\theta_\ell\|_2^2.
$
For pure mixtures, $S(G)=0$ regardless of atom separation, whereas greater separation can increase the scores of non-pure mixtures, making them easier to distinguish from near-pure mixtures. This does not contradict the results in \cref{fig:one_doc_score_error}, which concern the estimation accuracy of $\wh S(G)$ relative to $S(G)$ rather than its ability to distinguish near-pure from non-pure mixtures in the ensemble setting. %Thus, even when the estimation error is relatively insensitive to atom separation, the selection performance can improve as the underlying mixtures become more distinguishable.
 
\begin{table}[htbp]
\centering
\caption{Near-pure mixtures selection and topic recovery performance for different $\delta$. The numbers are averages over 70 replications.}
\label{tab:comparison_results}
\small
\begin{tabular}{c|ccc|ccc}
\hline
& \multicolumn{3}{c|}{E-$\mathrm{TLC}^2$} & \multicolumn{3}{c}{SPA} \\
$\delta$ & $\mathrm{Precision}$ & $\mathrm{TC}$ & $\mathrm{Err}_{\theta}$
& $\mathrm{Precision}$ & $\mathrm{TC}$ & $\mathrm{Err}_{\theta}$ \\
\hline
0.05 & 0.776 & 0.941 & 0.110 & 0.986 & 0.618 & 0.408 \\
0.10 & 0.780 & 0.952 & 0.099 & 0.991 & 0.657 & 0.364 \\
0.20 & 0.801 & 0.925 & 0.107 & 0.991 & 0.671 & 0.344 \\
0.50 & 0.865 & 0.952 & 0.076 & 1.000 & 0.838 & 0.183 \\
\hline
\end{tabular}
\end{table}

\section{Supplement to the real data analysis in \cref{sec_real_data}}\label{app_sec_supp_real_data}

We provide details of fitting the ensemble softmax mixture model using EM in \cref{subsec-fit-ensem-ssm}. \cref{subsec-msd-sensitivity} examines the sensitivity of our method to the overspecified number of components used in EM for subspace estimation.

\subsection{Fitting ensemble softmax mixtures}
\label{subsec-fit-ensem-ssm}
Let
$\omega=(\alpha^{(1)},\ldots,\alpha^{(D)},
\theta_1,\ldots,\theta_K)$
collect the ensemble parameters, where
$\alpha^{(d)}\in \Delta^{K-1}$, and let $\omega^*$ denote the true
parameter values. Write $\pi_j^{(d)}(\omega)$ for the
probabilities in \eqref{model_ensemble} and
$\pi_j^{(d)*}:=\pi_j^{(d)}(\omega^*)$.
For fixed shared features $\cX=\{x_1,\ldots,x_p\}$,
the population negative cross-entropy, weighted by document sample sizes, is
\[
 \ell(\omega)
:=
\frac{1}{\sum_{m=1}^D N_m}
\sum_{d=1}^D N_d
\sum_{j=1}^p
\pi_j^{(d)*}\log\pi_j^{(d)}(\omega).
\]
Nonnegativity of the Kullback--Leibler divergence implies
that $\ell(\omega)$ is maximized at $\omega=\omega^*$.
Replacing the unknown probabilities by their empirical
frequencies gives the criterion based on empirical negative cross-entropy 
\begin{equation}\label{loss_SMM}
\max_{\omega}\,\wh\ell(\omega),
\qquad
\wh\ell(\omega)
:=
\frac{1}{\sum_{m=1}^D N_m}
\sum_{d=1}^D N_d
\sum_{j=1}^p
\wh\pi_j^{(d)}\log\pi_j^{(d)}(\omega),
\end{equation}
where the maximization is over the admissible mixture parameters.

Because $\wh\ell$ is generally nonconcave, we use a
generalized EM algorithm based on the surrogate
\[
\begin{aligned}
\wh Q(\omega\mid\omega')
&:=
\frac{1}{\sum_{m=1}^D N_m}
\sum_{d=1}^D N_d
\sum_{j=1}^p \wh\pi_j^{(d)}
\sum_{k=1}^K g_{\omega'}^{(d)}(k,x_j)  
\left\{\log\alpha_k^{(d)}
+\log A_{\theta_k}(x_j)\right\},
\end{aligned}
\]
where the posterior component probabilities are
\begin{equation}\label{SMM_E}
g_{\omega}^{(d)}(k,x_j)
=
\frac{\alpha_k^{(d)}A_{\theta_k}(x_j)}
{\sum_{a=1}^K\alpha_a^{(d)}A_{\theta_a}(x_j)},
\qquad d\in[D],\quad j\in[p],\quad k\in[K].
\end{equation}
The EM inequality gives
\[
\wh\ell(\omega)-\wh\ell(\omega')
\ge
\wh Q(\omega\mid\omega')
-\wh Q(\omega'\mid\omega'),
\]
so any increase in the surrogate yields at least the same
increase in the log-likelihood.

For fixed $\omega'$, the surrogate is concave in $\omega$.
We maximize it over the mixing weights and take a gradient
step for each atom. Using square brackets to denote iteration
indices, the updates for $s\ge0$, $d\in[D]$ and $k\in[K]$ are
\begin{equation}\label{SMM_M}
\begin{aligned}
\wh\alpha_k^{(d)[s+1]}
&=
\sum_{j=1}^p
\wh\pi_j^{(d)}
g_{\wh\omega^{[s]}}^{(d)}(k,x_j),\\
\wh\theta_k^{[s+1]}
&=
\wh\theta_k^{[s]}
+
\frac{\eta_k}{\sum_{m=1}^D N_m}
\sum_{d=1}^D N_d
\sum_{j=1}^p
\wh\pi_j^{(d)}
g_{\wh\omega^{[s]}}^{(d)}(k,x_j) 
\left\{
x_j-
\sum_{\ell=1}^p
x_\ell A_{\wh\theta_k^{[s]}}(x_\ell)
\right\},
\end{aligned}
\end{equation} 
where $\eta_k>0$ is some chosen step size.  

To initialize the atoms, let $\wh M$ be the matrix defined
in \eqref{def_M_hat} in \cref{app_sec_SS_ensemble}, and let
$\wh V_K=(\wh v_1,\ldots,\wh v_K)$ contain its leading
$K$ eigenvectors.
For each random start, draw $u_1,\ldots,u_K$
independently and uniformly from the unit sphere
$\Sp^{L-1}$ and set 
\begin{equation}\label{def_theta_init}
\wh \alpha_k^{(d)[0]} = {1\over K},\qquad \wh\theta_k^{[0]}
=
\frac{\wh V_K\wh V_K^\T u_k}
{\|\wh V_K\wh V_K^\T u_k\|_2},
\qquad k\in[K].
\end{equation}
We run the algorithm from multiple random initializations
and retain the converged solution with the largest $\wh\ell$.

\subsection{Sensitivity to the overspecified number of components}
\label{subsec-msd-sensitivity}
\begin{figure}[ht]
    \centering
    \includegraphics[width=\linewidth]{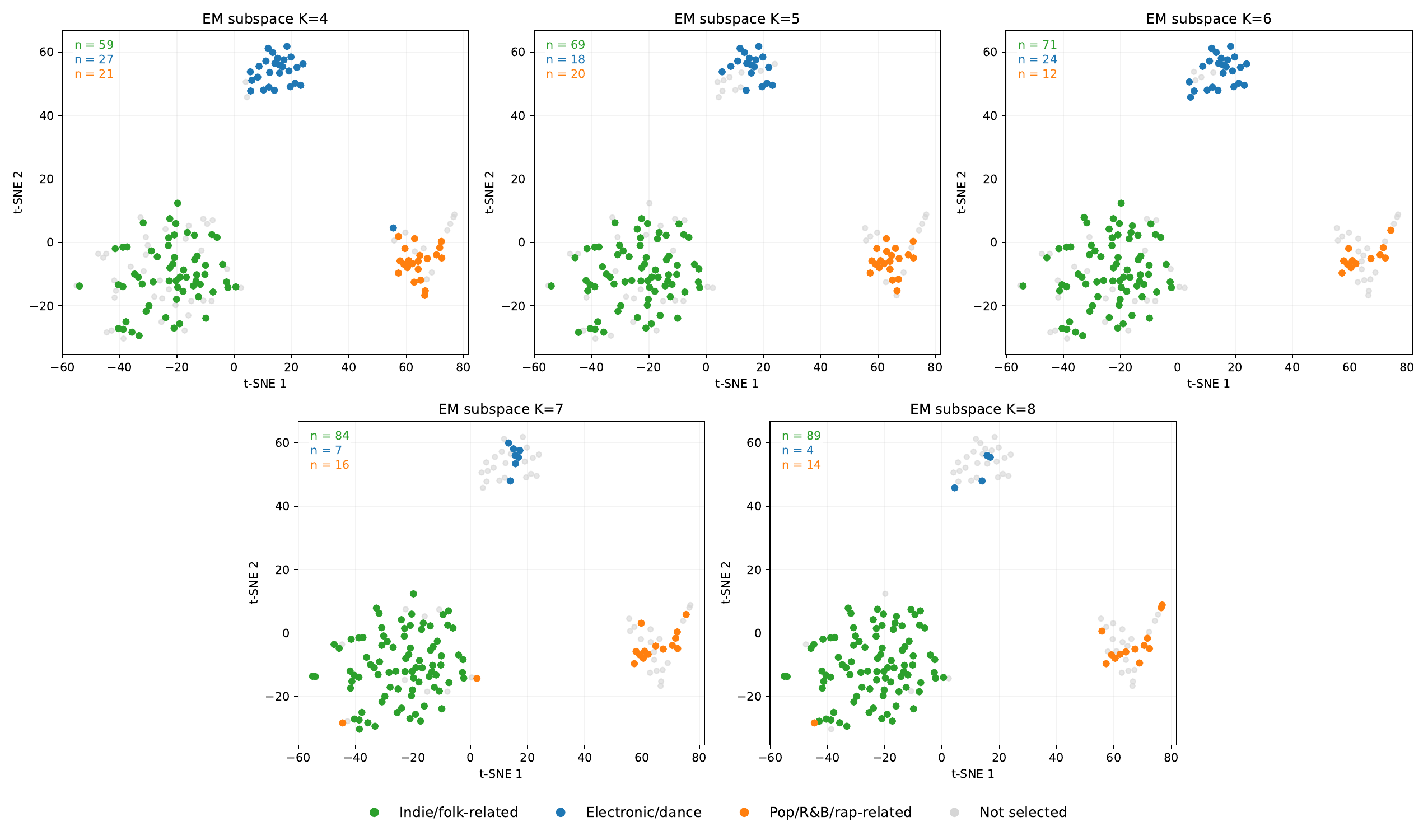}
    \caption{Comparison of user selection using EM-based subspaces with
$K\in\{4,5,6,7,8\}$. All panels share a single t-SNE embedding,
computed using cosine distance between the constrained MLEs of
the union of selected users. Colored points represent the 30\%
of users with the lowest scores for the corresponding $K$,
with colors indicating three-center $k$-means assignments fitted
separately in the original 53-dimensional parameter space.
Gray points represent users in the union who are not selected
under that $K$.}
    \label{fig:MSD}
\end{figure}

To assess sensitivity to the overspecified number of components $K$ used for subspace estimation in \cref{sec_real_data}, we repeat the analysis using subspaces estimated by ensemble EM with $K\in\{4,6,7,8\}$ and compare the results with the baseline choice $K=5$. For each estimated subspace, we compute TLC$^2$ for each user, select the 30\% of users with the smallest scores, and obtain their dominant-atom estimates. We then cluster these estimates using $k$-means with $K_0=3$. To facilitate comparison, we compute a single two-dimensional t-SNE embedding, based on cosine distances, of the dominant-atom estimates for the union of all selected users. In each panel of \cref{fig:MSD}, users selected under the corresponding choice of $K$ are colored by their $k$-means cluster assignments, while the remaining users are shown in gray. Each user therefore occupies the same position across all panels. The three clusters exhibit broadly similar structure across the different choices of $K$. The top ranked songs associated with each cluster center also yield interpretations consistent with our earlier findings: indie/folk, electronic/dance, and pop/R\&B listening profiles. These results suggest that our method is relatively insensitive to the overspecified $K$ used for subspace estimation.

\end{document}